\documentclass[preprint,prd,showpacs,preprintnumbers,amsmath,amssymb,aps,floatfix,nofootinbib]{revtex4-1}
\usepackage[colorlinks=true,linkcolor=blue,filecolor=blue,urlcolor=blue,citecolor=blue,pdftex=true,plainpages=false]{hyperref}
\usepackage{amsmath,amssymb,amsbsy,amsfonts,bm}
\usepackage{graphicx}
\usepackage{rotating}
\usepackage[mathscr]{euscript}
\usepackage{multirow}

\usepackage{color}
\usepackage{bm}
\definecolor{purple}{rgb}{1,0,1}
\definecolor{grey}{rgb}{0.6,0.6,0.6}
\newcommand{\strike}[1]{}  
\newcommand{\cut}[1]{{\textcolor{grey}{#1}}}            

\usepackage{graphicx}
\usepackage{subfigure}

\def\chpt{\raise0.4ex\hbox{$\chi$}PT}
\def\schpt{S\raise0.4ex\hbox{$\chi$}PT}
\def\rschpt{rS\raise0.4ex\hbox{$\chi$}PT}
\def\figref#1{Fig.~\ref{fig:#1}}
\def\Figref#1{Figure~\ref{fig:#1}}
\def\figrefs#1#2{Figs.~\ref{fig:#1} and \ref{fig:#2}}

\def\figrefthree#1#2#3{Figs.~\ref{fig:#1}, \ref{fig:#2}, and \ref{fig:#3}}

\def\secref#1{Sec.~\ref{sec:#1}}
\def\secrefs#1#2{Secs.~\ref{sec:#1} and \ref{sec:#2}}

\def\Secref#1{Section~\ref{sec:#1}}
\def\tabref#1{Table~\ref{tab:#1}}

\def\gtwid{{\,\raise.3ex\hbox{$>$\kern-.75em\lower1ex\hbox{$\sim$}}\,}}
\def\ltwid{{\,\raise.3ex\hbox{$<$\kern-.75em\lower1ex\hbox{$\sim$}}\,}}

\def\Tr{{\rm Tr}}

\def\ie{{\it i.e.},\ }
\def\eg{{\it e.g.},\ }
\def\etal{{\it et al.}}

\def\vs{{\it vs.}\ }

\def\cD{{\cal D}}

\def\cI{{\cal I}}

\def\cM{{\cal M}}

\def\cO{{\cal O}}

\def\cQ{{\cal Q}}

\def\cV{{\cal V}}

\def\rcite#1{Ref.~\cite{#1}}
\def\rcites#1{Refs.~\cite{#1}}
\def\Rcite#1{Reference~\cite{#1}}
\def\eqn#1{\label{eq:#1}}
\def\Equation#1{Equation~(\ref{eq:#1})}

\def\eq#1{Eq.~(\ref{eq:#1})}

\def\eqs#1#2{Eqs.~(\ref{eq:#1}) and (\ref{eq:#2})}
\def\eqsthree#1#2#3{Eqs.~(\ref{eq:#1}), (\ref{eq:#2}) and (\ref{eq:#3})}

\def\msbar{{\overline{\rm MS}}}

\newcommand{\qedl}{\ensuremath{{\rm QED}_{L}}}
\newcommand{\qedtl}{\ensuremath{{\rm QED}_{TL}}}

\newcommand{\bi}{\begin{itemize}}
\newcommand{\ei}{\end{itemize}}

\newcommand{\be}{\begin{equation}}
\newcommand{\ee}{\end{equation}}

\newcommand{\bea}{\begin{eqnarray}}
\newcommand{\eea}{\end{eqnarray}}

\newcommand{\bt}[1]{\begin{table}[!t] \begin{center}\begin{tabular}{#1} \hline\hline  \\[-0.5em]}
\newcommand{\et}[2]{\hline\hline \end{tabular} \end{center} \caption{#1} \label{#2} \end{table}}

\def\cpt{\raise0.4ex\hbox{$\chi$}PT}
\def\scpt{S\raise0.4ex\hbox{$\chi$}PT}
\def\rscpt{rS\raise0.4ex\hbox{$\chi$}PT}

\newcommand{\KEM}{\ensuremath{(M^2_{K^0})^\gamma}}
\newcommand{\ek}{\ensuremath{\epsilon_{K^0}}}
\newcommand{\pizEM}{\ensuremath{(M^2_{\pi^0})^\gamma}}
\newcommand{\uuEM}{\ensuremath{(M^2_{uu'})^\gamma}}
\newcommand{\ddEM}{\ensuremath{(M^2_{dd'})^\gamma}}
\newcommand{\DelK}{\ensuremath{\Delta M^2_{K^0}}}

\newcommand{\Deluu}{\ensuremath{\Delta M^2_{uu'}}}
\newcommand{\Deldd}{\ensuremath{\Delta M^2_{dd'}}}
\newcommand{\Delxy}{\ensuremath{\Delta M^2_{xy}}}
\newcommand{\KPZ}{\ensuremath{(M^2_{K^+}-M^2_{K^0})^\gamma}}
\newcommand{\pisplit}{\ensuremath{(M^2_{\pi^+}-M^2_{\pi^0})^{\textrm expt}}}
\newcommand{\pisplitEM}{\ensuremath{(M^2_{\pi^+}-M^2_{\rm{``}\pi^0\rm{"}})^\gamma}}

\newcommand{\deltae}{\ensuremath{\Delta_{\rm EM}}}

\newcommand{\dem}{\ensuremath{e^2{\Delta}_{\rm EM}}}

\begin{document}

\title{Lattice computation of the electromagnetic contributions to kaon and pion masses}

\author{S.~Basak}
\affiliation{School of Physical Sciences, NISER Bhubaneswar, Orissa 752050, India}      

\author{A.~Bazavov}
\affiliation{ Department of Computational Mathematics, Science and Engineering
and Department of Physics and Astronomy,
Michigan State University, East Lansing, MI 48824, USA }

\author{ C.~Bernard}
\email{cb@wustl.edu}
\affiliation{ Department of Physics, Washington University, St. Louis, MO 63130, USA}

\author{ C.~DeTar}
\author{ L.~Levkova}
\affiliation{ Department of Physics and Astronomy, University of Utah, Salt Lake City, UT 84112, USA}

\author{ E.~Freeland}
\affiliation{ Liberal Arts Department, School of the Art Institute of Chicago, Chicago, IL 60603, USA }

\author{ Steven~Gottlieb}
\email{sg@indiana.edu}
\author{ A.~Torok}
\altaffiliation[Present address:~]{ThermoFisher Scientific, Hillsboro, OR 97124, USA\;}
\affiliation{ Department of Physics, Indiana University, Bloomington, IN 47405, USA}

\author{ U.M.~Heller}
\affiliation{ American Physical Society, One Research Road, Ridge, NY 11961, USA}

\author{ J.~Laiho}
\affiliation{ Department of Physics, Syracuse University, Syracuse, NY  13244, USA}

\author{ J.~Osborn}
\affiliation{ ALCF, Argonne National Laboratory, Argonne, IL 60439, USA}

\author{ R.L.~Sugar}
\affiliation{ Physics Department, University of California, Santa Barbara, CA 93106, USA}

\author{ D.~Toussaint}
\affiliation{ Physics Department, University of Arizona Tucson, AZ 85721, USA}

\author{ R.S.~Van~de~Water}
\author{ R.~Zhou }
\affiliation{ Theoretical Physics Department, Fermi National Accelerator Laboratory, Batavia 60510, USA}

\collaboration{MILC Collaboration}
\date{\today}

\begin{abstract}
We present a lattice calculation of the electromagnetic (EM)
effects on the masses of light pseudoscalar mesons.  The simulations employ  2+1 dynamical flavors of asqtad QCD quarks,  and quenched photons. Lattice
spacings vary from $\approx\!0.12\;{\rm fm}$ to $\approx\!0.045\;{\rm fm}$.  
We compute the quantity $\epsilon$, which parameterizes the corrections
to Dashen's theorem for the $K^+$--$K^0$ EM mass splitting, as well
as \ek, which parameterizes the EM contribution to the mass of the $K^0$ itself.  
An extension of the nonperturbative EM renormalization
scheme introduced by the BMW group is used in separating EM effects from isospin-violating
quark mass effects.   We correct for leading finite-volume effects in 
our realization of lattice electrodynamics in chiral perturbation theory, and remaining
finite-volume errors are relatively small.
While electroquenched
effects are under control for $\epsilon$, they are 
estimated only qualitatively for \ek, and constitute one of the largest
sources of uncertainty for that quantity.
We find $\epsilon = 0.78(1)_{\rm stat}({}^{+\phantom{1}8}_{-11})_{\rm syst}$ and
$\ek=0.035(3)_{\rm stat}(20)_{\rm syst}$.
\strike{ $\KEM = 44(3)_{\rm stat}(25)_{\rm syst}\;({\rm MeV})^2$.}  We then use these results on
 2+1+1 flavor pure QCD highly improved staggered quark (HISQ) ensembles and find $m_u/m_d  = 0.4529(48)_\mathrm{stat}(
\null_{-\phantom{1}67}^{+150})_\mathrm{syst}$. 
\end{abstract}
\maketitle

\section{Introduction \label{sec:intro}}

The mass splitting between  the charged and neutral kaons, $K^\pm$ and $K^0$, arises from two effects that give
comparable contributions: the mass difference between up and down quarks,  and electromagnetism.
If the electromagnetic (EM)  contributions can be determined and removed from the experimental meson masses, 
the resulting pure-QCD
masses can then be used as input to a lattice QCD calculation to determine the light quark masses,
and in particular the ratio $m_u/m_d$, a fundamental parameter
of the standard model which measures the strength of strong isospin violations.

The size of the EM contributions to the  $K^\pm$--$K^0$ mass splitting is a long-standing issue.
Almost fifty years ago, Dashen \cite{Dashen:1969eg} showed that
the EM splitting of the charged and neutral kaons is equal to that of the pions in leading
order (LO) of chiral SU(3)$\times$SU(3) symmetry.  In other words, at LO,  
$(M^2_{K^\pm}-M^2_{K^0})^\gamma=(M^2_{\pi^\pm}-M^2_{\pi^0})^\gamma$,  where the 
superscript $\gamma$ denotes the EM contribution, \ie the difference between the quantity in the
real world and in a world where all quark charges are set to zero (keeping renormalized quark masses
unchanged). However, it  has been known for some time that the corrections
to this lowest order result are large; see, for example, \rcite{Aoki:2016frl} for a pedagogical review.  These
corrections can be estimated in a variety of continuum phenomenological models \cite{continuum-models}.
The model results differ considerably, however, and do not allow one to make controlled estimates of the
systematic errors.  Indeed, in lattice determinations of
$m_u/m_d$ that employ phenomenological estimates of EM contributions 
\cite{Aubin:2004fs,RMP,qrat,Laiho:2011np}, 
the error coming from the range of EM estimates dominates all other systematic errors.

Direct lattice calculations of the EM contribution to the kaon splittings  
 can greatly reduce the uncertainties.  This approach was pioneered by Duncan, Eichten, and Thacker
 \cite{Duncan:1996xy} in the quenched approximation of QCD, and has been applied in full QCD more recently
 by several groups \cite{Blum:2007cy,Blum:2010ym,BMW11,EM12,Basak:2013iw,Basak:2014vca,Basak:2016jnn,Horsley:2015eaa,Fodor:2016bgu}.
 Here we report on
our lattice QCD+QED computation of $(M^2_{K^\pm}-M^2_{K^0})^\gamma$. 
We then apply our result to compute $m_u/m_d$ in a pure QCD simulation.

There is an alternative approach to calculating EM effects on the lattice \cite{deDivitiis:2013xla, 
Giusti:2017dmp} in which one expands out QED and isospin-violating interactions to 
$\cO(\alpha_{\rm EM},m_u-m_d)$ (where $\alpha_{\rm EM}$ is the fine structure constant) and then computes the resulting matrix elements in isospin-conserving pure 
QCD.   We do not discuss this approach further here, but simply note that the existence of two independent methods 
makes possible important cross checks on the results and errors of both.  See \rcite{Portelli:2015wna}  for a review
that covers both approaches.


In lattice simulations of QCD+QED,  both the QCD and QED should in principle be {\it unquenched},
\ie include all contributions from virtual sea-quark loops.     However, Bijnens and Danielsson  \cite{Bijnens:2006mk} have shown
that QED quenching effects for mass differences such as $(M^2_{K^\pm}-M^2_{K^0})^\gamma$ 
are computable through next-to-leading order (NLO) in SU(3)$\times$SU(3) chiral perturbation theory, with no
dependence on unknown low energy constants (LECs).    In other words, the sea quarks may be taken to be electrically neutral
in the simulation, and the effects of their charges may be restored, correct to NLO, after the fact.  We take advantage of
this result here and simulate full, unquenched QCD + quenched QED (the {\it electroquenched approximation}) in order to determine the kaon EM splittings.
Since the QED part of the simulation is quenched, we need only to calculate
valence-quark propagators in a background consisting of pure unquenched QCD and quenched EM fields, which are
free fields and therefore easily generated.  For the pure QCD backgrounds, we use our large data set of ensembles
generated with 2+1 flavors of asqtad staggered quarks \cite{RMP}.  We have 
added a number of additional ensembles to better study finite-volume
effects.

One may parameterize the kaon EM splitting by \cite{Aoki:2016frl}
 \begin{equation}
\eqn{eps-def}
 \epsilon \equiv \frac{(M^2_{K^\pm}-M^2_{K^0})^\gamma - (M^2_{\pi^\pm}-M^2_{\pi^0})^\gamma}
 {(M^2_{\pi^\pm}-M^2_{\pi^0})^{\textrm{expt}}}\ ,
\end{equation}
where the experimental pion splitting is used in the denominator, rather than the EM pion splitting.  The two are equal up to 
isospin-violating effects, which are $\cO((m_u-m_d)^2)$,  and 
therefore small.  
Determining the EM contribution to the mass of the true $\pi^0$ is costly,  however, since it has quark-line disconnected
EM diagrams even in the isospin limit.  Instead, we drop the disconnected diagrams, which are expected to be 
small, and simply find the RMS average mass of $u\bar u$ and $d\bar d$ mesons. We call
the pion obtained in this manner the ``$\pi^0$.'' 
Both the true $(M^2_{\pi^0})^\gamma$ and our  $(M^2_{\rm{``}\pi^0\rm{"}})^\gamma$ are 
small because EM contributions to neutral mesons vanish in the chiral limit.  For the true $\pi^0$, this is required by
Dashen's arguments \cite{Dashen:1969eg}, and may be seen explicitly in chiral perturbation theory (\chpt) including
EM effects \cite{Urech:1994hd}.  For the ``$\pi^0$,''  a simple argument in 
partially quenched \chpt, given below in \secref{pi0}, shows that $(M^2_{\rm{``}\pi^0\rm{"}})^\gamma$  also vanishes in the chiral limit. 
This means that the disconnected EM contributions that we are neglecting are themselves small.  
(An alternative diagramatic proof of the small size  of the disconnected terms has been given previously in \rcite{deDivitiis:2013xla}.)
Further, Zweig's rule suggests that the mass contribution from the disconnected diagrams is in fact still smaller than
either $(M^2_{\pi^0})^\gamma$ or  $(M^2_{\rm{``}\pi^0\rm{"}})^\gamma$ separately.   

Summarizing, we use  \begin{equation}
\eqn{our-eps}
 \epsilon \cong \frac{(M^2_{K^\pm}-M^2_{K^0})^\gamma - (M^2_{\pi^\pm}-M^2_{\rm{``}\pi^0\rm{"}})^\gamma}
 {(M^2_{\pi^\pm}-M^2_{\pi^0})^{\textrm{expt}}}
\end{equation}
to compute $\epsilon$.  The systematic error coming from using the $\rm{``}\pi^0\rm{"}$ will of course need
to be estimated.

An alternative estimate of $\epsilon$ is also possible if we employ the experimental EM pion splitting in 
the numerator of \eq{eps-def} instead of our computed $\pi$ splitting.  This estimate is then independent of any assumptions about the disconnected diagrams in the  ``$\pi^0$."
For a test of  systematic effects in the calculation of $\epsilon$, we can therefore look at
\begin{equation}
\eqn{epsp-def}
 \epsilon'  \equiv \frac{(M^2_{K^\pm}-M^2_{K^0})^\gamma - (M^2_{\pi^\pm}-M^2_{\pi^0})^\textrm{expt}}
 {(M^2_{\pi^\pm}-M^2_{\pi^0})^{\textrm{expt}}}\ ,
 \end{equation}

 In \rcite{Aoki:2016frl},  the contribution to the pion splitting coming from quark masses (\ie the splitting
 that would be present in QCD alone) is defined to be $\epsilon_m (M^2_{\pi^\pm}-
 M^2_{\pi^0})^{\textrm{expt}}$.  
 Then \eqs{eps-def}{epsp-def} imply
 \begin{equation}
\eqn{eps-epsp-diff}
 \epsilon' = \epsilon-\epsilon_m\ .
 \end{equation}
 At NLO in \chpt, $\epsilon_m=0.04$ \cite{Gasser-Leutwyler}.   \Rcite{Aoki:2016frl} adds a conservative error and quotes  $\epsilon_m=0.04(2)$.  In our calculation, $\epsilon-\epsilon'$ appears to be positive. 
 However, because our systematic errors in both  $\epsilon$ and $\epsilon'$ are significantly larger than 
 0.04, we are only able to use  the difference $\epsilon-(\epsilon'+0.04)$ as one estimate of those errors, and have 
 nothing to  report about $\epsilon_m$ itself.

We also calculate the EM contribution to the squared mass of the neutral kaon,
\KEM.   It is convenient to express this quantity
in terms of the experimental pion splitting, just as we have done for the kaon splitting.  We follow \rcite{Aoki:2016frl} and define the dimensionless
quantity \ek\ by
\begin{equation}
\ek \equiv \frac{\KEM}{\pisplit}. \eqn{ek-def}
\end{equation}

The following is an outline of the remainder of the paper:
\Secref{lattice} gives the details of the 2+1 flavor asqtad staggered QCD ensembles \cite{RMP} on which we 
compute (quenched) EM effects.  In addition, we describe the pure QCD 2+1+1 highly improved staggered quark (HISQ) ensembles \cite{Bazavov:2012xda} on which we calculate
$m_u/m_d$, with input on EM effects from the asqtad simulations.
We discuss infinite volume \chpt\ in QCD+QED in \secref{chpt}. Modifications
for partial quenching \cite{Bijnens:2006mk} and staggered discretization errors \cite{Bernard:2010qd} are
detailed, and the staggered result for the meson masses at NLO is  presented and explained.
In \secref{QEDFV} we describe how we define QED in finite volume (FV). Finite-volume effects are then calculated at one loop in staggered \chpt\ in 
\secref{FVChPT}.  We  show that the resulting formulas give an excellent description of our lattice data over a 
wide range of volumes.
 We can therefore correct for FV effects,
with a small residual systematic error.   \Secref{fits} then presents a variety of chiral fits to the FV-corrected lattice
data, and \secref{results} describes our results and systematic errors for the EM contributions to the kaon massess, and the parameters $\epsilon$ and \ek.  Finally, in \secref{mumd} we use our EM results \cut{are} to
adjust the experimental kaon masses to their values in a pure-QCD world, which are then taken as input
to the calculation of $m_u/m_d$  
following \rcite{Bazavov:2014wgs}.

Our final results are:
 \begin{eqnarray*}
 \epsilon&=& 0.78(1)_{\rm stat}({}^{+\phantom{1}8}_{-11})_{\rm syst},\\
 \ek &=& 0.035(3)_{\rm stat}(20)_{\rm syst},
 \strike{ \KEM &=& 44(3)_{\rm stat}(25)_{\rm syst}\;({\rm MeV})^2 \; .}\\
 m_u/m_d &=& 0.4529(48)_\mathrm{stat}(\null_{-\phantom{1}67}^{+150})_\mathrm{syst}.
\end{eqnarray*}
Preliminary versions of this work have appeared in 
\rcites{basak,EM12,Basak:2013iw,Basak:2014vca,Basak:2016jnn}.

We note that $m_u/m_d$ may be computed on the lattice in other ways that do not depend on knowing the EM contributions
to the kaon masses.  In particular, \rcite{Durr:2010vn} uses a dispersive treatment of the experimental input from the decay $\rho\to3\pi$ instead of kaon splittings to obtain the ratio $m_u/m_d$ from their lattice determination of $m_s/m_l$, where
$m_l \equiv (m_u+m_d)/2$.  Since the $\rho\to3\pi$ decay violates isospin but is known to be fairly independent of EM
corrections, it gives a handle on $m_u/m_d$  that does not require EM input, at least to some level of accuracy.

\section{Lattice Details \label{sec:lattice}}
We calculate meson masses on the (2+1)-flavor MILC asqtad ensembles, with quenched photon fields, and with lattice spacings ranging from  
$\approx\!0.12\;{\rm fm}$ to $\approx\!0.045\;{\rm fm}$.  \tabref{ensembles} shows the ensembles employed.  
On all ensembles, we generate propagators for valence quarks that have charges 0, $\pm1/3 e$, or
$\pm2/3 e$, where $e \approx 0.303$ is the physical electron charge,
and we compute
the masses of mesons made from various combinations of these quarks.  On many ensembles we also 
have mesons made from quarks with charges greater than physical: $\pm e$ and $\pm 4/3 e$.  On some 
ensembles, we even have quarks with charges $\pm 2e$, although charges that high are not included
in the analysis at this time.

Quenched photon fields are generated in momentum space in the finite-volume Coulomb gauge \qedtl\ defined in detail in \secref{QEDFV}.  The momentum-space
distribution is Gaussian, and is  generated and Fourier transformed
to position space  by a serial program.  The spectrum program reads the
photon fields from disk and, for
each desired charge, converts the field to a $U(1)$ phase
factor with that charge.  The $SU(3)$ links are multiplied by the $U(1)$
links, and then the same gauge smearing that we use for $SU(3)$ alone is
applied.  This amounts to an $a^2$-improved action, but without any tadpole
improvement of $U(1)$.

\begin{table}
\begin{center}
\begin{small}
\begin{tabular}{|c|l|l|c|c|c|c|l|}
\hline\hline
$\approx\! a$[fm]& Volume
& $\beta$
& $m'_l/m'_s$& \# configs.  &$L$ (fm) & $m_\pi L$& \hspace{3mm}$r_1/a$ \\
\hline\hline
0.12 & $12^3\times64$ & 6.76& 0.01/0.05&  1000  & 1.4 & \phantom{1}2.7 & \hspace{3mm}--   \\
     & $16^3\times64$ & 6.76& 0.01/0.05&  1303  & 1.8 & \phantom{1}3.6  & \hspace{3mm}-- \\
     & $20^3\times64$ & 6.76& 0.01/0.05&  2254 & 2.3 &  \phantom{1}4.5 &2.739(12)  \\
      & $28^3\times64$ & 6.76& 0.01/0.05& \phantom{2}274 & 3.2& \phantom{1}6.3  & \hspace{3mm}--   \\
      & $40^3\times64$ & 6.76& 0.01/0.05& \phantom{2}115 & 4.6& \phantom{1}9.0  & \hspace{3mm}--   \\
      & $48^3\times64$ & 6.76& 0.01/0.05& \hspace{-5mm}132+52 & 5.5& 10.8  & \hspace{3mm}--   \\
      & $20^3\times64$ & 6.76& 0.007/0.05& 1261 & 2.3& \phantom{1}3.8 & 2.739(13)  \\
      & $24^3\times64$ & 6.76& 0.005/0.05& 2099 & 2.7& \phantom{1}3.8 & 2.739(13)  \\
\hline
0.09  & $28^3\times96$ &7.09& 0.0062/0.031& 1930 & 2.3&  \phantom{1}4.1& 3.789(6) \\
      & $40^3\times96$ & 7.08& 0.0031/0.031& 1015 & 3.3& \phantom{1}4.2  & 3.755(6) \\
\hline
0.06  & $48^3\times144$ & 7.47& 0.0036/0.018& \phantom{2}670 & 2.8& \phantom{1}4.5 & 5.353(12) \\
         & $56^3\times144$ & 7.465 & 0.0025/0.018& \phantom{2}798 & 3.3& \phantom{1}4.4 & 5.330(12)  \\
         & $64^3\times144$ & 7.46& 0.0018/0.018& \phantom{2}826 & 3.8& \phantom{1}4.3 &   5.307(12) \\
\hline
0.045  & $64^3\times192$ & 7.81& 0.0028/0.014& \phantom{2}801 & 2.8& \phantom{1}4.6 & 7.208(25)  \\
\hline\hline
\end{tabular}
\end{small}
\caption{Parameters of the (2+1)-flavor asqtad ensembles used in this study. 
The quark masses $m'_l$ and $m'_s$ are the light and strange dynamical masses used in the runs.  The  number of configurations listed  as `132+52' for the $a\!\approx\!0.12\:$fm, $48^3\times 64$ ensemble gives values for
two independent streams, the first in single precision, and the second in double.  We treat them as separate data, and do not 
average the results. The $r_1/a$ values are mass-independent, in that they are extrapolated to
physical quark masses, rather than the sea mass of the simulations. The errors listed  for $r_1/a$ are the sum in quadrature of the statistical errors and
the extrapolation errors.  We use  the $a\approx 0.12$, $m'_l=0.01$, $m'_s=0.05$ result for $r_1/a$
for those $a\approx 0.12$ ensembles where no $r_1/a$ value has been directly computed.
\label{tab:ensembles} }
\end{center}
\end{table} 

\subsection{New ensembles}
\label{sec:newensembles}

To study finite-volume errors, which were found to be quite important in our prior work, we
have generated a number of new ensembles that are not detailed in \rcite{RMP}.  Our prior
finite-volume work used two volumes corresponding to spatial size $L=20$
and 28.  We have added $L=12$, 16, 40, and 48 in order to have data on both
larger and smaller volumes.
For $L=12$, we have generated the ensemble using the R algorithm 
\cite{Gottlieb:1987mq,Duane:1985ym,Duane:1986iw}
in a single stream of 5200 time units of evolution.  Each trajectory
consists of 150 steps with a step size of 0.00667.
The first 200 time units are dropped and every
5th time unit is then archived for analysis, yielding 1000 configurations in the
ensemble.  For $L=16$, we have four separate streams.  Three of them use 
the RHMC \cite{Clark:2006fx,Duane:1987de,Sexton:1992nu,Kennedy:1998cu,Hasenbusch:2001ne}
algorithm with a 3G1F Omelyan integrator \cite{Omelyan:2002E1,Takaishi:2005tz}.
The step size is 0.05, and there are twenty steps per
trajectory.
Each of these streams has 334 or 335 configurations separated 
by 6 time units.  A fourth stream employs
the R algorithm with the same parameters as for $L=12$
and has 300 configurations separated by 5 time units.  For $L=40$, we
use the RHMC algorithm with 40 steps of size 0.025
and analyze 115 configurations separated by 6 time units.
All of the above ensembles are generated by single-precision code, except that accumulations
are done in double precision.  For $L=48$, we use two streams, one in single
precision (as above), and one in double precision.  In each case, archived
configurations are separated by 6 time units.  From the single-precision ensemble,
132 configurations are  used for the spectrum analysis, whereas 52 are analyzed 
from the double-precision ensemble.  These have not been  combined in the finite-volume 
study.  Table \ref{tab:newensembles} summarizes information about the new ensembles.

\begin{table}
\begin{center}
\begin{small}
\begin{tabular}{|c|l|c|c|c|l|}
\hline\hline
$L$&algorithm&$\delta t$&steps&trajectories&comment\\
\hline\hline
12 & R & 0.00667 & 150 &  5& \\
16 & R & 0.00667 & 150 &  5& \\
16 & RHMC & 0.05 & 20 &  6& 3 streams\\
40 & RHMC & 0.025 & 40 &  6& \\
48 & RHMC & 0.025 & 40 &  6& single precision\\
48 & RHMC & 0.025 & 40 &  6& double precision\\
\hline\hline
\end{tabular}
\end{small}
\caption{Characteristics of the new ensembles generated to study finite
volume effects.  Each ensemble has a volume of $L^3\times 64$ with the
value of $L$ in the first column.  The second column indicates the algorithm
used to generate the ensemble.  The third and fourth columns contain the
molecular dynamics step size and the number of steps in each trajectory,
respectively.  The fifth column indicates how many trajectories separate
archived lattices on which the spectrum analysis is done.  The last column
contains additional comments.
\label{tab:newensembles} }
\end{center}
\end{table}

\subsection{Spectrum Calculations}
\label{sec:spectrumQUDA}

In order to calculate the meson spectrum, we read an archived dynamical $SU(3)$
gauge configuration and a quenched $U(1)$ gauge configuration and proceed to 
cast quark propagators from a corner wall source.  We use a variety of valence quark charges
and masses.  A multi-shift solver is employed  so that for each desired charge all desired
masses are found with one iterative process.  

The calculation of the meson spectrum has been primarily done on GPU based computers at
the Texas Advanced Computing Center, National Center for Supercomputing Applications,
and Indiana University using the QUDA approach pioneered at Boston University \cite{quda}, but
enhanced to support staggered quarks 
\cite{Babich:2011np,Babich:2011zz,shi:ipdps11,shi:ipdps12}.

\begin{figure}
\vspace{-5mm}
\begin{center}\includegraphics[width=0.55\textwidth]{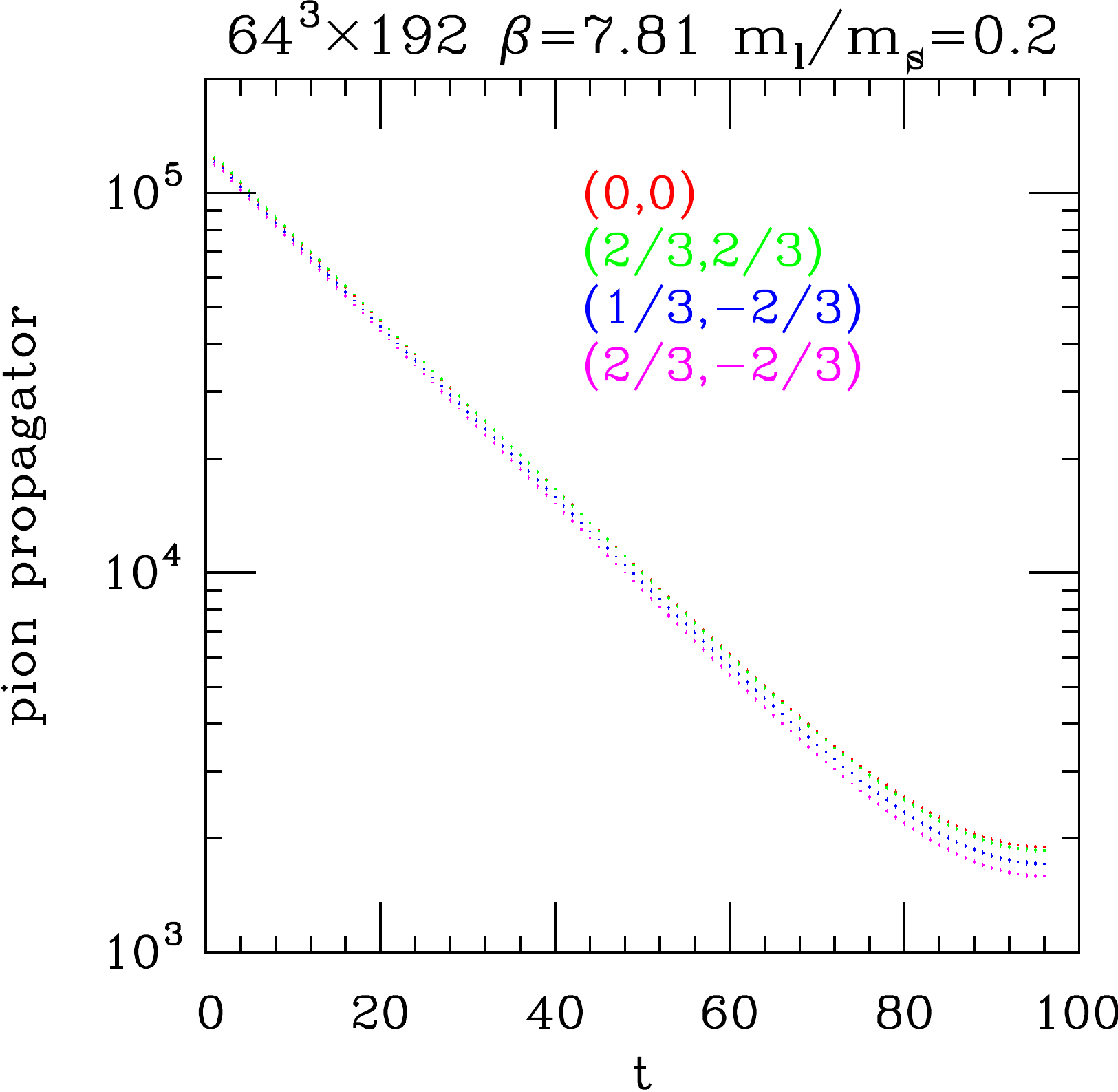}\end{center}
\vspace{-5mm}
\caption{\label{fig:propagators} 
The Goldstone pion propagator as a function of Euclidean time on the
$a\approx 0.045$ fm ensemble with $am'_l=0.0028$ and $am'_s=0.014$.
The grid size is $64^3\times 192$.  The meson propagators are periodic in
time and have been folded over, so the maximum time is 96.  Four charge
combinations are plotted:
$(0,0)$,
$(2/3,2/3)$, $(1/3,-2/3)$, and $(2/3,-2/3)$, with total charges $q_{xy}=0$, 0, 1
and $4/3$, respectively.  These charges are all in units of $e$.  The valence
quark and antiquark masses are both 0.0014 in lattice units.
}
\vspace{-0.15in}
\end{figure}

\begin{figure}[t!]
  \centering
  {\includegraphics[width=0.48\linewidth]
  {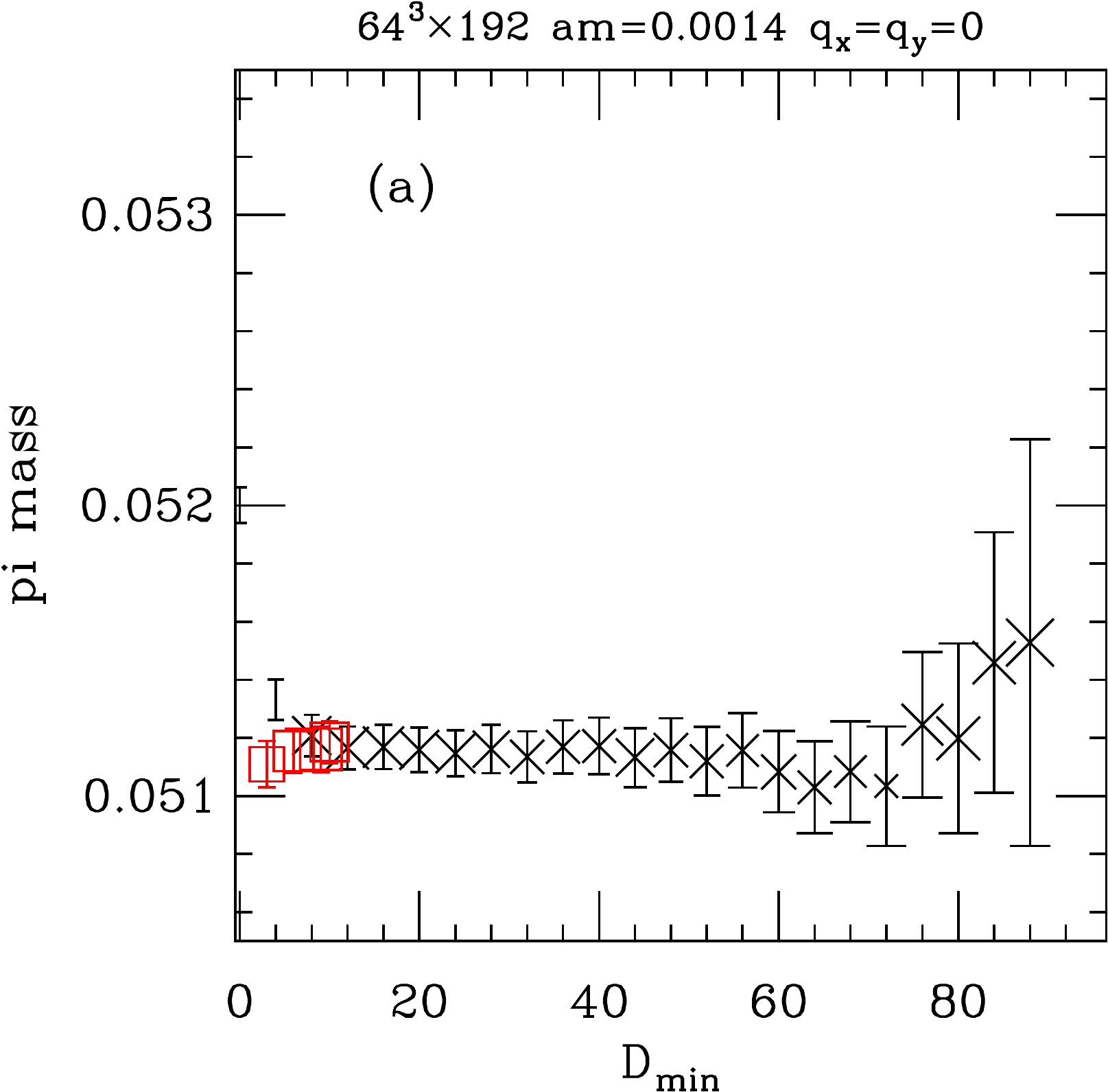}}
  \
  {\includegraphics[width=0.48\linewidth]
  {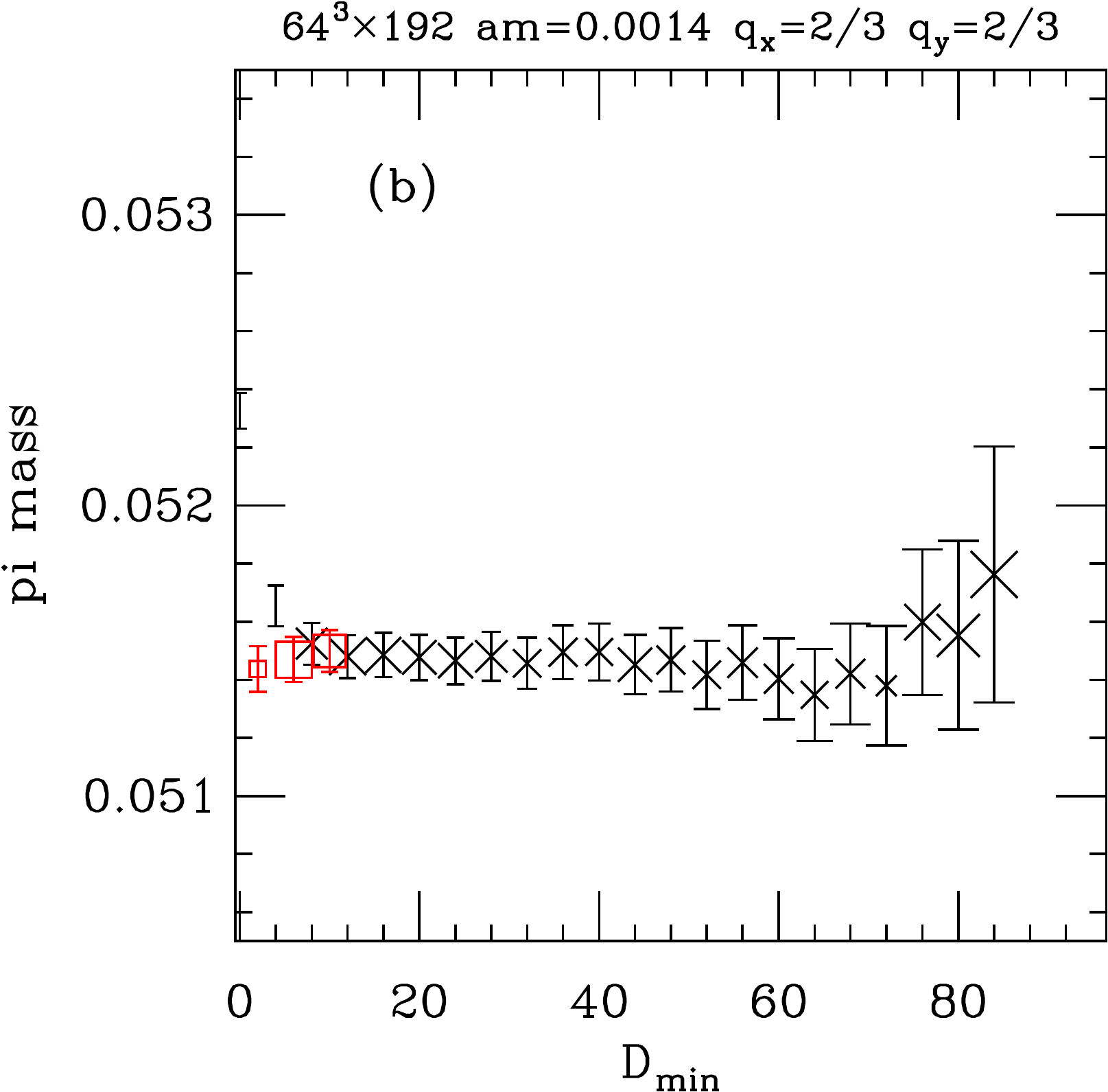}}
  \\
  \vspace{2mm}
  {\includegraphics[width=0.48\linewidth]
  {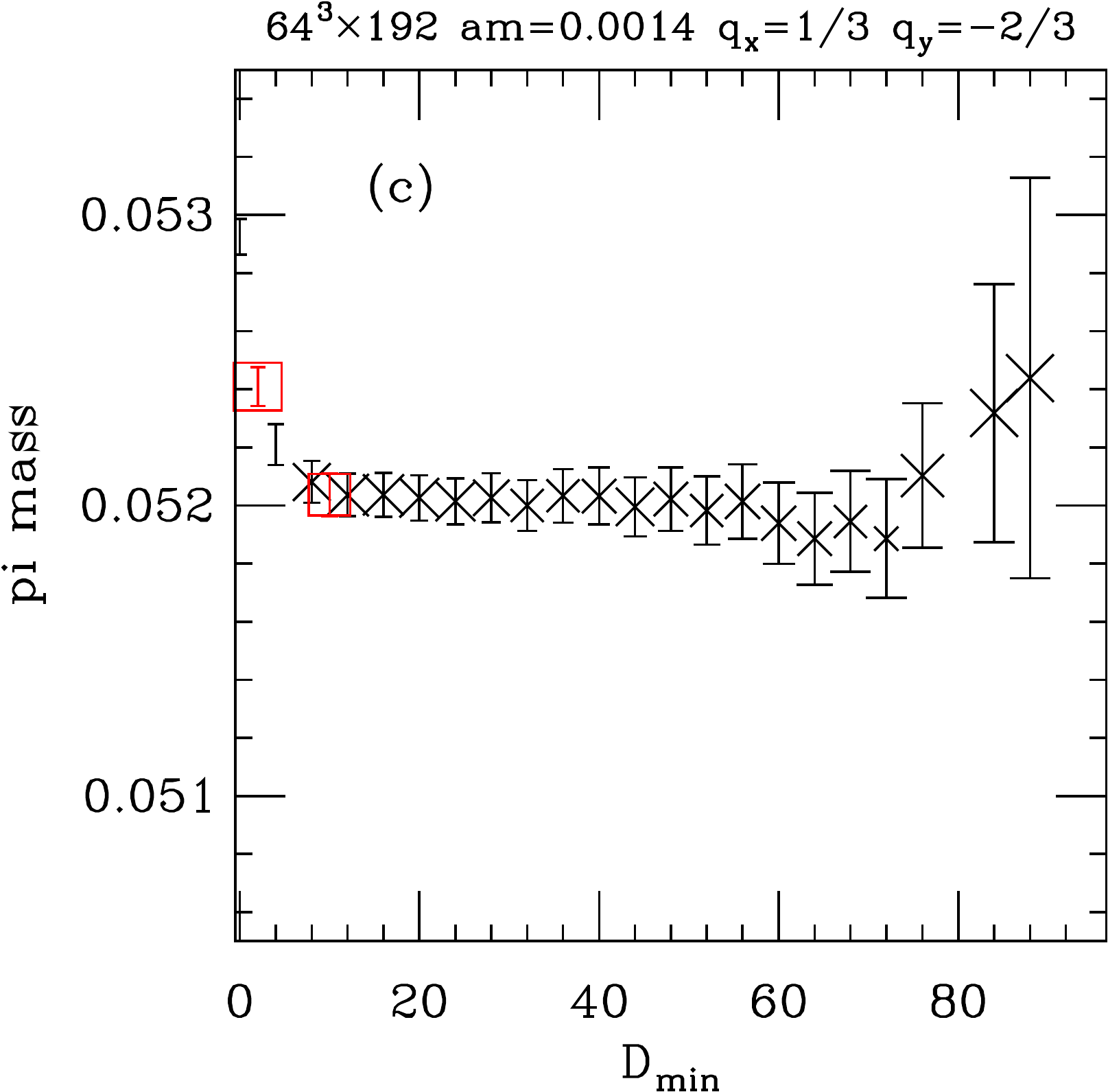}}
  \
  {\includegraphics[width=0.48\linewidth]
  {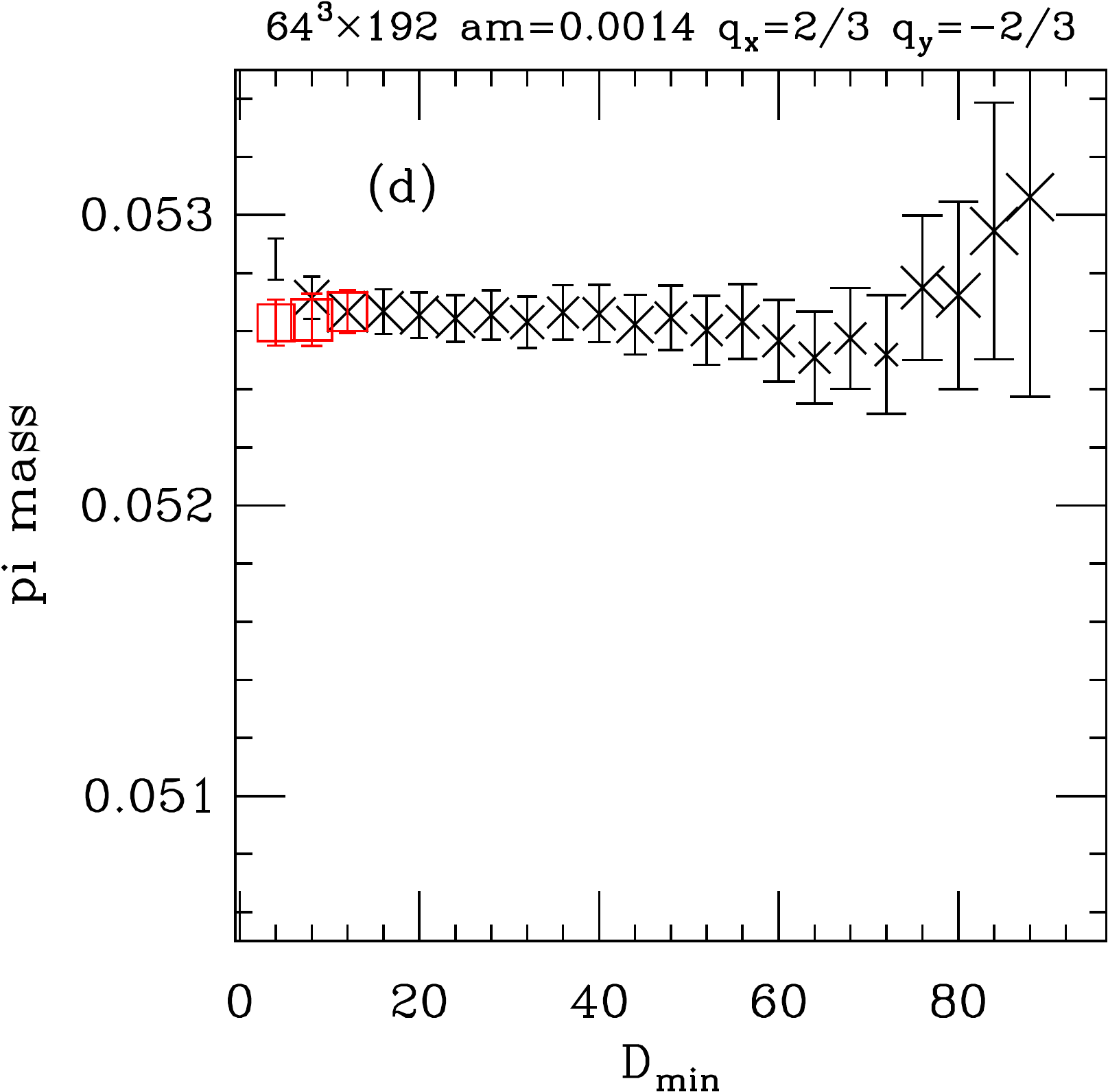}}
  \\
\vspace{-2mm}
\caption{\label{fig:dminfits}
Fits of the four pion propagators shown in \figref{propagators}.  Fits
are from $D_{min}$ to the center of the lattice.  The symbol sizes are 
proportional to the $p$ value of the fit.  Black crosses denote single-particle 
fits and red squares denote two-particle fits.
}
\vspace{-0.08in}
\end{figure}

\begin{figure}
\vspace{-5mm}
\begin{center}\includegraphics[width=0.55\textwidth]{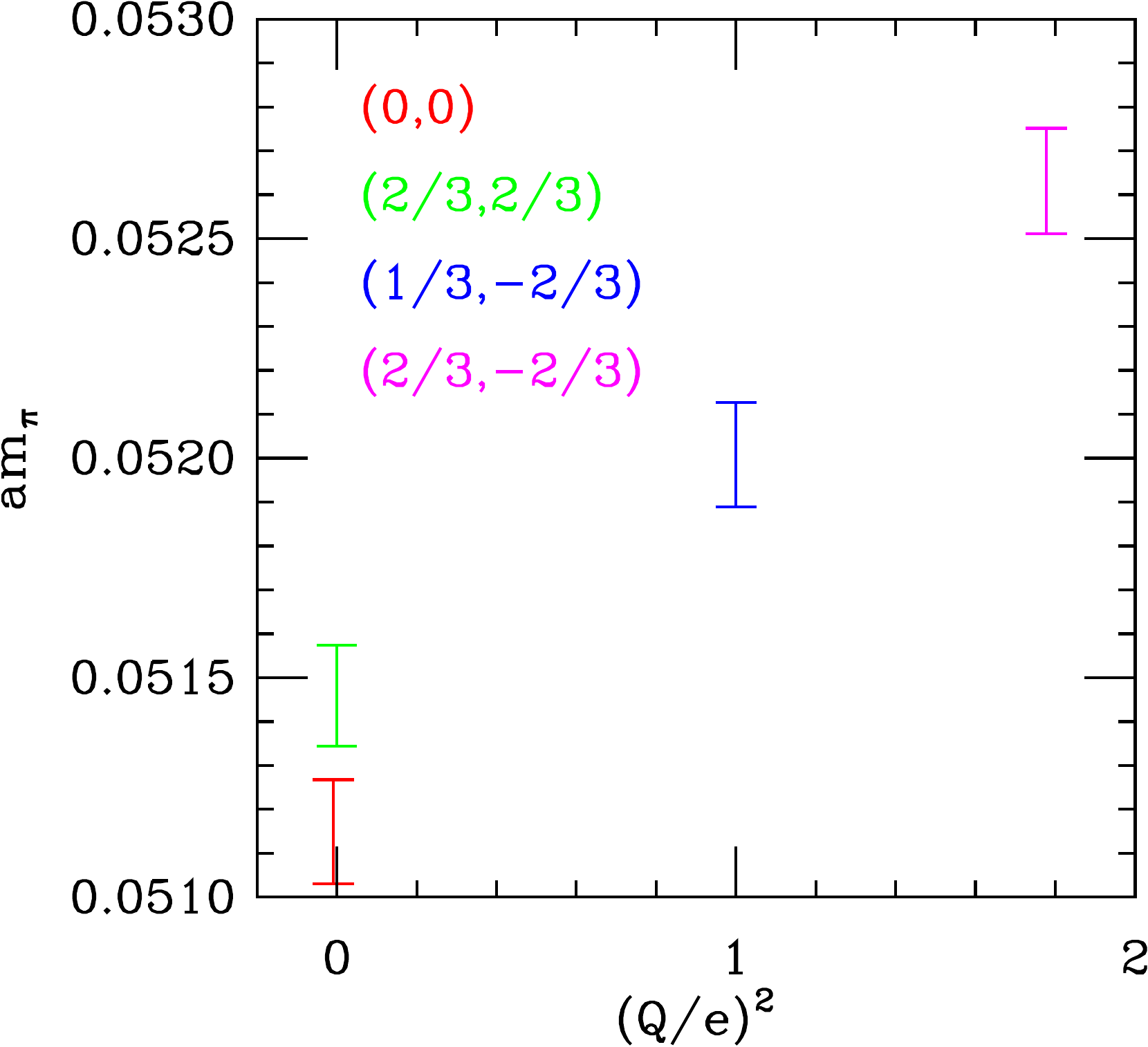}\end{center}
\vspace{-5mm}
\caption{\label{fig:massvsqsq} 
The Goldstone pion mass as a function of the square of the meson charge on
the $a\approx 0.045$ fm ensemble with $am'_l=0.0028$ and $am'_s=0.014$.
Four charge combinations are plotted: $(0,0)$,
$(2/3,2/3)$, $(1/3,-2/3)$, and $(2/3,-2/3)$, with total charges $Q=0$, 0, 1
and $4/3$, respectively.  These charges are all in units of $e$.  The valence
quark and antiquark masses are both 0.0014, in lattice units.
}
\vspace{-0.15in}
\end{figure}

In \tabref{valencemassescharges}, we summarize the quark charges, masses, and number of channels
we study on each ensemble.  \Figref{propagators} shows the Goldstone pion
propagators as a function of Euclidean time for the $a\approx 0.045$ fm
ensemble, which is our finest lattice spacing.  We show four charge combinations
for our lightest valence quark mass on that  ensemble.  Using the notation
further detailed in \secref{schiptem}, the quark charges are $q_x$ and $q_y$ in units of
the fundamental charge $e$, and
the meson charge $q_{xy}$ is $q_x - q_y$  since 
the meson is made from an $x$-quark
and $y$-antiquark.  The combinations $(q_x,q_y)$ we plot are $(0,0)$,
$(2/3,2/3)$, $(1/3,-2/3)$, and $(2/3,-2/3)$, with total charges $q_{xy}=0$, 0, 1
and $4/3$, respectively.  We see a nice 
linear decrease of the propagators in this semi-logarithmic plot over a large range of $t$, before the periodic boundary
conditions result in curvature at large $t$.
In \figref{dminfits}, we show the results of fitting the propagators in \figref{propagators}.
Each plot shows a series of fits starting from $D_{min}$ and extending to
the center of the lattice.  The symbol size is proportional to the 
$p$ value of the fit.  Crosses are fits with a single particle (two
free parameters),  and squares correspond to two particles (four parameters).
We see that there are many fits with good $p$ values, and that
 the meson masses depend significantly on the total charge.
We can even see a difference between the two cases of a
neutral meson, one with uncharged quarks and the other 
made from a quark and an antiquark whose charges cancel each other.
Much of this difference is unphysical, coming from the effect of EM quark-mass renormalization at fixed bare mass --- see \secref{renormalization}. Note that the quality of the plateaus for mesons with charged quarks is virtually identical
to that for the meson with uncharged quarks; we return to this point in \secref{FVChPT}.
The masses corresponding to fits with $D_{min}=50$, which is the value chosen
for this ensemble in our final analysis, are detailed in
\tabref{fourmasses}.  \Figref{massvsqsq} plots these masses versus the square of the meson
charge.

\begin{sidewaystable}
\begin{center}
\begin{small}
\begin{tabular}{|c|l|l|c|c|c|c|l|}
\hline\hline
$\approx\! a$[fm]& Volume
& $m'_l/m'_s$& charges & $am_v$ & channels \\
\hline\hline
0.12 & $12^3\times64$ & 0.01/0.05&$\pm 2/3, \pm 1/3, 0$ & 0.005, 0.007, 0.01, 0.02, 0.03, 0.04, 0.05 & 700 \\
     & $16^3\times64$ & 0.01/0.05&$\pm 2/3, \pm 1/3, 0$ & 0.005, 0.007, 0.01, 0.02, 0.03, 0.04, 0.05 & 700 \\
     & $20^3\times64$ & 0.01/0.05&  $\pm 2, \pm 4/3, \pm 1, \pm 2/3, \pm 1/3, 0$ & 0.005, 0.007, 0.01, 0.02, 0.03, 0.04, 0.05 & 532 \\
      & $28^3\times64$ & 0.01/0.05& $\pm 2, \pm 4/3, \pm 1, \pm 2/3, \pm 1/3, 0$& 0.005, 0.007, 0.01, 0.02, 0.03, 0.04, 0.05 & 532 \\
      & $40^3\times64$ & 0.01/0.05&$\pm 2/3, \pm 1/3, 0$& 0.005, 0.007, 0.01, 0.02, 0.03, 0.04, 0.05 & 700 \\
      & $48^3\times64$ & 0.01/0.05&$\pm 2/3, \pm 1/3, 0$& 0.005, 0.007, 0.01, 0.02, 0.03, 0.04, 0.05 & 700 \\
      & $20^3\times64$ & 0.007/0.05&$\pm 4/3, \pm 1, \pm 2/3, \pm 1/3, 0$& 0.005, 0.007, 0.01, 0.02, 0.03, 0.04, 0.05 & 364 \\
      & $24^3\times64$ & 0.005/0.05& $\pm 4/3, \pm 1, \pm 2/3, \pm 1/3, 0$& 0.005, 0.007, 0.01, 0.02, 0.03, 0.04, 0.05 & 364 \\
\hline
0.09  & $28^3\times96$ &0.0062/0.031& $\pm 4/3, \pm 1, \pm 2/3, \pm 1/3, 0$ & 0.0031, 0.0062, 0.0093, 0.0124, 0.0155, 0.0186, 0.031 & 364 \\
      & $40^3\times96$ & 0.0031/0.031& $\pm 2, \pm 4/3, \pm 1, \pm 2/3, \pm 1/3, 0$ & 0.0031, 0.0062, 0.0093, 0.0124, 0.0155, 0.0186, 0.031 & 532 \\ \hline
0.06  & $48^3\times144$ & 0.0036/0.018& $\pm 2, \pm 4/3, \pm 1, \pm 2/3, \pm 1/3, 0$ & 0.0036, 0.0054, 0.0072, 0.009, 0.0108, 0.0126, 0.018 & 532 \\
         & $56^3\times144$ & 0.0025/0.018& $\pm 2/3, \pm 1/3, 0$ &0.0018, 0.0025, 0.0036, 0.0044, 0.0054, 0.0072, 0.0108, 0.0144, 0.018 & 1125 \\
         & $64^3\times144$ & 0.0018/0.018& $\pm 2/3, \pm 1/3, 0$&0.0018, 0.0025, 0.0036, 0.0044, 0.0054, 0.0072, 0.0108, 0.0144, 0.018 & 1125 \\
\hline
0.045  & $64^3\times192$ & 0.0028/0.014& $\pm 2/3, \pm 1/3, 0$ & 0.0014, 0.0021, 0.0028, 0.0035, 0.0042, 0.0056, 0.0084, 0.0112, 0.014& 1125 \\
\hline\hline
\end{tabular}
\end{small}
\caption{Details of the charges and valence quark masses used for
the meson spectrum.  The last column indicates how many charge and mass
combinations were used to construct mesons.
\label{tab:valencemassescharges}}
\end{center}
\end{sidewaystable} 

\begin{table}
\begin{center}
\begin{tabular}{|c|c|c|c|c|c|}
\hline\hline
$q_x$ & $q_y$ & $q_{xy}$  & $am$ & $\chi^2$ & $p$ \\
\hline\hline
0 & 0 & 0 &        0.05115(12) & 39.26 & 0.713\\
2/3 & 2/3 & 0 &    0.05146(12) & 42.32 & 0.586\\
1/3 & -2/3 & 1 &   0.05201(12) & 39.43 &0.706\\
2/3 & -2/3 & 4/3 & 0.05263(12) & 40.27 & 0.672\\
\hline\hline
\end{tabular}
\caption{
The masses of the four mesons plotted in \figref{propagators}.  On this ensemble, we fit the
propagator from $t=50$ to the center of the lattice assuming a single particle.
The first two columns are the charges of the two quarks in units of $e$.
Since the meson is made from a quark and an antiquark, the meson charge
$q_{xy}$ in the third column is $q_x - q_y$.  The mass and its error are in the
fourth column.  Each fit has 45 degrees of freedom.  $\chi^2$ and the
$p$ value of the fit are in columns five and six, respectively.
\label{tab:fourmasses}}
\end{center}
\end{table}

In order to construct the correlations among the masses of
all the channels on a specific ensemble, we use a single
elimination jackknife fitting procedure.  On each ensemble, a single
value of $D_{min}$ is used for all channels. In the subsequent analysis, we subtract
the squared meson mass for $q=0$ quarks from the corresponding squared meson mass with
non-zero quark charges, properly taking in account the correlations. These correlations are  expected
to be very large, especially for mesons with the same valence quark masses but different valence quark charges, because the QCD contributions are identical in the two cases, and only the small QED effects are different. 
Because of this high degree of correlation, the errors of the subtracted
quantities are much smaller than one would find by the naive propagation of errors from
the masses themselves.  For example, the correlation
between the $q_x=0$, $q_y=0$ and the $q_x=1/3$, $q_y=-2/3$ masses in \tabref{fourmasses} 
is 0.998, and the error in the mass difference is 0.85\%.  If the error in the difference were propagated
naively, omitting the correlation, the error in the mass difference would be about 20\%.

\begin{figure}[t!]
  \hspace{0.48\linewidth}
  {\includegraphics[width=0.48\linewidth]
  {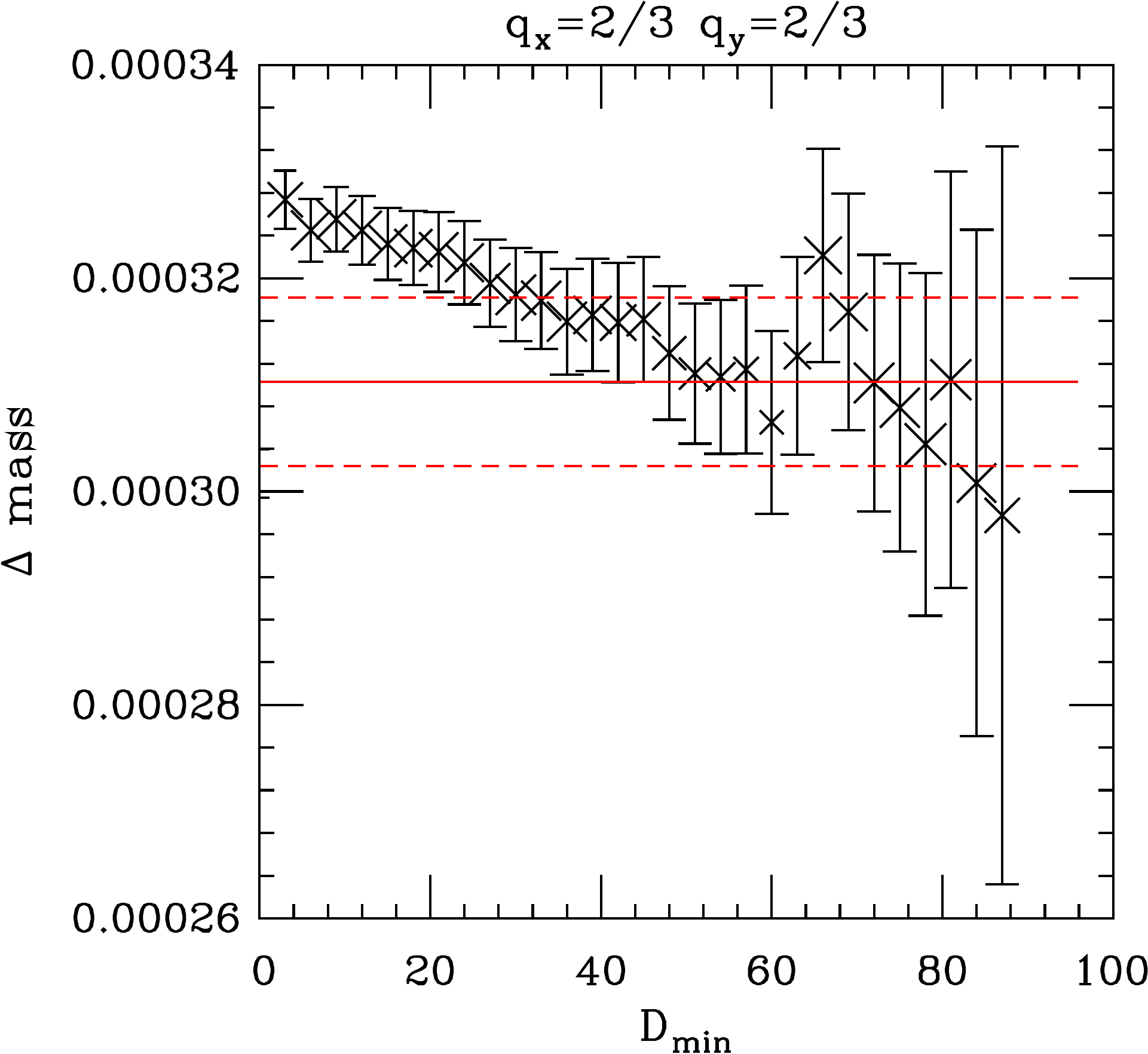}}
  \\
    \vspace{2mm}
   \centering
  {\includegraphics[width=0.48\linewidth]
  {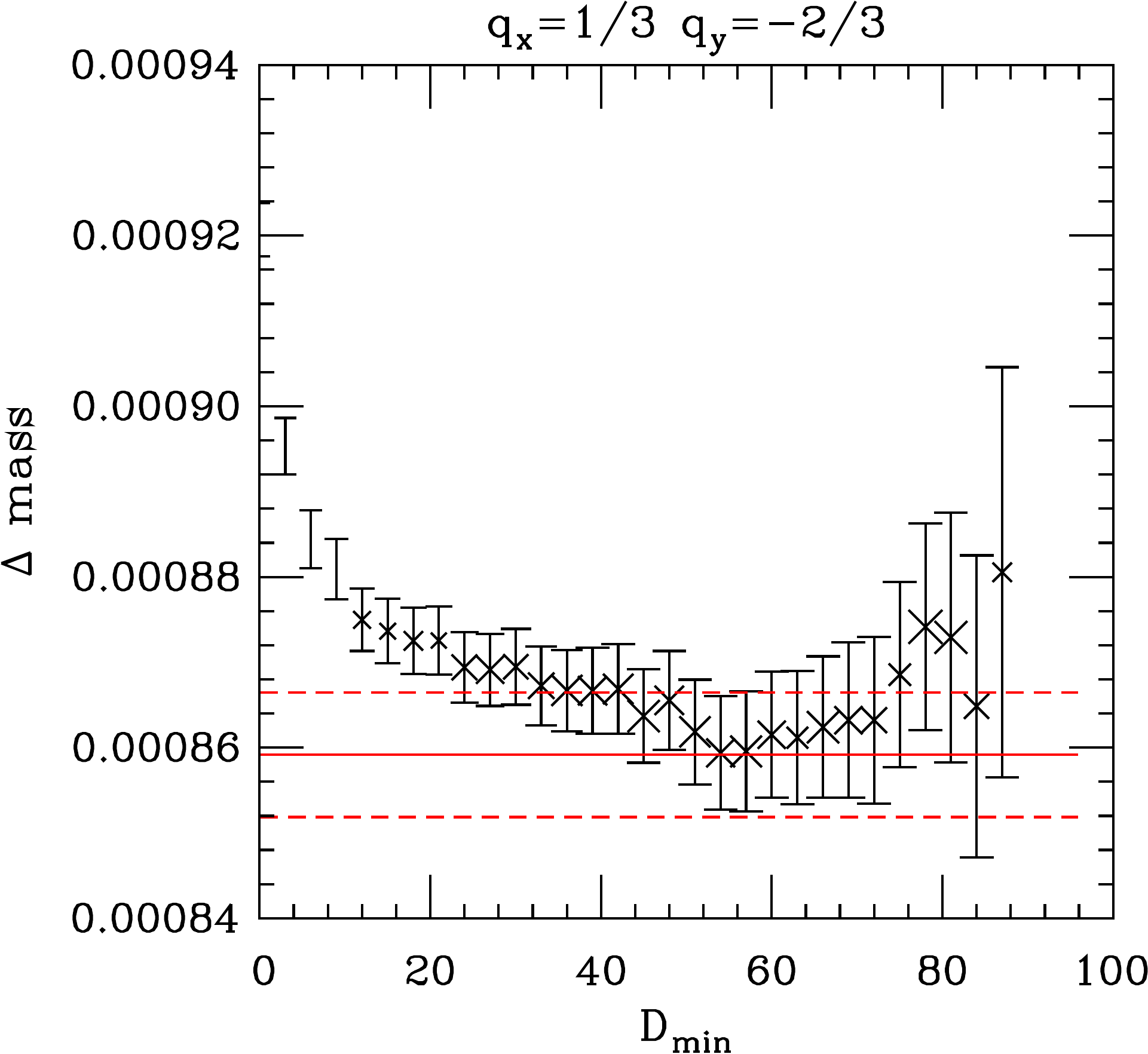}}
  \
  {\includegraphics[width=0.48\linewidth]
  {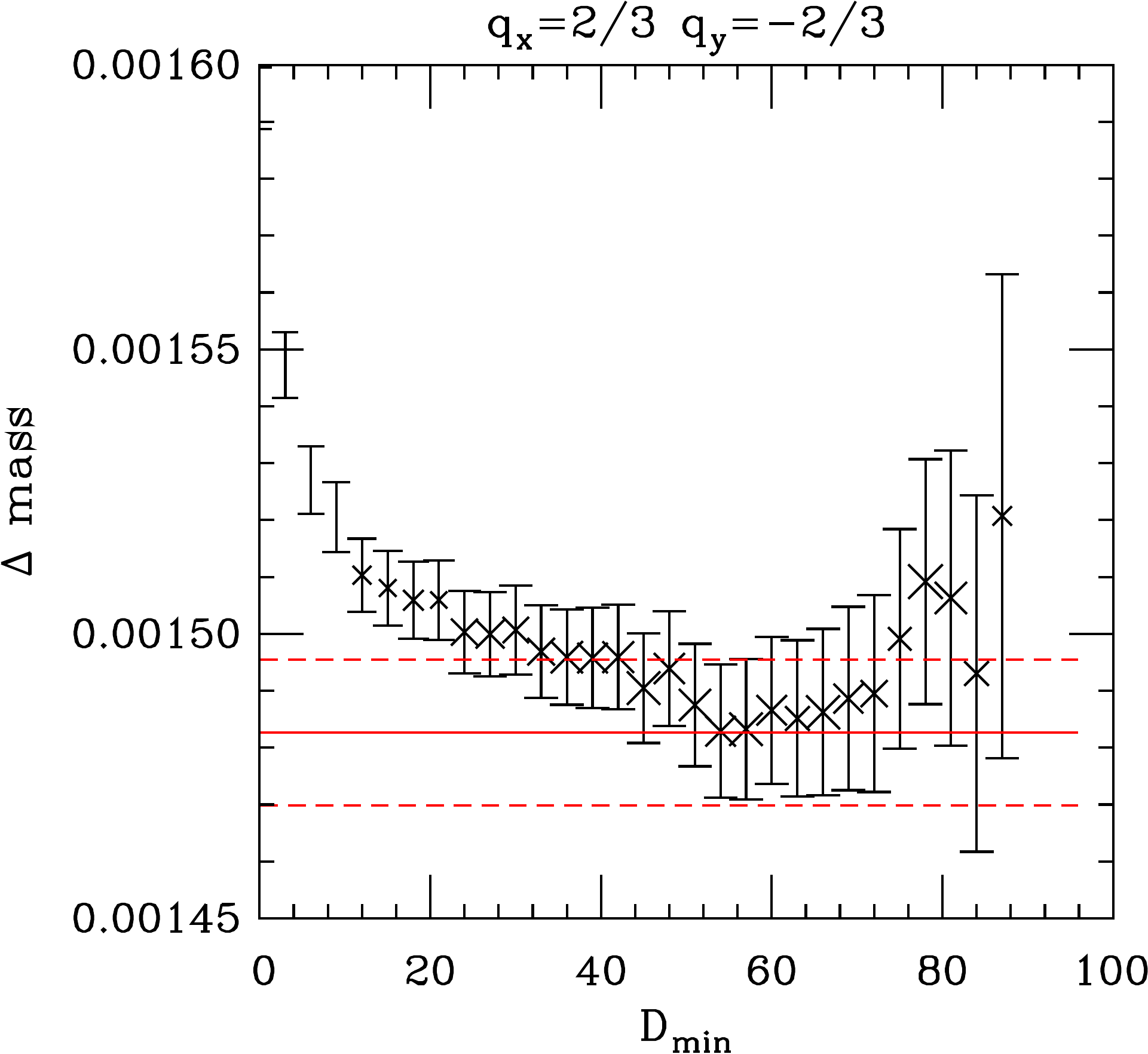}}
  \\
\vspace{-2mm}
\caption{\label{fig:sub_dminfits}
Fits of the ratio of the  propagators shown in \figref{propagators} for mesons with charged quarks,
divided by the propagator for the meson with neutral quarks.  The vertical axis gives the mass difference
between the two propagators in the ratio. Fits
are from $D_{min}$ to the center of the lattice.  The symbol sizes are 
proportional to the $p$ value of the fit.  The horizontal solid and dashed red lines show the
central value and error for the mass difference given by subtracting the masses in \tabref{fourmasses} and
propagating the errors using the covariance matrix determined by jackknife, as described in the text.}
\vspace{-0.08in}
\end{figure}

The small errors in the subtracted masses is illustrated by an alternative analysis shown in
\figref{sub_dminfits} for the same data as in \figrefs{propagators}{dminfits}.  Here, rather than fitting individual propagators, we fit the ratio of each propagator
for a meson made of charged quarks with the corresponding propagator for the meson made from
neutral quarks.  Because of the effect of the periodic boundary conditions in time, the ratio depends not
only on the meson mass difference, but also on the meson masses themselves.  The latter dependence 
is mild, but still nonnegligible, and makes fits with three unconstrained parameters (the mass difference,
the mass of $q_x=q_y=0$ meson, and the overall amplitude) somewhat unstable.  Instead,  we have constrained, with Bayesian priors,
the  mass of the $q_x=q_y=0$ meson to 0.05115(12), as given in \tabref{fourmasses}.  The plots show the resulting
mass differences as a function of the minimum distances in the fits, $D_{min}$.  The horizontal solid and 
dotted red lines show the mass differences and errors computed from the individual masses using
$D_{min}=50$.

\subsection{Scale setting}
\label{sec:scale}

We use the intermediate quantity $r_1$ \cite{Bernard:2000gd,Sommer:1993ce} to set the relative
scale of our ensembles, and take  $r_1=0.3117(22)$ fm \cite{Bazavov:2011aa} as the absolute scale.
From the smoothing fit to $r_1/a$ values described in \rcite{RMP}, we extrapolate the $r_1/a$ values
at the simulated quark masses to the physical quark masses (given below in \tabref{other-params}), 
holding $\beta$ fixed.
This defines a mass-independent
scale-setting scheme, which is needed in order to apply chiral perturbation theory.  The scheme is
mass independent because it gives an $r_1/a$ value that depends only on $\beta$ and not on the
simulated quark masses $m'_l$ and $m'_s$.  Mass-independent values of $r_1/a$ for our ensembles
are listed in \tabref{ensembles}.   The errors shown are a sum (in quadrature) of statistical errors
and errors of the extrapolation to the physical quark masses.

\section{Chiral Perturbation Theory with Electromagnetism\label{sec:chpt}}

\subsection{Continuum chiral theory for QCD+QED}   \label{sec:chiptem}

In the continuum,  the chiral effective theory for QCD+QED was worked out by Urech \cite{Urech:1994hd}.
Along with all hadrons heavier than the pseudoscalar mesons, high-momentum photons are integrated out
of the chiral theory, resulting in a single effective meson-interaction term at LO.  Photons and mesons with 
low momentum (less than the chiral cutoff $\Lambda_\chi$), are treated explicitly in this chiral perturbation 
theory (\cpt).   

The {\it partially quenched} version of the chiral theory is relevant here.  In partial 
quenching, the valence and sea quarks are treated as distinct; when EM is included, this means
that valence and sea quarks may have different electric charges and/or masses.%
\footnote{We will call
the limit where valence- and sea-quark masses and charges are equal {\it full}\/ 
QCD+QED. }
Bijnens and Danielsson~\cite{Bijnens:2006mk} have calculated the meson masses and decay constants at 
NLO (one loop) in partially-quenched \cpt\ in QCD+QED with three flavors of
sea quarks ($u$, $d$, $s$). 
A key insight of
\rcite{Bijnens:2006mk} is that sea-quark charges affect meson masses  in
particularly simple ways at NLO.  In analytic terms involving the sea-quark charges,  only the sums of the 
squared sea-quark charges appear --- there are no cross terms
between sea-quark and valence-quark charges.
(It is necessary to assume here that the sum of the three sea-quark charges vanishes, as it does
in the real world.)
Sea-quark  charges may also appear in the one-loop 
chiral logarithms, but these are completely determined in terms of the LO LECs.  
This implies that in the difference of squared mass of two mesons with the same valence quark masses
but different valence quark charges, the analytic terms depending on sea-quark charges cancel.  Thus the difference  may be reliably computed on the lattice with a simulation in which the sea quarks are 
uncharged (the electroquenched approximation).   The sea-quark charge dependence, which comes only 
from one-loop chiral logarithms, may be put in after the fact.  All dependence on unknown NLO LECs 
cancels in the difference.  

Note that the quantities we need to calculate to determine $\epsilon$ 
in \eq{our-eps}, namely $(M^2_{K^\pm}-M^2_{K^0})^\gamma$ and 
$(M^2_{\pi^\pm}-M^2_{\rm{``}\pi^0\rm{"}})^\gamma$, are squared-mass differences of the 
type required to make them 
reliably calculable with electroquenched simulations, in the sense described  in the previous paragraph.
We emphasize, however, that the calculability depends on using SU(3) (3-flavor) \cpt\ at 
NLO, which will have nonnegligible systematic corrections that need to be estimated.  The
alternative, treating only the $u$, $d$ quarks as light (2-flavor SU(2) \cpt),
is not necessarily an improvement, despite the fact that in SU(2) \cpt\ the errors are generically much 
smaller at a given order than in SU(3).  The reason is that the calculability of squared-mass 
differences
in the electroquenched approximation depends on the tracelessness of the quark charge matrix, which
holds in SU(3), but not in SU(2).    Thus, if SU(2) is used, the chiral errors are likely to be
smaller, but one must include a separate quenching error that needs to be estimated in some independent 
fashion.  That is the approach taken in \rcite{Fodor:2016bgu}.

In this paper, we compute the EM effect on the neutral kaon mass, $\KEM=\ek\pisplit$, in addition to $\epsilon$.
In this case, \chpt\ does not allow us to control the electroquenching error, because that error it is not computable at lowest nontrivial
\chpt\ order.  The quantity  \KEM\ is the difference between the squared mass of a neutral kaon made out
of charged valence quarks, with a charged sea,  and the squared mass of a kaon made out of 
neutral valence quarks,
{\it with a neutral sea}\/.   Even effects that depend on the sea-quark charges alone
do not cancel here.   Our estimate of the electroquenching error in \KEM\ is 
therefore based on
large-$N_c$ power counting only ($N_c=3$ is the number of QCD colors), and must be considered a rough guide only.

\subsection{Staggered chiral perturbation theory with EM}   \label{sec:schiptem}

With the staggered lattice action,  each quark flavor appears as four species, known as {\it tastes}.
This is a remnant of the 16-fold doubling of species of naive lattice fermions.  To obtain standard QCD
in the continuum limit, it is necessary to eliminate the unwanted taste degrees of freedom in the sea.
Our simulations accomplish this by taking the fourth-root of the fermion determinant for each quark flavor
\cite{Marinari:1981qf}.  Numerical and theoretical arguments for the validity of this procedure in the continuum
limit can be found in Refs.~\cite{Follana:2004sz,Durr:2004as,Shamir:2004zc,Durr:2004ta,Wong:2004nk,Prelovsek:2005rf,Bernard:2006zw,Bernard:2006ee,Durr:2006ze,Shamir:2006nj,Bernard:2007qf,Donald:2011if}.
 The appropriate chiral theory for staggered
quarks with the rooting procedure is called {\it rooted staggered \cpt} (\rschpt) 
\cite{Lee:1999zxa,Aubin:2003mg}. 
Starting with the staggered chiral Lagrangian of \rcite{Aubin:2003mg}, it is straightforward \cite{Bernard:2010qd} to include EM effects following \rcite{Urech:1994hd}.

At leading order, the Euclidean, staggered QCD+QED chiral Lagrangian is%
\footnote{\Rcite{Bernard:2010qd} used this Lagrangian but, because of space limitations, did not explicitly display it.}  
\bea
\mathcal{L}^{(\rm LO)} &=&	 \frac{1}{4}F_{\mu\nu}F_{\mu\nu}  
+\frac{\lambda}{2}(\partial_\mu A_\mu)^2
  +\frac{f^2}{8} \Tr(d_\mu \Sigma^\dag\, d_\mu \Sigma) -\frac{B_0f^2}{4}\Tr(\cM\Sigma + \cM\Sigma^\dagger) \nonumber \\
		&&		+  \frac{m_0^2}{24} (\Tr( \Phi ))^2  +a^2{\cal V} 
				- e^2C\; \Tr( Q \Sigma Q \Sigma^\dag ),
		\label{eq:Lagrangian_stag}
\eea
where $\Tr$ denotes a trace over flavor and staggered taste indices.
The quantities $A_\mu$, $F_{\mu\nu}$, and  $\lambda$ are the
photon gauge potential, the EM field strength, and the gauge-fixing parameter, respectively.
The meson fields are contained in 
\be	
	\Sigma = \exp (i \Phi / f),
	\qquad
	\Phi = 
		\begin{pmatrix}
			U	& \pi^+	& K^+ \\
			\pi^-	& D		& K^0 \\
			K^-	& \bar{K}^0	& S
		\end{pmatrix},
		\label{eq:sigma}
\ee
where diagonal entries $U$ $D$, and $S$, are the quark-antiquark pairs $u\bar{u}$, $d\bar{d}$, 
and $s\bar{s}$ respectively.
Each of the meson fields $U, \Pi^+, K^+,\dots$ in \eq{sigma} are composed of 16 tastes, as in
$\pi^+\equiv\sum_{b=1}^{16} \pi^+_b T_b$, where the $T_b$ are the Hermitian taste generators
 \begin{equation}\label{eq:T_b}
        T_b = \{ \xi_5, i\xi_{\mu5}, i\xi_{\mu\nu}\ (\mu<\nu), \xi_{\mu}, \xi_I\}.
\end{equation}
Here $\xi_{\mu}$ are a set of 4 Euclidean gamma matrices,
$\xi_{\mu\nu}\equiv \xi_{\mu}\xi_{\nu}$, $\xi_{\mu5}\equiv \xi_{\mu}\xi_5$, and   $\xi_I \equiv
I$ is the $4\times 4$ identity matrix. The term $a^2\cV$ in \eq{Lagrangian_stag}  is the taste-violating
potential \cite{Aubin:2003mg}, with $a$ the lattice spacing.
The anomaly term $ \frac{1}{24} m_0^2 \langle \Phi \rangle^2 $ gives mass to the $\eta'$, and causes
mixing of the flavor-neutral fields $U$, $D$, $S$ through  ``hairpin'' (quark-line disconnected) 
diagrams~\cite{Bernard:1993sv,Sharpe:2000bc}.
As usual in partially quenched and/or staggered calculations,  it is convenient to
keep this term and use the simple $U$, $D$, $S$ basis along the diagonal of $\Phi$.  At the end of
the calculation, we can take $m_0 \to \infty$ \cite{Sharpe:2001fh} and decouple the $\eta'$.

In \eq{Lagrangian_stag}, $\cM$ is the quark mass matrix, 
\be
\cM = {\rm diag}(m_u, m_d, m_s),
\ee
and 
$Q$ is the quark (electric) charge matrix
\be
Q = {\rm diag}(q_u, q_d, q_s)  =  {\rm diag}(2/3, -1/3, -1/3),
\ee
with the property $\Tr(Q)=0$.   The  covariant derivative $d_\mu$ is given by
 \be
	d_\mu \Sigma = \partial_\mu \Sigma - i eQ A_\mu \Sigma + i \Sigma eQ A_\mu ,
	\label{eq:covderiv}
\ee
where we have set vector and axial source terms to zero since they are not needed for present purposes.
Electromagnetic effects on the meson masses come both directly, from the low-energy photon field $A_\mu$,  and 
indirectly, through the term $e^2C \Tr( Q \Sigma Q \Sigma ^\dag )$  (with $e$ the fundamental electric 
charge and $C$ an LEC), which 
represents the effects of high-energy photons that have been integrated out.

With $p$ a typical meson 4-momentum, and $M$ and $m$ generic meson and quark masses, 
respectively, the standard power-counting scheme  of \rschpt\ is $p^2 \sim M^2 \sim m \sim a^2$, where 
factors of the chiral scale $\Lambda_\chi$ (to make the dimension of each quantity the same) are implicit.
Including EM, \chpt\ becomes a joint expansion in $p^2$ and $e^2$.   The Lagrangian of \eq{Lagrangian_stag} is LO in the sense that it includes the leading terms  both in $p^2$ and in $e^2$.
Even though EM corrections are in general smaller or much smaller  than typical SU(3) chiral corrections,%
\footnote{For example, a typical $\cO(p^2)$ chiral
correction  is $(f_K-f_\pi)_{\rm NLO}/f_\pi \sim20\%$, while the $\cO(e^2)$ (and higher) correction
to the charged pion mass is $(M^2_{\pi^\pm}-M^2_{\pi^0})^\gamma/
M^2_{\pi^0}\sim 5\%$, and is much less than that, on a percentage basis, for the kaon.
}
we are interested here in EM quantities, which start at $\cO(e^2)$, so $e^2$ terms are rightly included in the LO Lagrangian.  One-loop diagrams  from \eq{Lagrangian_stag} then produce
$\cO(e^2p^2)$ corrections, which we consider NLO.  Higher non-analytic (chiral log) corrections have not been computed in 
\rschpt, but it will be necessary to add higher-order analytic terms  ($\cO(e^2p^4)$  and sometimes $
\cO(e^4)$ and $\cO(e^4p^2)$) in order to get acceptable chiral
fits.   We will refer to $\cO(e^2p^4)$ and $
\cO(e^4)$ terms as next-to-next-to leading order (NNLO), and those of $\cO(e^4p^2)$ or $\cO(e^2 p^6)$ as N${}^3$LO; this counting treats $e^2\sim p^4$.
Terms that go like $e^4$ ultimately have negligible impact on our results for $\epsilon$, but can be 
necessary to describe small, but statistically significant, effects in our lattice data, especially when we 
include data for quarks with larger-than-physical charges.

We consider a generic pseudoscalar meson composed of two different valence quarks $x$ and $y$ with 
masses $m_x$ and $m_y$.  In units of $e$, the quark ({\em not} antiquark) charges are $q_x$ and $q_y$, so that the
meson charge is   $q_{xy} = q_x-q_y$.  
At LO, the squared mass of such a meson with taste $b$ is 
\begin{eqnarray}\eqn{LO-EM-mass}
M^2_{xy,b} &=&  \chi_{xy,b}+ q^2_{xy}\dem\,, \\
\chi_{xy,b} &=& B_0(m_x +m_y) + a^2\Delta_b\,, \eqn{chi-def}
\end{eqnarray}
where $\chi_{xy,b}$ is the LO squared mass without EM effects, $\Delta_b$ is the taste splitting coming from the staggered potential $\cV$, and
\be
	\deltae \equiv \frac{4C}{f^2} .
	\label{eq:dem}
\ee
Dashen's theorem is immediately evident from \eq{LO-EM-mass} since the LO EM contribution proportional
to $\deltae$ is independent of quark masses.

We remark that \eqs{LO-EM-mass}{chi-def} are in general complete LO masses (with and without EM) only when the meson is flavor charged ($x\not=y$). For $x=y$, there are additional contributions in the
taste-singlet case (coming from the anomaly, $m_0$, term) and the taste-vector or axial vector cases 
(coming from taste-violating hairpins in $\cV$).  As is standard in partially quenched or staggered \cpt\ 
calculations, such terms are treated as separate two-meson vertices, giving rise to disconnected 
contributions to flavor-neutral propagators.

Beyond LO,  the fourth-root procedure needs to be implemented.  This can be done systematically 
at the level of the chiral theory by using a replica trick \cite{Damgaard:2000gh} for the sea quarks: replicating them $n_r$ times and setting $n_r=1/4$ at the end of the calculation
\cite{AUBIN-BERNARD-REPLICA,Bernard:2006zw,Bernard:2006ee}.  
(Additional, un-replicated valence quarks, here called $x$ and $y$, must also be introduced.)  We do not show the replications 
explicitly in \eq{Lagrangian_stag}; in practice it is actually more convenient at the one-loop level to use quark-flow techniques \cite{Sharpe_quarkflow} to 
keep track of diagrams with sea-quark loops, and multiply them by hand by a factor of 1/4.
Since both the replica and the quark-flow approaches distinguish sea and valence quarks, it is
straightforward to take into account, in the chiral calculations, the fact that our simulations are 
partially quenched.  

From \eq{Lagrangian_stag}, it is straightforward to compute the squared mass of a pseudoscalar meson to order NLO ($\cO(p^4)$, one-loop).   We focus on the
the taste-$\xi_5$ (pseudoscalar taste) meson because it is
the valence meson that we have simulated.  The
taste-$\xi_5$ meson is a true Goldstone boson in the massless limit and in the absence of EM (for 
electrically charged
mesons).    From now on, we always mean the taste-$\xi_5$ meson if we do not otherwise specify the meson's
taste.

We are interested in the EM contribution to the squared mass,
\begin{equation}\eqn{gamma-def}
(M^2_{xy})^\gamma \equiv   M^2_{xy} - M^2_{xy}\Big\vert_{q_x=q_y=q_u=q_d=q_s=0}\qquad{\rm [fixed\  renorm.\ mass]},
\end{equation}
where the second term on the right-hand side is the squared mass in a world without EM, 
where all quark charges, both valence ($q_x$, $q_y$) and sea ($q_u$, $q_d$, $q_s$)
vanish. The difference should be taken at fixed renormalized quark masses, so that
only physically meaningful EM effects contribute to $(M^2_{xy})^\gamma$.  This is a nontrivial
requirement because the masses of quarks with different charges, \eg $u$ and $d$, have different 
EM renormalization.    It is much more convenient to work with an intermediate quantity
\begin{equation}\eqn{Delta-def}
\Delta M^2_{xy} \equiv   M^2_{xy} - M^2_{xy}\Big\vert_{q_x=q_y=q_u=q_d=q_s=0}\qquad{\rm [fixed\  bare\ mass]},
\end{equation}
where the two terms on the right-hand side are computed at the same values of the bare quark masses. 

On the lattice, we have computed $M_{xy}$ for various choices of valence quark charges, including vanishing
charges, for each valence bare quark mass studied.  This means that it is straightforward to construct the
quantity $\Delta M^2_{xy}$, as
well as 
its correlated errors with other choices of quark charges and valence masses.
   On the other hand, the construction of $(M^2_{xy})^\gamma$ would require theoretical assumptions about the
   EM mass renormalization,  coupled with interpolation or extrapolation of the data to adjust the bare masses
   in the subtraction in \eq{gamma-def}.   It is much easier to postpone the renormalization step until
   after the chiral fit, when we will have the ability to make these adjustments easily.   
   Fortunately, the functional form of the chiral fit that is  appropriate to the physical quantity   $(M^2_{xy})^\gamma$ 
 may also be applied to a fit of the unphysical intermediate quantity $\Delta M^2_{xy}$.  As we will see,
 the only consequence of fitting  $\Delta M^2_{xy}$ instead of $(M^2_{xy})^\gamma$  is that the former
 will have unphysical contributions to two LECs that are affected by EM renormalization.
 We therefore postpone detailed discussion of renormalization until \secref{renormalization}. 
 Except for some comments about the affected LECs, we ignore the difference between $\Delta M^2_{xy}$ and $(M^2_{xy})^\gamma$ in the current section.

Separating orders in the chiral expansion, we write the difference in \eq{Delta-def} as 
\begin{eqnarray}
\Delta M^2_{xy} &=&  \Delta_{\rm LO}M^2_{xy} + \Delta_{\rm NLO}M^2_{xy} + 
\Delta_{\rm NNLO}M^2_{xy} + \cdots, \eqn{Delta-orders} \\
\Delta_{\rm LO} M^2_{xy} &=&   q^2_{xy} \dem \eqn{Delta-LO}, \\
\Delta_{\rm NLO} M^2_{xy} &=&    \Delta^{\rm log}_{\rm NLO} M^2_{xy}+ \Delta^{\rm analytic}_{\rm NLO} M^2_{xy} \eqn{Delta-NLO},
\end{eqnarray}
where $\Delta_{\rm LO} M^2_{xy}$ is independent of taste. \Equation{Delta-LO} follows from \eq{LO-EM-mass}, 
and \eq{Delta-NLO} divides the NLO contribution into logarithmic (non-analytic) and analytic contributions.
For NNLO and higher orders, the chiral logarithms are not known; when such orders are
needed in the chiral fits, we therefore include
the analytic contributions only.
   
\begin{figure}[t!]
	\begin{center}
	\begin{tabular}{c c c c }
		\includegraphics[scale=0.1]{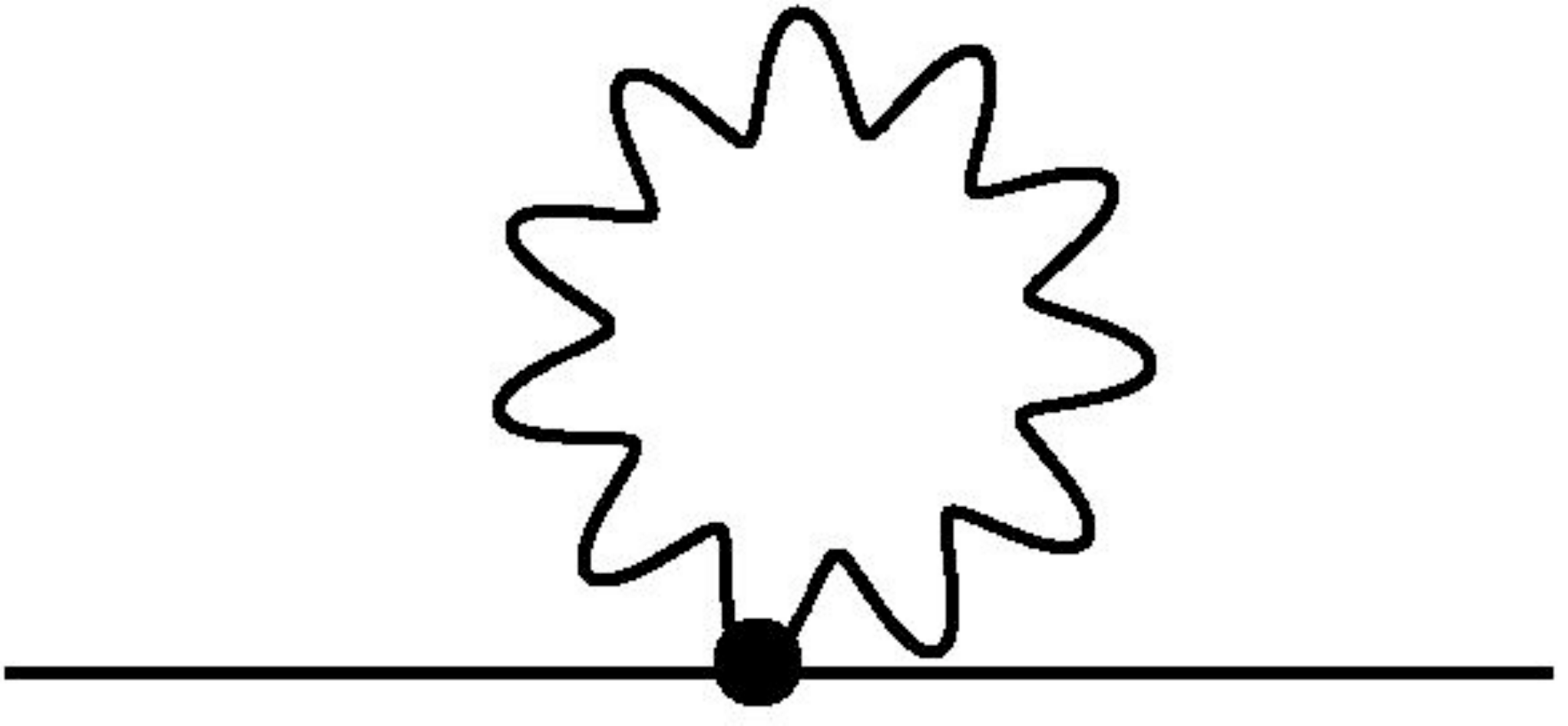}   
		&
		\includegraphics[scale=0.1]{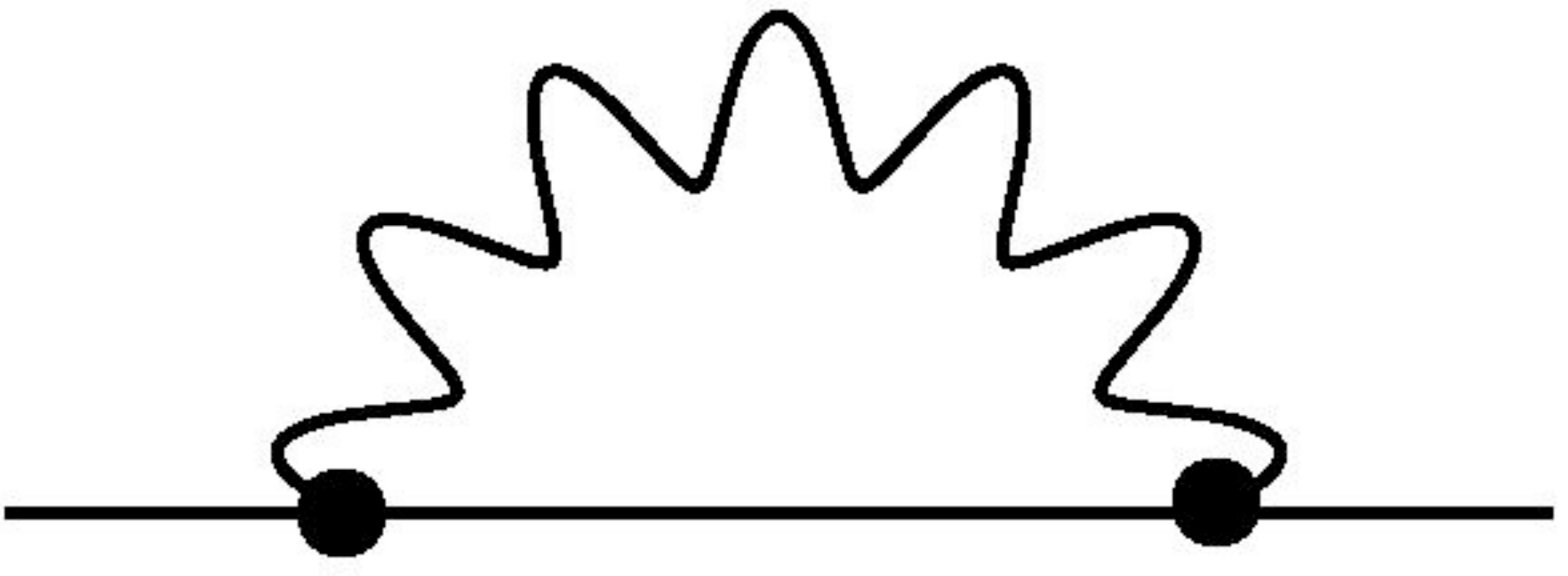}   
		&
		\includegraphics[scale=0.22]{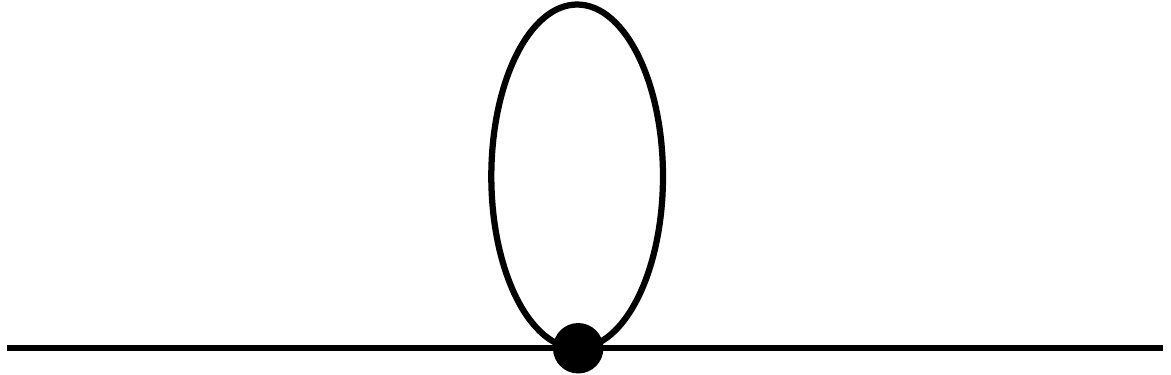}   
		&
		\includegraphics[scale=0.1]{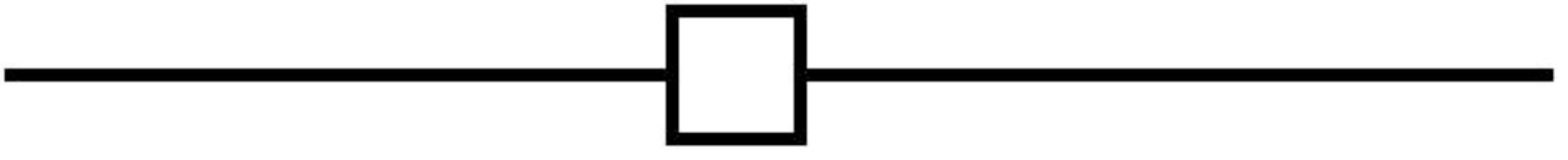}   
		 \\
		 (a)  &  (b)  &  (c)  &  (d) 
	\end{tabular}
	\end{center}
	\caption{ Feynman diagrams that contribute to the meson-mass at $O(p^4,e^2p^2)$.
	Straight lines are the pseudoscalar meson propagator and wiggly lines are the photon.
	A filled dot represents a vertex from the $O(p^2,e^2)$ Lagrangian, $\mathcal{L}^{({\rm LO})}$, while an open square represents an insertion of the $O(p^4,e^2p^2)$ Lagrangian, $\mathcal{L}^{({\rm NLO})}$.
	(a) photon tadpole;
	(b) photon sunset;
	(c) meson tadpole;
	(d) $O(p^4,e^2p^2)$ tree-level insertion.}
	\protect\label{fig:FeynDiag}
\end{figure}

The mass of the Goldstone meson has been
computed to NLO (one loop, $\cO(p^4,e^2p^2)$) in \rscpt\ with EM in \rcite{Bernard:2010qd}.
Figure~\ref{fig:FeynDiag} shows the NLO contributions to the meson mass.
The photon tadpole diagram does not contribute here since it vanishes in dimensional regularization; 
in FV, however, the momentum integral becomes a sum, and the photon tadpole is nonzero,
 as discussed in \secref{QEDFV}.
The photon sunset diagram is essentially the same as in the continuum, since the meson-photon vertex is 
taste-conserving, and the external pseudoscalar-taste meson is also the meson in the loop.
The calculation of the contribution from the meson tadpole, Fig.~\ref{fig:FeynDiag}(c), is very similar to 
that in \rcite{Aubin:2003mg}, with the addition of a new 4-meson vertex from the $C$ term in 
\eq{Lagrangian_stag}.   

The result of the calculation is that the NLO contribution to the squared mass splits into an
EM contribution proportional to $e^2$ and a non-EM contribution, which is identical to that in
Ref.~\cite{Aubin:2003mg}, and which cancels in the difference  $\Delta M^2_{xy,5}$, where we include
the subscript 5 to emphasize here that we are talking about the meson with taste $\xi_5$. The 
one-loop diagrams \figref{FeynDiag}(a)--(c) give
\bea
	\Delta_{\rm NLO}^{\rm log} M_{xy, 5}^2
	&=& - \frac{1}{16 \pi^2} \, e^2 q^2_{xy} \, \chi_{xy,5} \left[ 3 \ln (\chi_{xy,5}/\Lambda_\chi^2)   - 4  \right]   \nonumber \\
	 &&  -\frac{2 \dem}{16 \pi^2 f^2}  \left(\frac{1}{16}   \right) 
				  \sum_{\sigma,b}  \Big[   q_{x \sigma} q_{xy}  \,  
\ell(\chi_{x \sigma, b})
 				-   q_{y \sigma} q_{xy} \, \ell(\chi_{y \sigma, b})    \Big]  ,   \label{eq:schpt-logs}
\eea
where sea-quark flavors and the 16 meson tastes are labeled by $\sigma$ and $b$, respectively, $\Lambda_\chi$ is the chiral scale, and $\ell(\chi)$ is 
the renormalized loop integral 
\be
\int	\frac{d^4k}{\pi^2} \frac{1}{k^2 + \chi} \to \ell(\chi) \equiv \chi\ln(\chi/\Lambda^2_\chi).
\ee
The result in the first line in Eq.~(\ref{eq:schpt-logs}) is from the photon sunset 
diagram, \figref{FeynDiag}(b), 
and that in the second line is from the meson tadpole, \figref{FeynDiag}(c).   We have put the 
squared masses on the right-hand side to their values in the 
absence of EM  ($\chi_{xy,b}$), rather than
the full LO masses, \eq{LO-EM-mass}.  This change makes only a higher order, $\cO(e^4)$, difference. 
In chiral fits, we have also tried replacing $\chi_{xy,b}$ by the full LO masses in the one-loop terms; 
the small difference does not change either the quality of fits or the physical results significantly.

The contributions from Fig.~\ref{fig:FeynDiag}(d) lead to analytic contributions with unknown LECs.
It is useful to write these contributions in terms of natural  dimensionless variables of 
\rschpt\ \cite{Aubin:2004fs}
\begin{eqnarray}
\mu_i = \frac{2B_0m_i}{8 \pi^2 f_\pi^2}, \nonumber \\
\mu_{a^2} = \frac{a^2\bar \Delta}{8 \pi^2 f_\pi^2}, \eqn{x-def} 
\end{eqnarray}
where $i$ labels quark flavors, $i\in \{x,y,u,d,s\}$, and $\bar\Delta$ is the mean of the taste splittings $\Delta_b$ (weighted by multiplicities).
It is straightforward to find the possible analytic terms using the standard spurion approach, but it is much
quicker simply to write down all polynomials of a given order using the rules that follow from the symmetries:
\begin{enumerate}
\item Charge conjugation symmetry implies that a valence $x\bar y$ meson has the same mass
 as its antiparticle, the $y\bar x$ meson, so terms
 must be symmetric under the interchange  $q_x, \mu_x \leftrightarrow q_y, \mu_y$.
\item  In the absence of EM, the partially conserved staggered axial symmetry that rotates $x$ into $y$ quarks guarantees that $M^2_{xy,5}$ is proportional to  $m_x+m_y$ (times possible additional mass factors).  When EM is turned on, the symmetry is explicitly broken, but only for charged mesons ($q_x\not=q_y$).  Thus, when $q_x=q_y$, all terms must either vanish or be proportional to  $m_x+m_y$.     
\item  The fact that the sea quarks couple equally to valence quarks implies that terms must be
symmetric under sea-quark interchange:   $q_u, \mu_u \leftrightarrow q_d, \mu_d
 \leftrightarrow q_s, \mu_s \leftrightarrow q_u, \mu_u$.
 \item The sum of sea-quark charges vanishes in the two cases of interest here, the physical case and the electroquenched case.
 Therefore terms proportional to the sum $q_u+q_d+q_s$ may be dropped. 
\end{enumerate}
Given these rules, there are six independent analytic contributions possible at $\cO(e^2p^2)$ (NLO):
\begin{eqnarray}
 e^2 q^2_{xy} \mu_{a^2},\qquad e^2  q^2_{xy}(\mu_u + \mu_d+\mu_s),\qquad
 e^2(q^2_x+q^2_y)(\mu_x+\mu_y),\nonumber\\ 
 e^2 q^2_{xy}(\mu_x+\mu_y),\qquad e^2 (q^2_x\mu_x + q^2_y\mu_y), \qquad
 e^2 (q^2_u+q^2_d+q^2_s)(\mu_x+\mu_y).\eqn{list_NLO}
\end{eqnarray}
Of these, the last contribution will cancel for $\Delta M^2_{xy}$ since it is independent of the valence charges. 
The remaining contributions are independent of the sea-quark charges.  That means that sea-quark-charge dependence only enters at NLO in the chiral logarithms, \eq{schpt-logs}, and hence is computable,
as discovered by Bijnens and Danielsson \cite{Bijnens:2006mk}.%
\footnote{The  fact that $q_u+q_d+q_s=0$ is crucial to this conclusion.  If the sum of the sea-quark charges were not zero,
calculability of sea-quark charge dependence of $\Delta M^2_{xy}$ at NLO
would be spoiled, for example, by a term in $ M^2_{xy}$ 
proportional to $(q_x+q_y)(q_u+q_d+q_s)(\mu_x+\mu_y)$.  Such a term would be generated,
in the notation of  \rcite{Bijnens:2006mk}, by a contribution to the Lagrangian of the form $\Tr(\cQ_L)\Tr(\cQ_R u^\nu u_\nu) + (L\leftrightarrow R)$.}

The result of an $\cO(1)$ shift in the scale of the chiral logarithms suggests that an appropriate scale 
for the analytic contributions in \eq{list_NLO} is $f^2_\pi$ (from the first line in \eq{schpt-logs}) or $\deltae$
(from the second line in \eq{schpt-logs}).  In fact, these two quantities are the same order of magnitude, 
as can be seen by estimating
$\deltae$ by assuming that the experimental $\pi^+$--$\pi^0$ mass splitting comes entirely from the leading order contribution $\dem$.  We therefore choose the scale $f^2_\pi$ for all the NLO analytic contributions.
In addition, we find it helpful to include mean taste splittings  in analytic terms that absorb the chiral-scale
dependence coming from the meson tadpole, which has an average over tastes in
the second line in \eq{schpt-logs}.  As in \rcite{Aubin:2004fs}, this definition of the LECs at nonzero lattice
spacing simplifies the chiral-scale dependence of the LECs, and also tends to capture much
of the lattice-spacing dependence of the lattice data, reducing the size of the pure discretization
term (proportional to $\mu_{a^2}$) in the fit.   The NLO analytic contribution to
$\Delta M^2_{xy}$ is then
\begin{eqnarray}
\Delta^{\rm analytic}_{\rm NLO} M^2_{xy} = e^2f_\pi^2\;\Big[&&\kappa_1\, q^2_{xy} \mu_{a^2} + \kappa_2\,   q^2_{xy}(\mu_u \!+\! \mu_d\!+\!\mu_s\!+\!3\mu_{a^2}) 
+\kappa_3\, (q^2_x+q^2_y)(\mu_x+\mu_y)+ \nonumber\\
&&\  + \kappa_4\,  q^2_{xy}(\mu_x+\mu_y) 
+\kappa_5\, (q^2_x\mu_x+q^2_y\mu_y+q^2_{xy}\mu_{a^2})\Big].\eqn{anal_NLO}
\end{eqnarray}

The usual expectation would be that the dimensionless LECs $\kappa_i$ are  $
\cO(1)$.   However, several features of the current problem indicate that the 
expectation may be violated.  First of all, previous work, both in the continuum 
\cite{continuum-models}  
and on the lattice 
\cite{Blum:2010ym,BMW11,EM12,Basak:2013iw,deDivitiis:2013xla,Basak:2014vca,
Horsley:2015eaa,Fodor:2016bgu}, suggests that $\epsilon$ is large ($
\cO(1)$, rather than $\ll 1$), which would imply that the NLO terms produce $\cO(1)$ 
corrections to the LO result, and hence that at least some of the NLO LECs may be expected to
be significantly larger than 1.  
 A second issue arises from the nature of our data set.  Because the ensembles we study
 here all have a strange quark mass tuned to near the physical value ($m'_s\approx m_s$), and a light quark mass
 significantly lighter than that ($m'_l \le 0.2 m'_s$), the $\kappa_2$ term in \eq{anal_NLO} is approximately
 a constant up to discretization errors, and may therefore compete in the fit with the LO term $q^2_{xy}\dem$.
 In most fits, in fact, $\kappa_2$ has a tendency to get large and $\deltae$ to get small --- even negative in some cases!  This is a typical problem that occurs with SU(3) fits to data sets in which $m'_s$ does not take a
 significant range of values less than $m_s$.  Fortunately, the final results for physical quantities depend only
 mildly on the relative sizes of the LO and $\kappa_2$ terms.  In most of our fits, including
 the central fit, we simply set $\kappa_2=0$, but leave $\deltae$ unconstrained.   However we also consider fits where both $\deltae$ and $\kappa_2$ are unconstrained, as well as ones in which $\kappa_2$ is constrained
 by a prior that enforces $\kappa_2\ltwid1$.  Differences between results of these fits and the central one
 are included in an estimate of the systematic error of the chiral extrapolation.
 
A final complication is the fact that $\Delta M^2_{xy,5}$, the quantity we are fitting, includes unphysical
contributions because it has not been adjusted for
the effects of EM quark-mass renormalization.  In particular, the term multiplied by $\kappa_5$ 
in \eq{anal_NLO} is 
precisely of the form that would be induced by  the $\cO(\alpha_{\rm EM})$ EM renormalization of the quark 
masses $m_x$ and $m_y$, so $\kappa_5$ will have an unphysical 
renormalization contribution.  Indeed all fits that do not include an additional correction for renormalization
give $\kappa_5\approx 12$, with $\kappa_5=12.2(2)$ in the central fit.  After renormalization is
taken into account in some way, this effective value of $\kappa_5$ is significantly reduced.  On our central fit, the
 preferred nonperturbative scheme
described in \secref{renormalization} is nearly equivalent to simply
setting $\kappa_5=0$ after the fit.  With an $\msbar$ scheme and a perturbative determination of the
renormalization constant at one loop, $\kappa_5$ is reduced, effectively, by a factor of 2 but remains
clearly nonzero.

Beyond NLO, the \schpt\ logarithms have not been calculated, so we are unable to continue the
chiral expansion in a systematic fashion.  However, for acceptable chiral fits to the lattice data, we
must include some or all of the NNLO analytic terms, and at least one N${}^3$LO term.  
Following the symmetry rules above, the independent NNLO terms (for vanishing sea-quark charges) are
\begin{eqnarray}
\Delta^{\rm analytic}_{\rm NNLO} M^2_{xy} = e^2f_\pi^2\;\Big[&&\rho_1\, q^2_{xy} \mu^2_{a^2} +\rho_2\,   q^2_{xy} \mu_{a^2}(\mu_u \!+\! \mu_d\!+\!\mu_s\!+\!3\mu_{a^2}) 
+\rho_3\,(q^2_x+q^2_y) \mu_{a^2}(\mu_x+\mu_y)+ \nonumber\\
&&\hspace{-7mm}+ \rho_4\,  q^2_{xy} \mu_{a^2}(\mu_x+\mu_y) 
+\rho_5\, q_{xy}\mu_{a^2}[q_x(\mu_x+ \mu_{a^2})-q_y(\mu_y+\mu_{a^2})]+ \nonumber\\
&&\hspace{-7mm}+\rho_6\,q^2_{xy}(\mu_u \!+\! \mu_d\!+\!\mu_s\!+\!3\mu_{a^2})^2 +\rho_7\,q^2_{xy}(\mu^2_u + \mu^2_d+\mu^2_s)+
\nonumber\\
&&\hspace{-7mm}+\rho_8\,q^2_{xy}(\mu_x+\mu_y)(\mu_u \!+\! \mu_d\!+\!\mu_s\!+\!3\mu_{a^2})+ \nonumber\\
&&\hspace{-7mm}+\rho_9\, q_{xy}(\mu_u \!+\! \mu_d\!+\!\mu_s\!+\!3\mu_{a^2})[q_x(\mu_x+ \mu_{a^2})-q_y(\mu_y+\mu_{a^2})]+ \nonumber\\
&&\hspace{-7mm}+\rho_{10}\, (q^2_x+q^2_y)(\mu_x+\mu_y) (\mu_u \!+\! \mu_d\!+\!\mu_s\!+\!3\mu_{a^2})
+\rho_{11}\, (q^2_x+q^2_y)(\mu_x+\mu_y)^2+\nonumber\\
&&\hspace{-7mm}+\rho_{12}\, (q^2_x-q^2_y)(\mu^2_x-\mu^2_y)
+\rho_{13}\, q^2_{xy}(\mu_x+\mu_y)^2+\rho_{14}\, q^2_{xy}(\mu^2_x+\mu^2_y)\Big]+
\nonumber\\
+e^4f_\pi^2\;\Big[&&\rho'_1\, q^2_{xy}(q^2_x+q^2_y) + \rho'_2\, (q^2_x-q^2_y)^2
\Big],
\eqn{anal_NNLO}
\end{eqnarray}
where the terms with $\rho_i$ coefficients are $\cO(e^2p^4)$, and those 
 with $\rho'_i$ coefficients are $\cO(e^4)$.   Taste-splitting terms ($\mu_{a^2}$) have been added
 to mass terms ($\mu_j$) in plausible ways based on the example of the NLO chiral logarithms, but
 of course these choices are merely guesses of how best to absorb discretization errors into the mass
 terms.  
 
 \Equation{anal_NNLO} includes taste-violating analytic terms, such as the term 
 multiplied by $\rho_1$, that arise naturally in \rschpt.  
 However, lattice-spacing dependence can also arise simply from ``generic''
 discretization effects that break no continuum symmetries and therefore produce no new LECs.  Rather,
 they induce $a$-dependence in the LECs that are already present. While the leading taste violations  in QCD 
 with asqtad quarks are $\cO(\alpha^2_{\rm S} a^2)$, the leading generic errors are $\cO(\alpha_{\rm S} a^2)$.  
The quark couplings to EM do not change the leading generic errors because the combination of paths in the
asqtad action removes $\cO(a^2)$ terms as always.  However, the EM gauge action we use is unimproved
and therefore induces $\cO(a^2)$ generic errors.\footnote{The fact that the $a^2$ errors occur in the EM sector,
and therefore automatically come with a factor of $\alpha_{EM}$ in quark quantities, does not help here 
because we are focusing on EM quantities, which have that same overall $\alpha_{EM}$ factor.}

   Generic discretization errors
  of the   NLO analytic parameters $\kappa_1,\dots,\kappa_5$ in \eq{anal_NLO} may produce effects
  of a size comparable to that from the NNLO parameters, so should be included.  Even more important, a generic error on
  the LO parameter $\Delta_{\rm EM}$ may induce effects comparable to NLO and is therefore required in our fits. 
  We thus include six generic variation parameters $\psi_0,\dots,\psi_5$ that give $a$-dependence to the LO
  and NLO LECs:
  \begin{eqnarray}
  \Delta_{\rm EM}(a)  & = & \Delta_{\rm EM}\left(1 + \psi_0\,  \frac{a^2}{r^2_1} \right), \eqn{generic_LO}\\
  \kappa_i(a)  & = & \kappa_i\left(1 + \psi_i\, \frac{a^2}{r^2_1}\right),\qquad (i=1,\dots,5). \eqn{generic_NLO}
  \end{eqnarray}
  The parameters $ \Delta_{\rm EM}$ and $\kappa_i$ on the right-hand side here are the continuum ($a=0$) 
  values.  In \eqs{generic_LO}{generic_NLO} we have assumed $\cO(a^2)$ generic errors.
  However, we also make fits assuming $\cO(\alpha_{\rm S} a^2)$ errors, and include the results of
  those fits in our systematic error estimates.  In practice, it makes little difference whether we assume
 $\cO( a^2)$ or  $\cO(\alpha_{\rm S} a^2)$ generic errors. Fits with the former actually tend to  have slightly lower $p$
 values and slightly larger statistical errors.  Nevertheless, they are preferred because the leading errors
 are $\cO(a^2)$.

 At N${}^3$LO, possible terms are $\cO(e^4 p^2)$ or $\cO(e^2 p^6)$.  The latter are not necessary for 
 good fits on any subsets of our data that we have considered, and we do not discuss them further here.
 The former are necessary, especially when we include data with quark charges larger than their physical
 values. The  independent N${}^3$LO $\cO(e^4 p^2)$ terms are
 \begin{eqnarray}
\Delta^{\rm analytic}_{\rm N{}^3LO} M^2_{xy} = e^4f_\pi^2\;\Big[&&
\lambda_1\, q^2_{xy}(q^2_x+q^2_y) \mu_{a^2}+ \lambda_2\, (q^2_x-q^2_y)^2\mu_{a^2}+\nonumber \\
&&+\lambda_3\, q^2_{xy}(q^2_x+q^2_y) (\mu_u \!+\! \mu_d\!+\!\mu_s\!+\!3\mu_{a^2})+\nonumber \\
&&+ \lambda_4\, (q^2_x-q^2_y)^2(\mu_u \!+\! \mu_d\!+\!\mu_s\!+\!3\mu_{a^2})
                      +\lambda_5\, (q^4_x+q^4_y) (\mu_x+\mu_y)+ \nonumber \\
&&+ \lambda_6\, (q^4_x\mu_x+q^4_y\mu_y)+\lambda_7\, q_x q_y(q^2_x+q^2_y) (\mu_x+\mu_y)+\nonumber\\ 
&&+ \lambda_8\, q_x q_y(q^2_x\mu_x+q^2_y\mu_y)+ \lambda_9\, q^4_{xy} (\mu_x+\mu_y) 
\Big].
\eqn{anal_NNNLO}
\end{eqnarray}
When the charges of the quarks in the mesons are limited to physical values or smaller ($\pm 2e/3$,
$\pm e/3$, or $0$), only the $\lambda_6$ term is necessary for acceptable fits, and its value is
$\approx\!4$. (The central fit gives $\lambda_6=4.1(1)$.)   Note that this term has the form of an $\cO(e^4)$ quark mass renormalization.  This implies
that $\lambda_6$, like the NLO LEC $\kappa_5$ (\eq{anal_NLO}), has an unphysical renormalization
contribution.  We note that, even though fits with $\lambda_6$ set to zero have very low $p$ values, 
$< 10^{-10}$, the term has little effect on the physical quantities studied here.  In particular, if we
simply set $\lambda_6=0$ after the fit, these quantities change by amounts less than or equal to their
statistical errors, and much less than their total (systematic plus statistical) errors.

\subsection{Electromagnetic quark-mass renormalization}
\label{sec:renormalization}

In this section, we discuss the renormalization of quark masses due to EM effects, \ie $\cO(e^2)$ or
higher.  This is important because the multiplicative renormalization factor $Z_m$ is different for quarks with different EM charges, and thus affects how we separate ``true'' EM effects from quark mass effects such as isospin violations.
Because we are not interested here in determining absolute, physical quark masses (\eg $\msbar$ quark masses in MeV, say),  renormalization due to the strong interactions alone can be ignored since the 
corresponding $Z_m$ is the same for all quark flavors.  Therefore, when we refer in this paper to ``renormalized''
or ``bare'' quark masses, we mean renormalized or bare with respect to EM.  All quark masses discussed
are bare as far as the strong interactions are concerned.

It is instructive first to estimate the size of the EM renormalization effect on the determination of $\epsilon$. 
 At fixed lattice spacing $a$, let $\delta u$ and $\delta d$ be the fractional shift in the $u$ and $d$ bare masses such that their renormalized
 EM masses are both equal to $m_l$.  At $\cO(e^2)$, we have
\begin{equation}\eqn{renorm-est1}
m_l(\delta u - \delta d) = C\; \frac{(q_ue)^2-(q_de)^2}{4\pi}\, m_l = C\; \frac{\alpha_{\rm EM}}{3}\, m_l.
\end{equation}
Assuming that the size of any logarithms in $a\mu$ remains modest  ($\mu$ is the scale of the
renormalized masses),  the constant  $C$ is expected to be of order 1.  With $\alpha_{\rm EM}\sim 0.01$, this gives 
$ \delta u- \delta d \sim 0.003$.  Compared to the experimental pion splitting, the induced 
mass-squared splitting between a $K^+$ and a $K^0$ is then approximately 
\begin{equation}\eqn{renorm-est2}
\frac {B_0\,m_l\,(\delta u- \delta d)}{m_{\pi^+}^2 -m_{\pi^0}^2} \sim \frac {0.003\; m_{\pi}^2/2}{m_{\pi^+}^2 -m_{\pi^0}^2} \sim 0.02\;.
\end{equation}

Our estimate of the EM renormalization effect on $\epsilon$ is thus quite small, $0.02$.  
 The reason the effect is small is that the residual chiral symmetry of staggered
quarks guarantees that the renormalization is multiplicative, so that the shifts in the $u$- and $d$-quark
masses are small.  The shift in the $s$-quark mass is much larger; however, its effect cancels in
$\epsilon$ between $M^2_{K^+}$ and $M^2_{K^0}$.   On the other hand, for quantities
such as $(M^2_{K^0})^\gamma$, the EM effect on the squared $K^0$ mass itself, the fractional systematic
error from not including renormalization effects is at least an order of magnitude larger
than for $\epsilon$.  One must  also keep
in mind that the estimate in \eq{renorm-est2} is qualitative, and could easily be off by a factor of 3 or more
if $C$ is larger or smaller than naively expected.

We now proceed to more detailed discussion of perturbative renormalization,
which converts bare quark masses to  $\msbar$ renormalized masses at
some convenient scale, here taken to be $\mu=2\;$GeV.  Only a one-loop determination is available in
the literature.
For staggered quarks in QCD, the renormalized $\msbar$ mass is given at this order
 in terms of the bare mass $m(a)$ at lattice spacing $a$
by \cite{Aubin:2004ck}
\begin{eqnarray}\eqn{QCD-mass-renorm}
m^\msbar(\mu) &=& Z_m \; m(a) = \left(1 +\alpha_V(q^*)Z_m^{(2)}(a\mu) \right) m(a),\\
Z_m^{(2)}(a\mu) &=&  b-\frac{4}{3\pi} -\frac{2}{\pi}\ln(a\mu),
\eqn{Z-def}
\end{eqnarray}
where  $\alpha_V(q^*)$ is the strong coupling
in the $V$ scheme  \cite{Lepage:1992xa} evaluated at scale $q^*$, and $b$
is a constant depending on the details of the staggered action. We have neglected 
discretization corrections of $\cO((am)^2)$. 

In order to find the corresponding EM renormalization
for staggered quarks, we merely have to remove the overall
SU(3) Casimir  factor of $4/3$ from $Z^{(2)}_m$ and to
replace  $\alpha_V(q^*)$ with  $\alpha_{\rm EM}=e^2/(4\pi)$.
Issues such as the proper scheme and scale $q^*$ for $\alpha_{\rm EM}$ are irrelevant 
since $\alpha_{\rm EM}$ is so small compared to $\alpha_{\rm S}$, and hence runs very slowly.  Because
we do not include EM corrections to the QCD tadpole factors in the asqtad action, 
we take  $b=2.27$ \cite{Aubin:2004ck}, which corresponds to the
case of asqtad-like smearing without tadpole improvement.  
The one-loop EM renormalization is then
\begin{equation}\eqn{EM-mass-renorm}
\delta m \equiv \left(m^\msbar(\mu)  -m(a)\right)_{\rm EM} =  q^2e^2m(a)\left(c -\frac{3}{8\pi^2}\ln(a\mu)\right),
\qquad c=0.110,
\end{equation}
where $q$ is the charge of the quark in units of $e$.   

The EM renormalization first affects $\Delta M^2_{xy}$ at NLO in \chpt.  
To include one-loop renormalization in the chiral fit at this order, we simply add
\begin{equation}\eqn{renorm-term}
\Delta_{\rm renorm} M^2_{xy} =
B_0\left(\delta m_x + \delta m_y  \right)
\end{equation}
to \eq{Delta-orders}.
Note that changes in $\mu$ can then be absorbed in the chiral fits by changes in the 
NLO LECs:   $\kappa_5$ and (if discretization
effects are important) $\kappa_1$,  \eq{anal_NLO}.
After the fit, the effect of \eq{renorm-term} is removed from the result.  This procedure  is equivalent
to readjusting the bare quark masses so that the renormalized masses have the desired value,
so that, in particular, $m_u^\msbar(\mu)=m_d^\msbar(\mu)$.  As discussed below in
\secref{results}, the net result is that including
the one-loop EM renormalization would shift $\epsilon$ by 0.03, with small variations
depending on the
details of the fit.
This is consistent with (but somewhat larger than) the order-of-magnitude estimate of the effect made
above.
Based on this small shift, which is significantly less than the other systematic errors in our result, our approach in
preliminary calculations \cite{basak,EM12,Basak:2013iw,Basak:2014vca} was to omit renormalization in the central value, and simply include
an estimate of the effect in the systematic errors.   However, \eq{EM-mass-renorm} will get strong 
corrections starting at two loops, \ie $\cO(\alpha_{\rm S} e^2)$, and  experience
from pure-QCD quark mass renormalization
suggests  that
we would need the corrections through $\cO(\alpha_{\rm S}^2e^2)$ to be able
to be confident of the coefficient of $e^2$ at the few percent level.%
\footnote{Compare for example the one-loop result for the strange
quark mass in
\rcite{Aubin:2004ck} with the two loop result of \rcite{Mason:2005bj}.} 
We are thus only able to  take \eq{EM-mass-renorm} as a qualitative estimate of the EM renormalization effect in the
$\msbar$ scheme. 

In the absence of high-order perturbative calculations,  a nonperturbative determination
of the EM renormalization is necessary to get reliable results.  As we will see below, such a nonperturbative
approach yields an estimate for the effect of EM renormalization on $\epsilon$ of approximately 0.07,
a bit more than twice as large as the one-loop perturbative estimate.

The nonperturbative method we use has been proposed by the BMW Collaboration \cite{Borsanyi:2013lga}.
The idea is to compare the masses of neutral $\pi^0$-like mesons constructed from $u\bar u$ quarks
and $d\bar d$ quarks with quark-line connected propagators only (no intermediate states with only gluons
and/or photons are allowed).  

We first introduce the needed connected correlators for
arbitrary valence quarks $x$ and $y$. The connected $x\bar x$ and
 $y\bar y$ correlators are explicitly constructed in PQQCD by introducing additional
valence flavors $x'$ and $y'$  with $q_{x'}=q_x$, $m_{x'}=m_x$ and $q_{y'}=q_y$, $m_{y'}=m_y$.  The
connected correlators are then
\begin{eqnarray}\eqn{connected-xx}
G_{xx'}(t) &=& \frac{1}{2}\sum_{\vec z}\langle \bar x(t,\vec z)\gamma_5 x'(t,\vec z)\ \bar x'(0)\gamma_5 x(0)\rangle, \\
G_{yy'}(t) &=& \frac{1}{2}\sum_{\vec z}\langle \bar y(t,\vec z)\gamma_5 y'(t,\vec z)\ \bar y'(0)\gamma_5 y(0)\rangle,  \eqn{connected-yy} 
\end{eqnarray}
where disconnected contributions are absent since $x$ and $x'$ are different quarks, so $x$ cannot 
contract with $\bar x'$ (and similarly for $y$ and $\bar y'$).   We let  $M_{xx'}$ and $M_{yy'}$ be their masses.
These mesons are each of the form discussed in rule 2 above \eq{list_NLO}:  neutral mesons
composed of two different, but equally charged, quarks. The EM contributions to $M_{xx'}$ and $M_{yy'}$ must
therefore be proportional to $B_0\,q_x^2\,e^2(m_x+m_{x'})$ and $B_0\,q_y^2\, e^2(m_y+m_{y'})$, respectively, where
we have inserted the factor of $B_0$ to put these contributions in units of squared meson mass.  For $q_x = 2/3$,
$m_x \!\sim\! m_u$, and $q_y=-1/3$, $m_y \!\sim\!  m_d$, the contributions are of order
 $\alpha_{\rm EM} M_\pi^2$.    This is much smaller than the effect of isospin violation on the squared mass difference  
 $M^2_{xx'} - M_{yy'}^2$, which is
$B_0(m_x-m_y) \!\sim\! M_\pi^2\;$  for approximately physical mass of the quarks, since the quark mass difference
is of the same order as the masses themselves.   

To lowest nontrivial order in $\alpha_{\rm EM}$, we may therefore define an isospin limit by adjusting the bare masses
$m_u$ and $m_d$ such that $M^2_{uu'} = M^2_{dd'}$ \cite{Borsanyi:2013lga}.  This is not by itself a sufficient renormalization condition, however,
since it does not fix the overall scale of the light quark masses.  We can do that by demanding that the
renormalized mass of the $u$ and $d$ quarks is the same as their mass in  pure isospin-symmetric QCD,
the theory onto which we are matching our QCD+(quenched)QED theory.  
Since chiral symmetry requires that the EM effects on the
mass of the physical $\pi^0$ are also of order $\alpha_{\rm EM} M_\pi^2$, the pion mass in pure QCD may be taken to have
the experimental mass of the $\pi^0$, $M_{\pi^0, {\rm expt}}$.
This leads to a nonperturbative EM renormalization condition.  In the QCD+(quenched)QED theory  we adjust the bare masses $m_u$ and $m_d$ to enforce  
\begin{equation}\eqn{isospin-limit-def}
M^2_{uu'} = M^2_{dd'} =  (M^2_{\pi})^{\rm QCD} \equiv M^2_{\pi^0, {\rm expt}}.
\end{equation}
We call the renormalization scheme defined by this condition the ``BMW scheme."
A related ``Dashen scheme'' has been introduced by the QCDSF 
Collaboration \cite{Horsley:2015eaa}.  In their scheme, the masses of connected $u\bar u$, $d\bar d$ and $s\bar s$ mesons are all set equal at a symmetric point.

We define the mass $m_l$  as the common $u$, $d$ mass such that the charged pion 
in our pure QCD simulations has
mass $(M_{\pi})^{\rm QCD}$.  Therefore, \eq{isospin-limit-def} may be enforced by setting
\begin{equation}\eqn{mu_md}
m_u = m_l ( 1- \delta_u),\qquad m_d = m_l ( 1- \delta_d)
\end{equation}
and choosing $\delta_u$ and $\delta_d$ so that the EM contributions to $M^2_{uu'}$ and $M^2_{dd'}$
vanish:
\begin{equation}\eqn{isospin-limit1}
\uuEM = 0 = \ddEM.
\end{equation}
Recall that $(M^2)^\gamma$ is defined as the difference between the squared mass of the meson composed of
charged quarks with that composed of uncharged quarks, but with the  same {\em renormalized masses}\/. 
In \eq{isospin-limit1} the EM renormalized mass is $m_l$, so that the neutral-quark (pure QCD) subtraction
terms in the definition of \uuEM\ and \ddEM\  (see \eq{gamma-def}) are equal to $(M^2_{\pi})^{\rm QCD}$.
Thus \eq{mu_md} should be interpreted as defining the bare masses $m_u$ and $m_d$ such that the
EM renormalized mass of each quark is $m_l$. 

The condition \eq{isospin-limit1} must then be rewritten in terms of \Deluu\ and \Deldd, the EM effects at fixed bare mass (see \eq{Delta-def}),
which are the quantities we directly compute and fit in our simulations.  With the bare mass fixed at
$m_u$ in \Deluu, and at $m_d$ in \Deldd, the charged-quark terms in \Deluu\ and \Deldd\ are the same
as in \uuEM\ and \ddEM, respectively, but the neutral quark subtraction terms are different.  Within the
approximation that $M^2_{xy} = B(m_x+m_y)$ in pure QCD, we may easily correct for the changed
subtraction terms and rewrite \eq{isospin-limit1} as
 \begin{equation}\eqn{isospin-limit2}
\Deluu(m_u)- 2B\,m_l\,\delta_u =  0 = \Deldd(m_d) - 2B\, m_l \,\delta_d.
\end{equation}
After a chiral fit to the data for $\Delta M^2_{xy}(m_x,m_y)$, we solve these conditions iteratively for $\delta_u$ and $\delta_d$ at each lattice spacing, or in the fit extrapolated to the continuum.  Iteration is in 
principle necessary because $\Deluu$ and $\Deldd$ depend nonlinearly on $\delta_u$ and $\delta_d$, respectively, at
fixed $m_l$.   However, since $\delta_u$ and $\delta_d$ are $\cO(\alpha_{\rm EM})$, one could simply evaluate \Deluu\ and \Deldd\ in \eq{isospin-limit2} at $m_l$ with negligible changes
to our final results.    For $B$,  we use the derivative with
respect to $2m_l$ of the NLO SU(2) \chpt\ result for $M^2_\pi$ in QCD:
\begin{equation}\eqn{B}
B = \frac{(M^2_\pi)}{2m_l}\left( 1 - \bar\ell_3\frac{M^2_\pi}{16\pi^2 f_\pi^2}\right)
\end{equation}
with $\bar\ell_3 = 2.81(64)$ \cite{Aoki:2016frl}.  Systematic errors associated with the value of $B$
are included in our error analysis in \secref{results}. 

The residual chiral symmetry of staggered quarks implies that quark mass normalization is multiplicative.
 That means that once we know $\delta_d$, we can use it to renormalize
 any charge-$1/3$ quark.  In particular, in this scheme the bare strange quark mass $m_S$ whose
EM renormalized mass is $m_s$, the known physical strange mass in pure QCD, is
\begin{equation}\eqn{ms}
m_S = m_s ( 1- \delta_d) .
\end{equation}
Once the strange quark mass has been renormalized, we may compute \KEM, the  EM effect on the neutral kaon,
from 
\begin{equation}\eqn{get-KEM}
\KEM = \DelK - B_s(m_s - m_S) -B_l(m_l-m_d),
\end{equation}
where $B_s$ and $B_l$ are the derivatives of $(M_K^2)^{\rm QCD}$ with respect to $m_s$ and $m_l$,
respectively.  
 Unfortunately, because a large fraction of \DelK\ is unphysical, and removed when constructing
 \KEM\ in the  renormalization step, the resulting systematic error in \KEM\ (or equivalently
 \ek, \eq{ek-def})  is relatively large ($\sim\!35\%$). 
 The result is particularly sensitive to the uncertainty in the derivative $B_s$.
 
We emphasize here two contrasting points about our renormalization scheme.  On the one hand, if we keep $\delta_u$ and $\delta_d$ in \eq{mu_md} fixed, we can replace $m_l$, the average physical $u,d$ mass, with any mass $m'_l$, and thereby find bare masses $m_u$ and $m_d$ that both have renormalized masses equal to $m'_l$.  On the other hand,  for masses $m'_l > m_l$ it is {\em not}\/ true that the resulting EM
contributions to $\uuEM$  and $\ddEM$ vanish or even remain equal to each other.   The condition in
\eq{isospin-limit2} may only be enforced at one value of $m'_l$, and it is only when we enforce it at or near
$m'_l=m_l$, as we do, that the terms we set to zero are necessarily small, of second order 
in a joint expansion in $\alpha_{\rm EM}$ and isospin violations. As a numerical test of the latter point,
we computed $\uuEM$  and $\ddEM$ for $m'_l=m_s/2$, \ie for a heavy pion with mass approximately
equal to the mass of the kaon.  We obtain  $\uuEM\approx 82\ ({\rm MeV})^2$ and 
$\ddEM\approx 21\ ({\rm MeV})^2$, which are nonnegligible and of the same order of magnitude as our
result for  \KEM.

A final renormalization scheme that we have tried consists of simply setting to zero after the chiral fit 
the two LECs, $\kappa_5$
and $\lambda_6$, that are dominated by unphysical renormalization effects at $\cO(\alpha_{EM})$ and
$\cO(\alpha^2_{EM})$, respectively.  Interestingly, this ``LEC scheme'' gives results for the central fit 
that are extremely close to those obtained
from the BMW scheme:   $\epsilon$ differs only by 0.03\%; \ek, by 0.2\%.
However, the results from different chiral fits vary much more with the LEC scheme than with the BMW one;
this is especially true of \ek, which can differ by more than 100\% as we change
the details of the fit, or the ranges of
valence masses and charges included.  For this reason we do not consider the LEC scheme further here.

\subsection{The Neutral Pion}   \label{sec:pi0}

The mass of the (partially quenched) $\pi^0$ comes from the correlator
\begin{equation}\eqn{truepi0}
G_{\pi^0}(t) = \frac{1}{2}\sum_{\vec z}\langle [\bar x(t,\vec z)\gamma_5 x(t,\vec z)- \bar y(t,\vec z)\gamma_5 y(t,\vec z)]\;[\bar x(0)\gamma_5 x(0)- \bar y(0)\gamma_5 y(0)]\rangle,
\end{equation}
where $x$ is an up-type valence quark with $q_x=2/3$,  $y$ is a down-type valence quark with $q_y=-1/3$, and we work in the isospin limit $m_x=m_y$.  (For simplicity, all quark masses in this subsection should
be interpreted as renormalized masses.)
This true $\pi^0$ has quark-line disconnected EM contributions because 
$q_x\not=q_y$. As mentioned in the introduction, such disconnected contributions would be costly to 
compute numerically, so we drop them. We define the squared mass ``$M^2_{\pi^0}$'' as a simple
average of the squared masses coming from the two connected correlators,
one for $x$ and one for $y$, 
obtained from 
\eqs{connected-xx}{connected-yy}, respectively.  We can now define
\begin{equation}\eqn{fakeMpi0}
M^2_{{\rm ``}\pi^0{\rm "}} = \frac{1}{2}(M^2_{xx'} + M^2_{yy'}).
\end{equation}

It is then easy to see that chiral symmetry implies that
$M^2_{{\rm ``}\pi^0{\rm "}}$ vanishes in the (two-flavor) chiral limit.  That is because both
$M^2_{xx'}$ and $M^2_{yy'}$ are of the form discussed in rule 2 above \eq{list_NLO}:  neutral mesons
composed of two different, but equally charged, quarks.  The EM contributions to their masses must
therefore be proportional to $e^2(m_x+m_{x'})=e^2(m_y+m_{y'})\propto e^2 M_\pi^2$.  Chiral symmetry
also implies that the EM contributions
to the true $M^2_{\pi^0}$ must be proportional to $e^2 M_\pi^2$, but the reasoning is slightly
different because $M^2_{\pi^0}$ is not of the form $M^2_{xx'}$ with $x$ and $x'$ different flavors.  
The spontaneously broken chiral symmetry associated with the $\pi^0$ is diagonal
and is not broken explicitly by the also-diagonal quark-charge matrix $Q$.  Hence the EM
contribution to its mass must vanish as usual in the two-flavor chiral limit.  We may
make a rough estimate of the size of $(M^2_{\pi^0})^\gamma$ by using the chiral logarithm contribution
calculated in \cite{Urech:1994hd},
$e^2\deltae M_\pi^2 (\ln (M_\pi^2/\Lambda^2_\chi) +1)/(8 \pi^2 f^2)$, and
taking $\deltae= 4C/f^2 \approx 4C/f_\pi^2\cong9900\, ({\rm MeV})^2$ from \cite{Bijnens:2006mk} and $\Lambda_\chi=m_\rho=0.77$ GeV.  This gives a magnitude of about
$30\; {\rm MeV}^2$.

The $\pi^+$ has totally different behavior from either the ``$\pi^0$'' or the $\pi^0$. Since its chiral symmetry
is broken explicitly by the quark charges, $\Delta M^2_{\pi^+}$ is nonvanishing in the 
two-flavor
chiral limit at leading order, and equal to $e^2\deltae$.  At NLO, \eqs{schpt-logs}{anal_NLO} 
show that there are both 
a chiral log and an analytic contribution  (from the $\kappa_2$ term) proportional to 
$e^2M_K^2$.  We may estimate the size of $\Delta M^2_{\pi^+}$ from the LO term,
\eq{Delta-LO}, and the NLO chiral logarithm contribution proportional to $e^2M_K^2$  in the continuum 
limit.  This gives $\Delta M^2_{\pi^+} \approx 1050 \;  {\rm MeV}^2$.  
Alternatively, since  $\Delta M^2_{\pi^+}$ is so much larger than  $\Delta 
M^2_{\pi^0}$, and since the $u$--$d$ quark mass difference contributes so little
to the $\pi^+$--$\pi^0$ splitting, we may simply use the experimental splitting
$M_{\pi^+}^2-M^2_{\pi^0}= 1261 \;  {\rm MeV}^2$ as an estimate
of  $\Delta M^2_{\pi^+}$.    Either way,  it is clear that  $\Delta M^2_{\pi^+}\gg
 \Delta M^2_{\pi^0}$.   

 Since both  $\Delta M^2_{\pi^0}$ and  $\Delta M^2_{{\rm ``}\pi^0{\rm "}}$
 are $\cO(\alpha_{\rm EM}M_\pi^2)$, the error due to the simulation of the ``$\pi^0$"
 rather than the $\pi^0$ is also $\cO(\alpha_{\rm EM}M_\pi^2)$.  We estimate the size of this systematic error in
 \secref{pi0-error}.

\section{QED in Finite Volume \label{sec:QEDFV}}

With the noncompact realization of QED on the lattice, which we use, it is necessary to drop some zero-modes in a finite volume in order to have 
a convergent path integral. In particular, the action in Coulomb gauge for the zero component of the vector potential, $A_0$, is
$\frac{1}{2}\int\left(\partial_i A_0\right)^2$.  Since the $A_0$ mode with spatial momentum $\vec k=0$ has vanishing action, it must be dropped.
Similarly, the action for the spatial components $A_i$ is $\frac{1}{2}\int\left[\left(\partial_0 A_i\right)^2+\left(\partial_j A_i\right)^2\right]$.  Here
only the mode with 4-momentum $k_\mu=0$ must be dropped, and that is what we do.
This version of QED in FV was first introduced by Duncan, Eichten and Thacker 
\cite{Duncan:1996xy}; 
following the nomenclature in Borsanyi \etal\ \cite{Borsanyi:2014jba}, we call the resulting theory 
\qedtl.   Summarizing, \qedtl\ is defined in Coulomb gauge by
\begin{eqnarray}
A_0(k_0,\vec k=0) &=& 0,\ \ \forall k_0,\nonumber\\
\vec k\cdot \vec A(k_0,\vec k) &=&0, \ \ \forall k_0,\vec k, \qquad [\qedtl] \nonumber \\
\vec A(k_0=0,\vec k=0) &=& 0.\eqn{qedtl-def}
\end{eqnarray}

Hayakawa and Uno, in their calculation of EM FV effects in \chpt\ \cite{Hayakawa:2008an}, introduce
a different FV action, called QED$_L$, in which they drop all all  modes with $\vec k=0$, both 
for  $A_0$ and for $A_i$.  Again in Coulomb gauge, \qedl\ is defined by
\begin{eqnarray}
A_0(k_0,\vec k=0) &=& 0,\ \ \forall k_0,\nonumber\\
\vec k\cdot \vec A(k_0,\vec k) &=&0, \ \ \forall k_0,\vec k,\qquad [\qedl] \nonumber \\
\vec A(k_0,\vec k=0) &=& 0, \ \ \forall k_0. \eqn{qedl-def}
\end{eqnarray}
The difference  between \eqs{qedtl-def}{qedl-def} is solely in the last line of each, in the treatment of $\vec A$
when $\vec k=0$.  
This difference implies that
the FV effects in the MILC calculations are different from those computed in \rcite{Hayakawa:2008an}.

To make explicit the difference between our set-up (\qedtl) and that of Ref.~\cite{Hayakawa:2008an} (\qedl), we give the Coulomb gauge photon propagator in each case:
\vspace{-1mm}
\begin{eqnarray}
\cD_{ij}(k)\equiv\langle A_i (k) A_j(-k)\rangle &= 
 &\begin{cases}\frac{1}{k^2}\left(\delta_{ij}-\frac{k_ik_j}{\vec k^2}\right), & \vec k \not=0;\\
0\ , & \vec k=0.
\end{cases} \hspace{22mm}[\qedl] \eqn{H-U-prop}\\
\cD_{ij}(k)\equiv\langle A_i (k) A_j(-k)\rangle &=
 &\begin{cases}
\frac{1}{k^2}
 \left(\delta_{ij}-\frac{k_ik_j}{\vec k^2}\right), & \vec k \not=0; \\
\frac {1}{k^2}\delta_{ij}\ , & \vec k=0,\ k_0\not=0;\hspace{10mm}[\qedtl]\\
0\ , & \vec k=0,\ k_0=0.
\end{cases}\eqn{MILC-prop}\\
\cD_{00}(k)\equiv\langle A_0(k) A_0(-k)\rangle &= 
 &\begin{cases}\frac{1}{\vec k^2}, & \vec k \not=0;\\
0\ , & \vec k=0.
\end{cases} \hspace{22mm}[\qedl\ {\rm and}\ \qedtl]  \eqn{A0-prop}
\end{eqnarray}
 The violation of Gauss's Law induced by the absence of the $\vec k=0$ $A_0$ mode makes it possible to have net charges on a FV torus with periodic boundary 
 conditions \cite{Hayakawa:2008an}.  But Gauss's Law has no implications for the spatial modes $A_i$, so does not distinguish 
 between \eqs{H-U-prop}{MILC-prop}.
 
 Borsanyi \etal\ \cite{Borsanyi:2014jba}  have independently 
 studied  QED in FV, using both the \qedl\ and  \qedtl\  versions.  They
 define \qedl\ by 
  \begin{equation}\eqn{QED_L}
 \sum_{\vec x} A_{\mu,x_0,\vec x} = 0,\ \ \forall x_0,\mu.
 \end{equation}
 This is in fact a partial gauge specification, because spatially-independent, 
 but time-dependent, gauge transformations would violate the $\mu\!=\!0$ condition  
$A_0(k_0,\vec k\!=\!0)=0$ (written here in momentum space). 
One can bring any EM gauge field that 
satisfies \eq{QED_L} into Coulomb gauge, as was assumed in writing \eq{H-U-prop}. 
The necessary
 gauge transformation is, in momentum space:
\begin{eqnarray}
	A_\mu(k_0,\vec k) &\to &A_\mu(k_0,\vec k) -ik_\mu \Lambda(k_0,\vec k)\eqn{gauge-transf}\\
	\Lambda(k_0,\vec k) &=& 
	\begin{cases}
-\frac{i\vec k\cdot\vec A(k_0,\vec k)}{\vec k^2}\ ,
 & \vec k \not=0; \\
0\ ,  & \vec k=0.\vspace{-2.5mm}
\end{cases},\eqn{GT-QED_L}
\end{eqnarray}

 Borsanyi \etal\  define  \qedtl\ by
 \begin{equation}\eqn{QED_TL}
 \sum_x A_{\mu,x} = 0,\ \ \forall \mu.
 \end{equation}
 Unlike \eq{QED_L}, this definition is gauge invariant, as can be immediately seen from \eq{gauge-transf}.   \Equation{QED_TL} can be put into a special Coulomb gauge that satisfies \eq{qedtl-def} by the transformation:
 \begin{equation}
\Lambda(k_0,\vec k) =
	\begin{cases}
-\frac{i\vec k\cdot\vec A(k_0,\vec k)}{\vec k^2}\ ,
 & \vec k \not=0; \\
 -\frac{i A_0(k_0,\vec k)}{k_0}\ ,
 & \vec k =0,\ k_0\not=0; \\
\ \ 0\ ,  & k_0=0,\ \vec k=0.\vspace{-2.5mm}
\end{cases}\eqn{GT-QED_TL}
\end{equation}
Thus the two definitions of \qedtl, \eqs{qedtl-def}{QED_TL}, are equivalent.

\section{Finite Volume Effects in Chiral Perturbation Theory \label{sec:FVChPT}}

Before discussing the FV calculations, it is important to make some remarks on the literature.  The first
calculation for the FV EM effects on pseudoscalar 
meson masses that we are aware of was performed by
Hayakawa and Uno \cite{Hayakawa:2008an}.  They worked in \qedl\ exclusively, and used \chpt\ at one-loop.  
Again for \qedl, Davoudi and Savage \cite{Davoudi:2014qua} showed, using nonrelativistic effective field theory, that the leading $1/L$ and 
$1/L^2$ terms found in \rcite{Hayakawa:2008an} are in fact universal, independent of the internal structure of the 
particle of interest. They related higher order terms directly to the structure, parameterized in terms of  EM
multipole moments and polarizabilities, and extended the calculations to 
include spin-1/2, as well as spin-0, particles. Shortly after 
\rcite{Davoudi:2014qua} appeared, Borsanyi  {\it et al.} \cite{Borsanyi:2014jba}, and our own work 
\cite{Basak:2014vca} independently completed  the FV calculations for \qedtl.    Where they overlap, the 
results of \rcite{Borsanyi:2014jba} and 
\rcite{Basak:2014vca} agree.  However we have focused only on pseudoscalar mesons, and have not worked out the analytic form of the asymptotic expansions
in powers of $L$ and $T$, which \rcite{Borsanyi:2014jba} does very nicely for both \qedtl\ and \qedl.  Further, 
Borsanyi {\it et al.} found a discrepancy with the results of \rcite{Davoudi:2014qua} for the first 
non-universal ($1/L^3$) terms for spin-1/2.  The issue involved is in fact quite subtle, but it seems to have been
resolved \cite{Fodor:2015pna,Lee:2015rua} in favor of the result in \cite{Borsanyi:2014jba}.

In FV, defined here by spatial extent $L$ and temporal extent $T$, the momentum components take on
discrete values
\begin{equation}\eqn{FVmom}
k_i = \frac{2\pi n_i}{L}, \qquad k_0 = \frac{2\pi n_0}{T},
\end{equation}
with $n_i$ ($i=1,2,3$) and $n_0$ integers.
Through NLO in \chpt, the meson  mass squared in  FV may then be calculated simply by replacing the momentum integrals in the diagrams of \figref{FeynDiag} by sums:
\begin{equation}\eqn{mom-sums}
\int \frac{d^4 k}{(2\pi)^4}  \to \frac{1}{L^3T}\sum_{n_\mu}.
\end{equation}
Because the Feynman diagrams are divergent, it is as usual convenient to perform the renormalization in 
infinite volume, and, in FV, 
calculate only the difference between the momentum sums and the integrals.  This difference, if treated carefully, 
is finite and does not require renormalization.  We thus stipulate that the EM effect $\Delta M^2_{xy}$ defined in
\eq{Delta-def} is the appropriately  renormalized infinite-volume result, and write
\begin{eqnarray}\eqn{deltaFV-def}
(\Delta M^2_{xy})_{\rm FV} &=& \Delta M^2_{xy} +\delta_{\rm FV},\\
\delta_{\rm FV} &=& \delta^{\rm meson}_{\rm FV} 
+e^2 q^2_{xy} m^2\delta^{\gamma}_{\rm FV}(mL,mT) \eqn{delta-photon}
\end{eqnarray}
where $\delta_{\rm FV}$ is the complete NLO FV correction, 
$\delta^{\rm meson}_{\rm FV}$ is the 
contribution from the meson tadpole, \figref{FeynDiag}(c), and  $\delta^{\gamma}_{\rm FV}$ is
the contribution from photon loops, \figref{FeynDiag}(a,b).  The factors
 $e^2q_{xy}^2 m^2$ have been
taken out of $\delta^{\gamma}_{\rm FV}$  for convenience.  For notational simplicity in this section, $m$ will denote the tree-level mass of the 
meson of interest in the absence of EM; ultimately we put $m^2 = \chi_{xy,5}$ in the results.   With the factor
of $m^2$ removed, $\delta^{\gamma}_{\rm FV}$ is dimensionless, and hence is a function of $mL$ and 
$mT$ (or $T/L$) only, rather
than $m,L,T$ separately.

The FV effects from the meson tadpole come from pions that  loop around the 
volume, and hence the effect is suppressed by a factor of $\exp(-mL)$.  Because of this suppression,
$\delta^{\rm meson}_{\rm FV}$ is of negligible size on our ensembles, $\ltwid 0.2\%$.  
However, since the calculation of the effect  is completely standard, it is straightforward to include it.
In the notation  of \rcite{Bernard:2001yj}, we just have to make the substitution $\ln(m^2/\Lambda^2)
\to \delta_1(mL)$, where $\delta_1$ is a sum over Bessel functions, to obtain the FV correction.
From \eq{schpt-logs}, this gives
\begin{equation}
\delta^{\rm meson}_{\rm FV} = \frac{-2 \dem}{16 \pi^2 f^2}  \left(\frac{1}{16}   \right) 
				  \sum_{\sigma,b}  \Big[   q_{x \sigma} q_{xy}  \,  
\chi_{x \sigma, b}\,\delta_1(\sqrt{\chi_{x \sigma, b}}L)
 				-   q_{y \sigma} q_{xy} \, \chi_{y \sigma, b}\,\delta_1(\sqrt{\chi_{y \sigma, b}}L) \Big]  .  \label{eq:delta-meson}
\end{equation}

In contrast to the meson tadpole effects, the FV effects from photon diagrams, parameterized by
$\delta^{\gamma}_{\rm FV}$, are large: $\sim\! 5$--$20\%$, depending on the ensemble
and valence masses. Since the results are nontrivial, we describe the calculation in some detail, starting with 
the sunset diagram, \figref{FeynDiag}(b). We work in Coulomb gauge, choose the external meson to be
at rest ($p=(p_0,0,0,0)$), and route the loop momentum $k$  along the interior meson line, with
momentum $p-k$ on the photon line.\footnote{The final result is of course independent of the momentum routing.
However,  when $T$ is not infinite, there are interesting subtleties, which can lead to apparent routing-dependence if
treated incorrectly.  See
the Appendix for a discussion.}
Because spatial $\vec p=0$ and $k_i \cD_{ij}=0$ (for both \qedl\ and \qedtl), only the 00 component of the photon
propagator contributes to the sunset diagram.  This diagram's contribution to the self-energy  then has integrand
(summand)
\begin{equation}\eqn{sunset-integrand-alone}
\cI_{\rm s} = - \frac{k_0^2 +p_0^2}{\vec k^2(k^2 +m^2)}, \qquad [\vec k\not=0],
\end{equation} 
where we have omitted an overall factor of $e^2q_{xy}^2$.  A linear term in $k_0$ in the numerator has been 
dropped because $k_0$ and $-k_0$ contributions
cancel for both  the infinite-volume integral and the FV sum.    

Since $\cI_{\rm s}$ goes to a constant as $k_0\to\infty$, the 
difference between the sum and integral over $k_0$ (not to mention the integral itself) is divergent, so the 
FV effect from this diagram alone (in Coulomb gauge) is not well defined.  However, once this diagram 
is combined with the photon tadpole, the problem goes away.  What is needed is in fact only the  $\cD_{00}$
contribution to the tadpole, which has the integrand $1/\vec k^2$.   Adding this to \eq{sunset-integrand-alone},
gives
\begin{equation}\eqn{sunset-integrand}
\cI_{\hat {\rm s}} = \frac{\vec k^2+ m^2 - p_0^2}{\vec k^2(k^2 +m^2)}, \qquad [\vec k\not=0],
\end{equation}
where the ``hat'' on the subscript s indicates that the sunset diagram has been modified by a piece of the
photon tadpole.
It is useful to keep the rest of the tadpole
separate, because it gives different contributions in the \qedl\ and \qedtl\ cases, unlike $\cI_{\hat {\rm s}}$.

The FV effect on the self energy coming from \eq{sunset-integrand} is
\begin{equation}\eqn{sunset}
m^2\delta^{\hat {\rm s}}_{\rm FV}(p_0/m,mL,mT) = \frac{1}{L^3T} \sum\limits'_{k_0,\vec k}\; \cI_{\hat {\rm s}} - \int \frac{d^4 k}{(2\pi)^4}\; \cI_{\hat {\rm s}},
\end{equation}
where the prime on the summation symbol means that the $\vec k=0$ term is dropped, but there is no restriction on $k_0$.
As in \eq{delta-photon}, we take out a factor of $m^2$ to make $\delta^{\hat {\rm s}}_{\rm FV}$ 
dimensionless.

From \eq{H-U-prop}, the remaining (spatial) components of the photon tadpole in \qedl\ give the integrand
and corresponding FV effect
\begin{eqnarray}\eqn{tadpole-integrand-qedl}
\cI_{\rm t,\qedl} &=&  \frac{2}{k^2}, \qquad [\vec k\not=0], \\
m^2\delta^{{\rm t},\qedl}_{\rm FV}(mL,mT) &=& \frac{1}{L^3T} \sum'\limits_{k_0,\vec k}\; \cI_{{\rm t},\qedl} - \int \frac{d^4 k}{(2\pi)^4}\; \cI_{{\rm t},\qedl}. \eqn{tadpole-qedl}
\end{eqnarray}
In \qedtl, there is an extra contribution coming from the nonzero value of $\cD_{ij}$ when $\vec k=0$ but
$k_0\not=0$, see \eq{MILC-prop}.  
\begin{eqnarray}\eqn{tadpole-qedtl}
m^2\delta^{{\rm t},\qedtl}_{\rm FV}(mL,mT) &=& m^2\delta^{{\rm t},\qedl}_{\rm FV}(mL,mT) +m^2\delta^{{\rm t},+}_{\rm FV}(mL,mT).\\ 
\delta^{{\rm t},+}_{\rm FV}(mL,mT) &=&  \frac{1}{m^2L^3T} \sum\limits_{k_0\not=0}\frac{3}{k_0^2}\nonumber\\
&=&  \frac{mT}{(mL)^3} \,\frac{3}{4\pi^2}\;2\sum\limits^\infty_{n=1}\frac{1}{n^2} = \frac{mT}{4(mL)^3},
\eqn{tadpole+}
\end{eqnarray}
where we have used the well-known result $\sum^\infty_{n=1}1/{n^2}=\pi^2/6$ \cite{Basel}. 

When $T$ is infinite,  we can obtain the correction to the meson mass-squared by evaluating the self energy
at $p_0 =i m$.  For finite $T$, however, this prescription is not obviously correct, and indeed is 
wrong in some cases.   Here, we will simply assume that we may use the prescription, and leave it
to the Appendix to explain the point in detail and show that plugging $p_0=im$ into the integrand in
\eq{sunset-integrand} gives the desired answer.  The complete contributions from the photon diagrams to the FV effect on the meson mass-squared are then
\begin{eqnarray}\eqn{dgamma-qedl}
\delta^{\gamma,\qedl}_{\rm FV}(mL,mT) &= & \delta^{\hat s}_{\rm FV}(i,mL,mT)  + \delta^{t,\qedl}_{\rm FV}
(mL,mT) \\
 \delta^{\gamma,\qedtl}_{\rm FV}(mL,mT) &= & \delta^{\gamma,\qedl}_{\rm FV}(mL,mT) + \frac{mT}{4(mL)^3}.
\eqn{dgamma-qedtl}
\end{eqnarray}

It now is necessary only to evaluate the difference of sums and integrals given in \eqs{sunset}{tadpole-qedl}.
This can be done straightforwardly using an importance-sampling integration program such as
VEGAS \cite{VEGAS}.  The sum may be treated as an integral by defining the  ``finite-volume integrand'' at 
the arbitrary point $k$ as the average of the infinite-volume integrand  at the 16 corners of the FV hypercube containing $k$, weighted appropriately by the distances in each direction to the corners.  
For example, if $\tilde k$ is the closest point in the sum ``below'' $k$ ($\tilde k_\mu < k_\mu$) then the weight
of the integrand at $\tilde k$  is $[1-(k_0-\tilde k_0)T/(2\pi)]\prod_{i=1}^3  [1-(k_i-\tilde k_i)L/(2\pi)]$.   When a corner is a special point (\eg $\vec k=0$, $k_0$ arbitrary) that
should be dropped from the sum, we simply  put in 0 for the integrand there.
One could
also use  the value at the closest corner of the FV hypercube rather than the weighted average, but the resulting integrand has discontinuities on the midplanes of
the hypercube, and the numerical integration therefore has larger errors.  

\begin{figure}
\vspace{-5mm}
\begin{center}\includegraphics[width=0.55\textwidth]{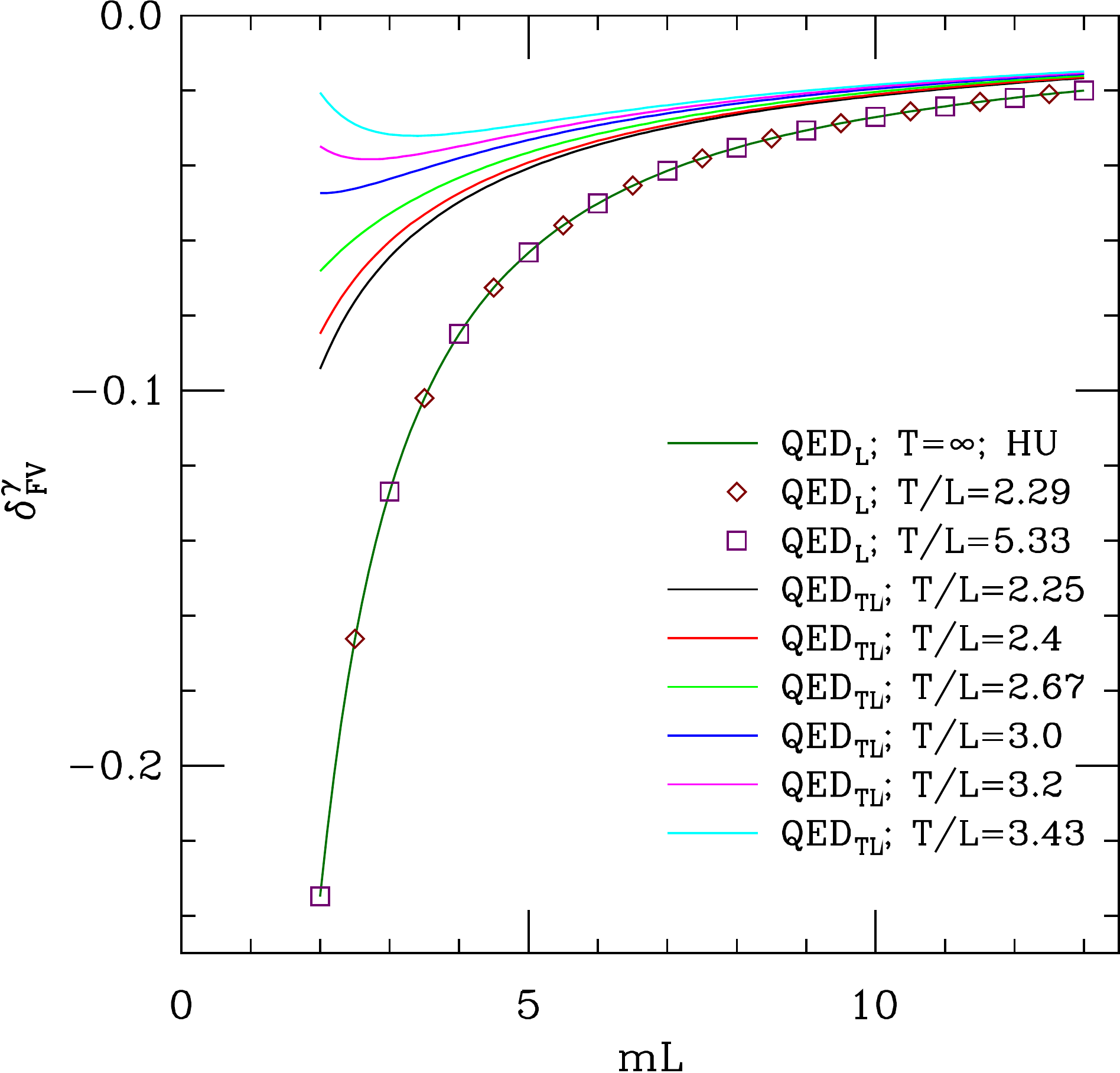}\end{center}
\vspace{-5mm}
\caption{\label{fig:dgamma} The FV effect from photon diagrams,  $\delta^\gamma_{\rm FV}$ for
\qedl\ and \qedtl, as a function of $mL$.   In the \qedl\ case, the dark green line shows the result when $T=\infty$ 
from  Hayakawa and Uno \protect{\cite{Hayakawa:2008an}}, while the  dark red diamonds  and purple 
squares show our
evaluation at $T/L=2.29$ and $T/L=5.33$, respectively. For \qedtl, the lines give our results for
six values of $T/L$ ranging between 2.25 and 3.43, which are the values relevant to the bulk of our data.  The numerical errors in the points and lines are too
small to be seen on this scale. }
\vspace{-0.15in}
\end{figure}

We have checked that our result for $\delta^\gamma_{\rm FV}$, the sum of the sunset and the photon tadpole 
diagrams, agrees with that of 
\rcite{Hayakawa:2008an} in the \qedl\ case.  In \figref{dgamma}, we plot in dark green the result calculated from 
the results of Hayakawa and Uno \cite{Hayakawa:2008an}, and superimpose points calculated by us at representative values of $mL$.  Hayakawa and Uno work at infinite $T$, whereas our points have been computed
at $T/L=2.29$ and $T/L=5.33$. It is clear that for such values of $T/L$ the finite-$T$ effects are negligible in
\qedl.  (See the Appendix for further discussion.)

The difference in the \qedtl\ case is the extra term $\delta^{t,+}_{\rm FV}$ in \eq{tadpole+}.   \Figref{dgamma}
also shows our \qedtl\ results for ranges in values of $mL$ and $T/L$ that cover all of our
data used in the final analysis;  data
with more extreme $T/L$ values ($\ge 4$ and $\le 1.6$) are used later in this section in testing the
applicability of our formulas.

Unlike \rcite{Hayakawa:2008an} and the present calculation,
Davoudi and Savage \cite{Davoudi:2014qua} and Borsanyi {\it et al}\/ \cite{Borsanyi:2014jba} do not compute the 
FV effects in the context of \chpt, but instead work first with the universal terms that describe
a point-like particle, and then consider corrections coming from the particle structure.  Aside from the
contribution from the meson tadpole, \eq{delta-meson}, which is suppressed by $\exp(-mL)$, the
one-loop \chpt\ calculation is in fact identical to the point-like approximation of \rcites{Davoudi:2014qua,Borsanyi:2014jba} because there are no corrections to the photon-meson vertices or internal meson lines in
\figref{FeynDiag}(a), (b).  In \qedtl, for point-like mesons, \rcite{Borsanyi:2014jba} finds
\be\eqn{BMW-asymptotic}
\delta^\gamma_{\rm FV} \underset{T,L\to\infty}{\sim}
-\frac{\kappa}{4\pi mL} -\frac{\kappa}{2\pi (mL)^2}+\frac{mT}{4(mL)^3}\ ,
\ee
where the last term is what we call $\delta^{t,+}_{\rm FV}$, \eq{tadpole+}, and the other terms come
from the asymptotic expansion of $\delta^{\gamma}_{\rm FV}$ in \qedl.  The  constant $\kappa$ is defined
in \rcite{Borsanyi:2014jba} by
\be
\kappa \equiv \int_0^\infty \frac{d\lambda}{\lambda^{3/2}}\left\{\lambda^{3/2}+1-
\left[\mathcal{\theta}_3(0,e^{-\frac\pi\lambda})\right]^3\right\}\ ,
\eqn{kappa}
\ee
where $\mathcal{\theta}_3(u,q)=\sum_{n=-\infty}^{+\infty}
q^{n^2}e^{i2nu}$ is a Jacobi theta function.   By numerical integration, one finds
 $\kappa\approx2.8373$.  An equivalent definition of $\kappa$ in \rcite{Hayakawa:2008an} is
\be
\kappa = \int_0^\infty \frac{d\lambda}{\lambda^{2}}\left\{\lambda^{3/2}+1-
\left[\mathcal{\theta}_3(0,e^{-\frac\pi\lambda})\right]^3\right\} \ .
\eqn{kappap}
\ee
The equivalence of \eqs{kappa}{kappap}  follows from the identity \cite{Wolfram}  
\be\eqn{Jacobi-identity}
\mathcal{\theta}_3(0,e^{-\pi x}) = \frac{1}{\sqrt{x}}\;  \mathcal{\theta}_3(0,e^{-\pi/ x}),
\ee
which can easily be proved using the Poisson summation formula.

\begin{figure}
\begin{center}\includegraphics[width=0.55\textwidth]{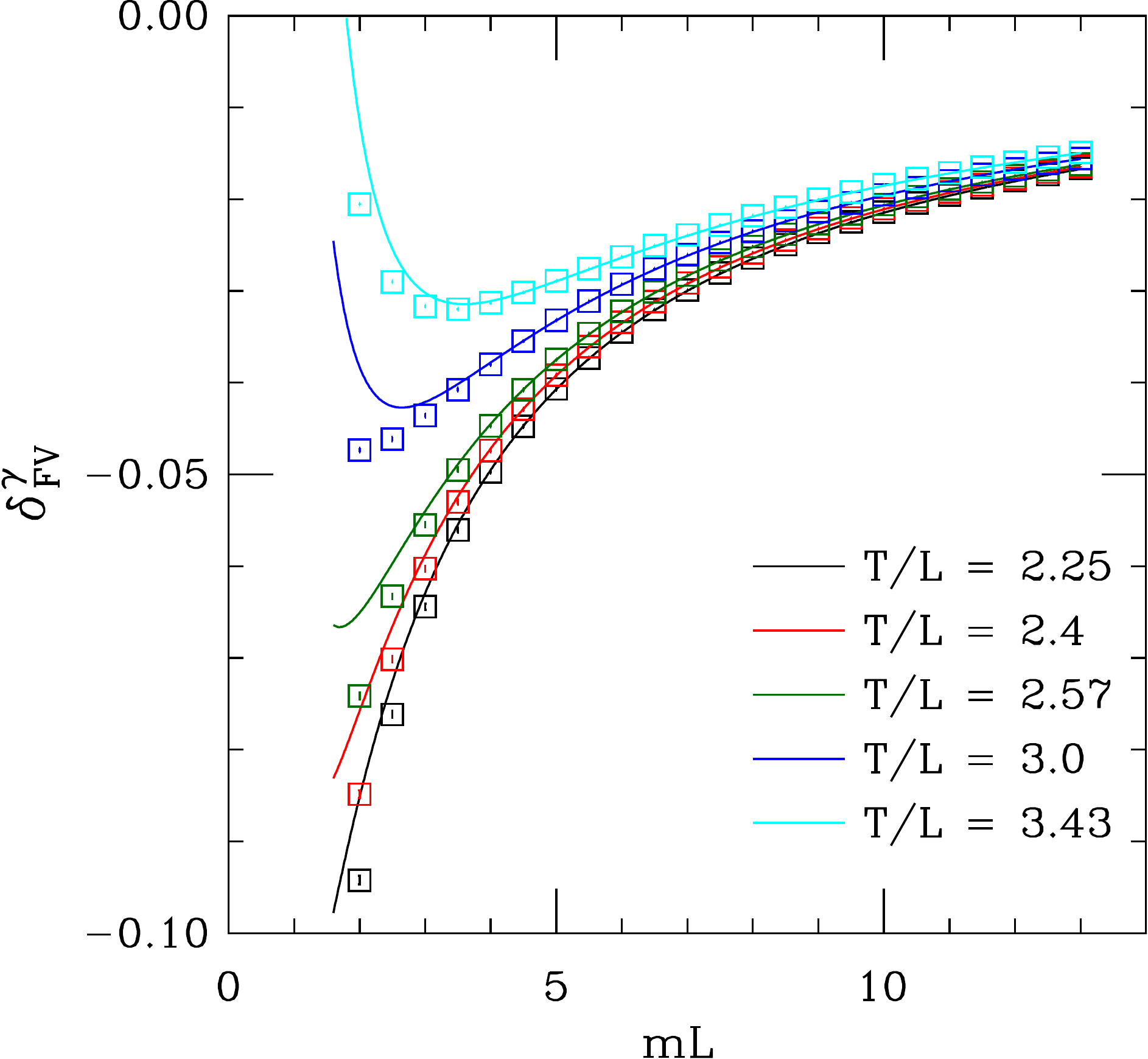}\end{center}
\vspace{-5mm}
\caption{\label{fig:FV-asymptotic} A comparison of the full FV effect for \qedtl, coming from 
one-loop photon 
diagrams with a point-like meson,  and the corresponding asymptotic forms determined in
\rcite{Borsanyi:2014jba}, for various values of $T/L$.  The squares show our calculations of the full effect, while the lines
are the asymptotic forms.    The numerical errors in the points are small and are just barely visible in some
of the points at the left.}
\end{figure}

In \figref{FV-asymptotic} we compare our results for \qedtl\ with the asymptotic form \eq{BMW-asymptotic}.
For $mL\ge 3.8$, which describes the unitary points in our data used in the final analysis, the differences
with the asymptotic form are negligible.  However, a few valence points in that analysis have $mL\gtwid 2.9$, 
for which the differences ($\ltwid 6\%$) are important to include.  In our test
of FV effects described later in the section, we have points as low as $mL=2.7$ and
aspect ratio of $T/L=5.33$ for which the 
differences are a bit bigger, $\approx 7\%$.   For convenience, we use our full results everywhere in the
analysis, even where the differences with the asymptotic form are negligible.

\begin{figure}[t!]
  \centering
  {\includegraphics[width=0.48\linewidth]
  {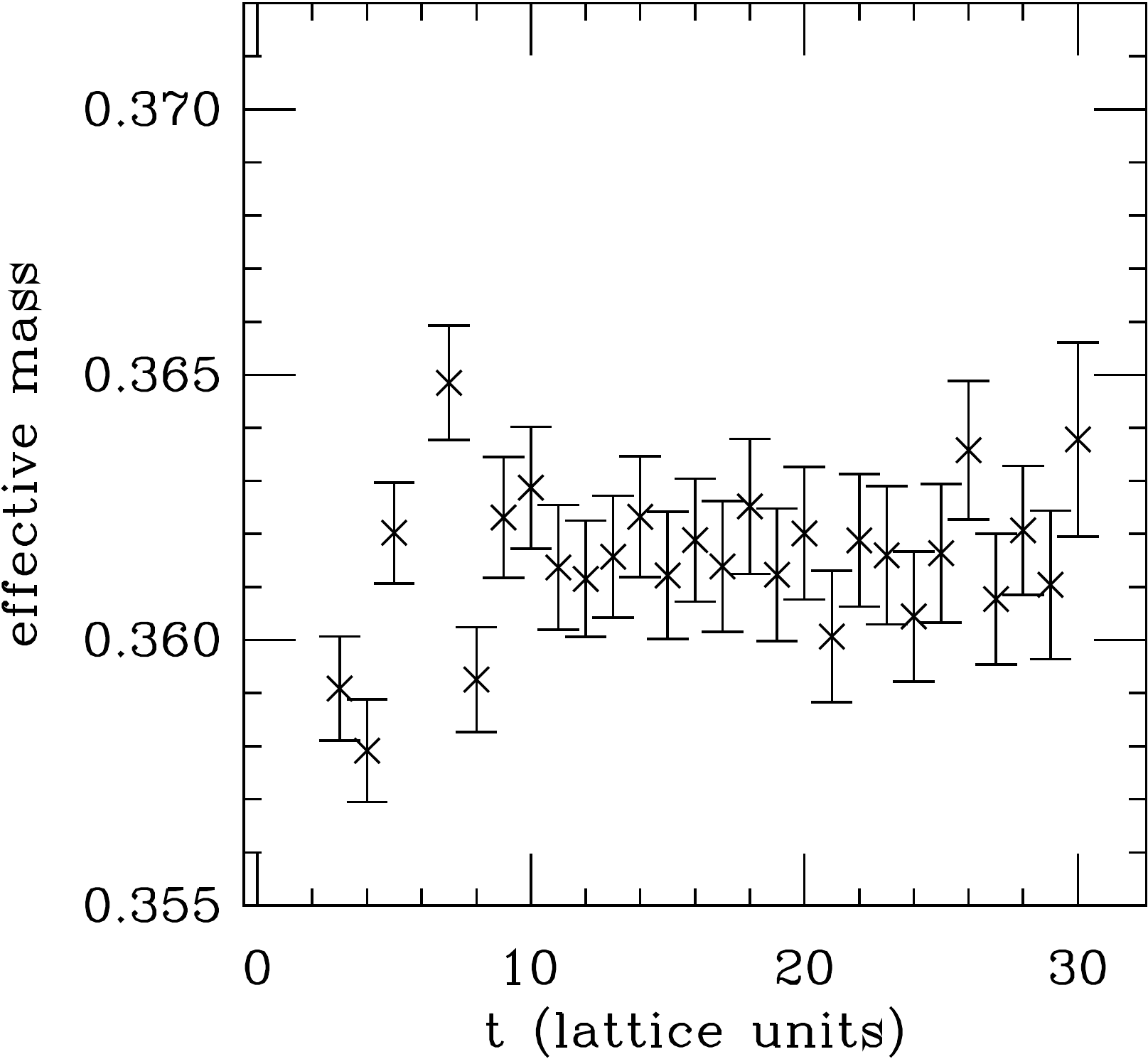}}
  \
  {\includegraphics[width=0.48\linewidth]
  {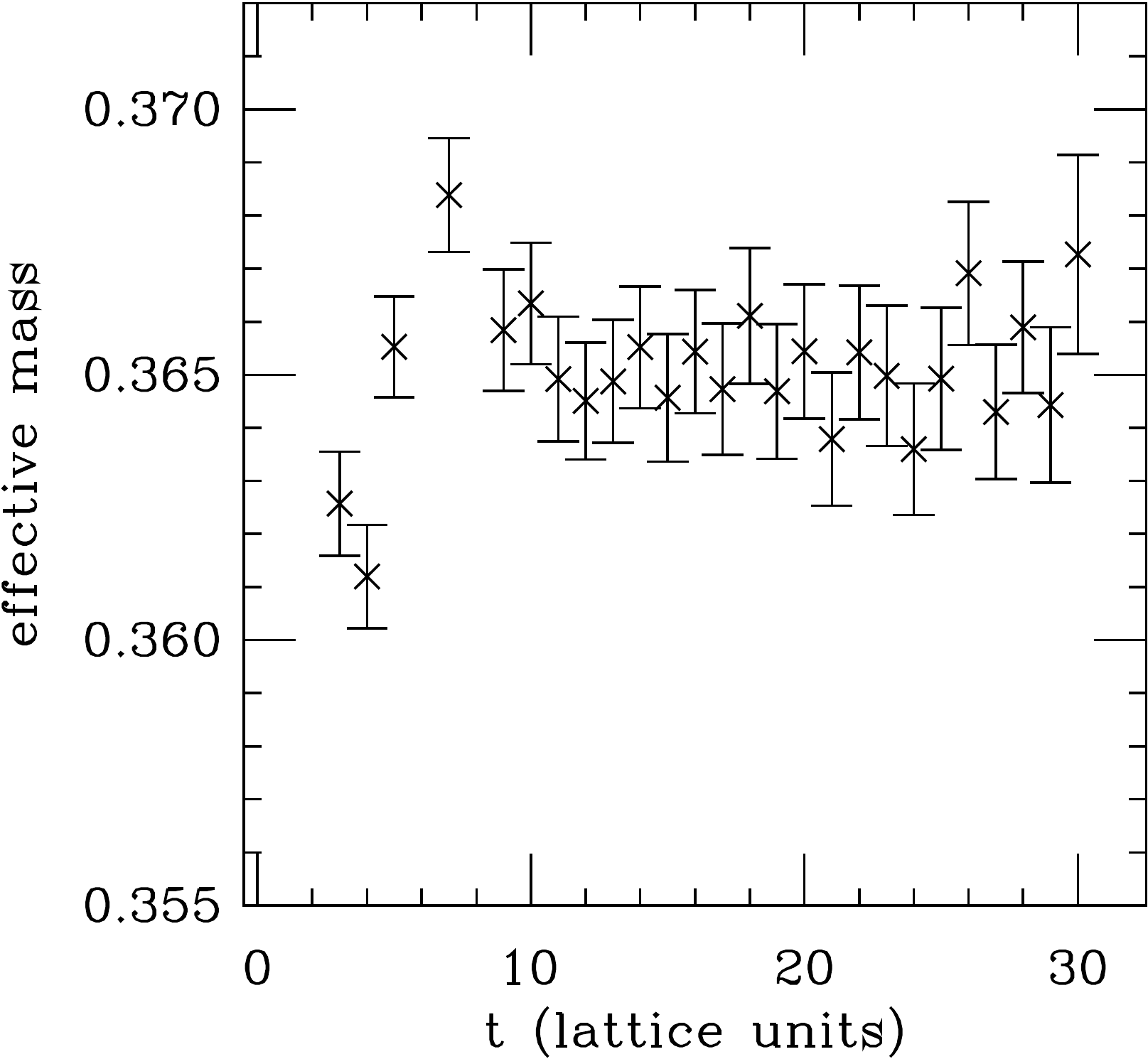}}
  \\
\vspace{-2mm}
\caption{\label{fig:plateaus}
Effective mass plots for a ``$K^+$ meson'' in pure QCD (left) and in QCD+quenched\qedtl\ (right).  The data is from the ensemble listed first in \tabref{ensembles}, with $a\approx 0.12\;$fm, $L/a=12$ and $T/a=64$.  The valence
masses are 0.01 and 0.04.}
\vspace{-0.08in}
\end{figure}

We emphasize here that the term $\delta^{t,+}_{\rm FV}= mT/(4(mL)^3)$  in \eq{tadpole+} indicates that the large-volume limit
is rather subtle in \qedtl. The result is acceptable if the limit $L\to\infty$ is taken before
$T\to\infty$, or if the limits are taken together at fixed aspect ratio $T/L$, but not if the limit $T\to\infty$ is taken first.  In other words, the 
\qedtl\ set-up is not well defined in finite spatial volume at zero temperature.  This fact has also been pointed out by
Borsanyi  {\it et al}\/.\ \cite{Borsanyi:2014jba}.  They make the further point  that  \qedtl\ violates reflection positivity because the 
constraint required to set the single $k_\mu=0$ mode of $A_i$ to zero involves the square of the integral over all space-time of $A_i$.  
Although many actions used in lattice QCD violate reflection positivity,  one might worry that in this case the violation leads to problems 
with defining or isolating the lowest states in correlation functions.  \Rcite{Borsanyi:2014jba} did have problems
from close excited states in extracting masses in pure quenched \qedtl.  In
our QCD plus quenched \qedtl\ simulations, however, this does not seem to be a problem. As illustrated in \figref{plateaus},
we find no significant differences between the qualities of plateaus in correlation functions in  QCD+\qedtl\ versus those for QCD alone.  The example shown is for a putative ``worst case'' in our data
because the aspect ratio $T/L=5.33$ has the largest value, and $L$ is the smallest.  
See also the plots for our ensemble with $a\approx\!0.045$ fm and $T/L=3$,
shown in \figref{dminfits}.  Again, no significant differences in plateau quality  between 
QCD+\qedtl\ and pure QCD are visible.

Despite that fact that we have not found any evidence of problems due to the lack of reflection
positivity in \qedtl, the reader may wonder why we did not just use \qedl\ or a massive-photon infrared
regulator \cite{Endres:2015gda}, both of which are reflection-positive.  The reason is straightforward:  When this
project was begun \cite{basak}, and by the time most of the numerical computations were completed \cite{EM12}, the
issues with \qedtl\ were not known.   We simply followed the \qedtl\ approach of the original paper on the subject,
\rcite{Duncan:1996xy}.   The fact that \qedtl\ has smaller FV corrections than \qedl\ in the relevant range
of parameters (as seen in \figref{dgamma}) is a nice accidental benefit of our choice of \qedtl, but it was also not known when this project
started and therefore had no influence on the choice.

\begin{figure}
\begin{center}\includegraphics[width=0.55\textwidth]{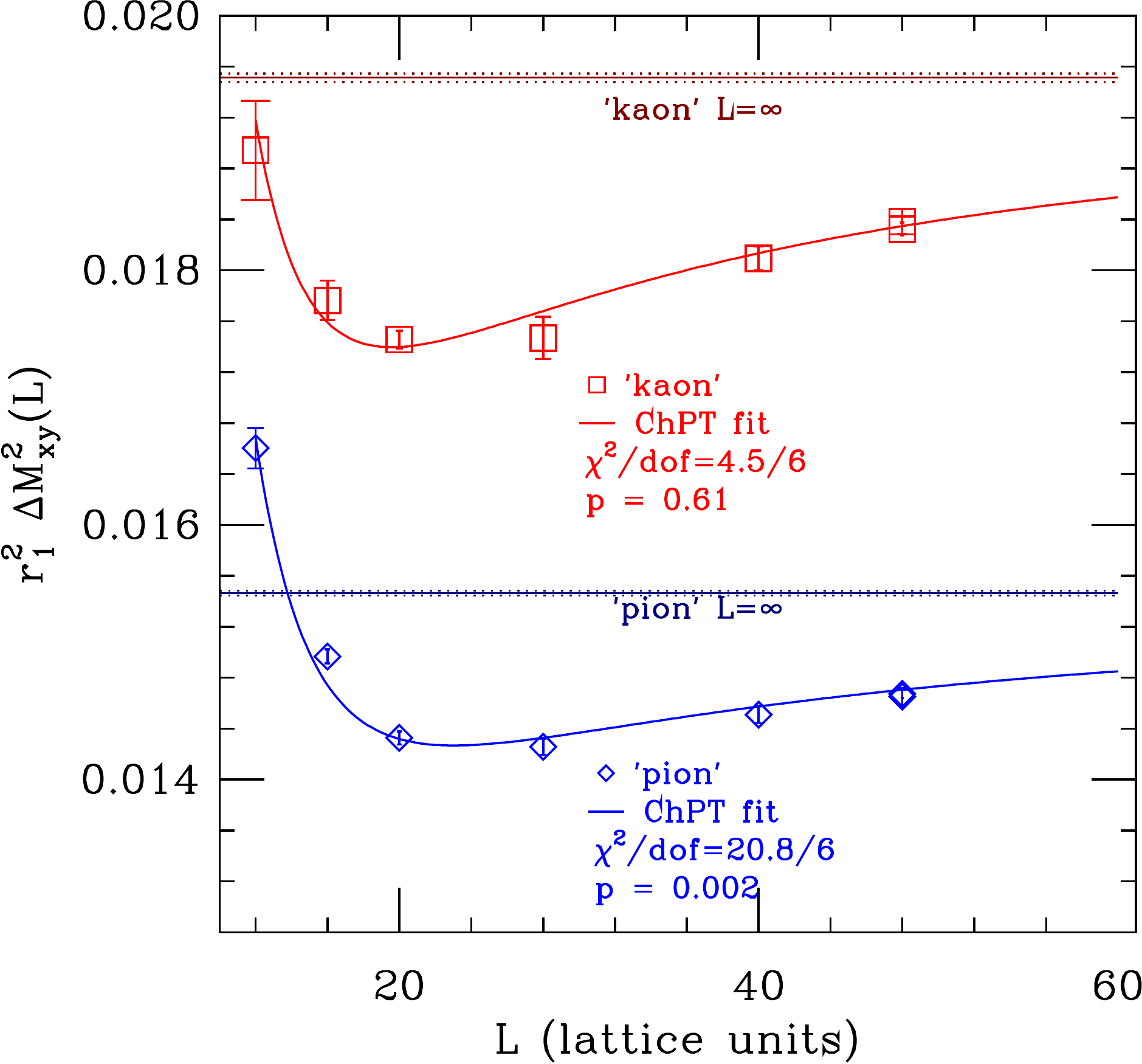}\end{center}
\vspace{-5mm}
\caption{\label{fig:FV} Finite volume effects at $a\approx0.12$ fm  and $am'_l=0.01, am'_s=0.05$ as a function of spatial lattice length $L$ for
two different meson masses: a unitary `pion' (blue) with degenerate valence masses $m_x=m_y=m'_l$, and a `kaon' (red) with valence masses
$m_x=m'_l$ and $am_y=0.04$, close to the physical strange quark mass.  The fit lines are to our FV form for \qedtl\ (omitting the negligible meson tadpole term), and have one free parameter each, the infinite volume value (shown by horizontal solid lines with dotted lines for errors).
}
\end{figure}

To test our understanding of the FV effects, we have generated ensembles with a wide range of spatial sizes
at $\beta=6.76$ ($a\approx0.12\;$fm) with sea-quark masses $m'_l=0.01$ and 
$m'_s=0.05$ (see \tabref{ensembles}).
In \figref{FV} we show fits, for two different meson masses on these ensembles, to our calculated FV correction,
given by \eqs{deltaFV-def}{delta-photon}, with $\delta^\gamma_{\rm FV}=\delta^{\gamma,\qedtl}_{\rm FV}$, \eq{dgamma-qedtl}. We neglect the meson tadpole
term in \eq{deltaFV-def} for convenience, since its effect is not visible on this scale.  This means that the
 FV correction used here is the same as in the point-like approximation for the mesons.
 The shape of the fit curves are completely determined by the FV calculation;  the only free parameter 
in each fit is the overall 
height of the curve given by the value in infinite volume. The theory gives a  good description of the data, and we use it to correct the data for FV effects.  We estimate the remaining systematic error associated
 with FV effects in \secref{FV-error}.  

One can now understand why it was difficult to observe FV effects directly in the data set available in \rcite{EM12}.  At that 
time, we had only the $L=20$ and $L=28$ ensembles to compare.  From \figref{FV}, one sees that the minima of the curves are  
in this region of $L$ or close to it, and therefore the difference expected between these volumes is small compared to the statistical errors in the data.

\section{Chiral-discretization fits and chiral-continuum extrapolations}

In this section, we first discuss the quantities that have been determined from pure QCD computations, and are used here
as inputs to the chiral-discretization fits.  We then show (a small subset of) the data
we fit, both before and after FV corrections.  Finally, we describe the fits themselves.

\subsection{Inputs}
\label{sec:inputs}

In addition to the lattice values of $r_1/a$ that set the relative scales, we need other lattice-dependent quantities as inputs to 
the QCD+QED calculations.    \tabref{other-params} lists the values of these quantities for one ensemble
at each of our (approximate) lattice spacings. The first three columns serve to identify the 
ensembles. Columns four and five give the light
and strange physical quark masses in $r_1$ units, which are determined from chiral fits to pure QCD lattice 
data\cite{RMP}.  These masses are ``physical'' in the sense that they have been determined by
demanding that the $\pi$ and $K$ mesons take their (isospin-averaged) experimental values 
in absence of EM.%
\footnote{There is an apparent circularity here, in that we are computing in this paper the EM effects on the
$\pi$ and $K$ masses.  In practice, we have used earlier, phenomenological estimates of EM effects 
(see \rcite{Aubin:2004he})
to remove them at this stage.  We can iterate to make the calculation self-consistent, but it is unnecessary, because
the EM effects make only a small change in the estimates of the strange and isospin-averaged light masses.}
They are, however, bare masses, in that no renormalization (perturbative or otherwise) has been applied.
 
\begin{table}
\begin{center}
\begin{small}
\begin{tabular}{|c|l|c|c|c|c|c|c|c|c|}
\hline\hline
$\approx\!a\,$[fm]& $\beta$ 
& $m'_l/m'_s$& $r_1 m_l $ & $r_1 m_s$ & $r_1 B_0$& $r_1^2 a^2 \Delta_A$ & $r_1^2 a^2 \Delta_T$& $r_1^2 a^2 \Delta_V$& $r_1^2 a^2 \Delta_I$\\
\hline\hline
0.12   & 6.76& 0.005/0.05& 0.00333(6)(5) &  0.0919(16)(13) & 6.832(4) & 0.230(2) & 0.371(5) & 0.487(6) & 0.609(17) \\\hline
0.09   & 7.08& 0.0031/0.031& 0.00338(6)(5)  &  0.0927(16)(13) & 6.639(6) & 0.075(5) & 0.124(6) & 0.160(10) & 0.222(18) \\\hline
0.06  & 7.46& 0.0018/0.018& 0.00343(7)(5)  &  0.0937(16)(13) & 6.487(6) & 0.027(1) & 0.044(2) & 0.058(2) & 0.071(3) \\\hline
0.045   & 7.81& 0.0028/0.014& 0.00342(6)(5)  &  0.0936(16)(13) & 6.417(6) & 0.010(2) & 0.017(3) & 0.023(3) & 0.028(3) \\\hline
cont.  &   7.08& --- & 0.00361(7)(5)  &  0.0990(17)(14) & 6.015(6) & 0 & 0 & 0 & 0 \\
\hline\hline
\end{tabular}
\end{small}
\caption{Quantities used as inputs in the chiral-discretization fits and/or their extrapolation.
The first three columns identify the ensemble, and then we list, in $r_1$ units, the physical values of
the light quark  and strange quark mass, the slope $B_0$ (\eq{chi-def}), and the taste splittings
for axial, tensor, vector, and singlet tastes, respectively.  
The last row is labeled ``cont.'' for ``continuum;'' see text for how this is defined.
The errors for the quark masses are from the
chiral extrapolation and the absolute scale, respectively; statistical errors are negligible.  For the
other quantities the errors given are statistical only.
\label{tab:other-params} }
\end{center}
\end{table}

Two errors are shown for the masses. The first is the systematic error coming from the chiral extrapolation.
It is determined by comparing the results of fits that include chiral logarithms through NNLO (plus higher
order analytic terms) and those that include the chiral logarithms only through NLO.  Other changes in the
fits give similar estimates of the errors.  The second error in the masses comes from the uncertainty in the
absolute scale, \ie the error in the physical value of $r_1$.

In the final row of the table, ``cont.'' stands for ``continuum.''   It is convenient for us to view the continuum
not as the $\beta\to\infty$, $a\to0$ limit, but as another ensemble with fixed $\beta$ and $a$, in which
all discretization effects 
have been extrapolated away.  In other words, we view the continuum as a 
lattice with a perfect action. This allows us to continue to employ bare lattice masses to describe
the physical point, just as we do at nonzero lattice spacing.  Here we have chosen the continuum to have $\beta=7.08$, the same as
the $a\approx0.09$ ensemble with simulation masses $m'_l/m'_s = 0.0031/0.031$.   The scale
of the two is however slightly different, since extrapolating away the discretization effects changes the
estimates of the physical quark masses.  This in turn affects the $r_1/a$ value, which is adjusted to be at
physical masses.  The  $0.0031/0.031$ ensemble has $r1/a=3.755$, while $r1/a=3.744$ for the
``continuum,''  a 0.3\% difference.  This difference shows that the discretization effects in our
mass-independent scale-setting scheme are small.

The LEC $B_0$ is given in column six
of \tabref{other-params}.  It is obtained from a fit of the squared masses of the Goldstone (taste $\xi_5$) mesons to \eq{chi-def} .   The fit is performed
for each lattice spacing
over the full range of meson masses that enter this analysis.     This LO result is used for the
meson masses in the NLO (and higher order) expressions in  \Delxy, \eq{Delta-orders}; 
$\Delta_{\rm LO} M^2_{xy}$ is of course mass independent.
Like the quark masses, the $B_0$ values shown here are bare (unrenormalized). 

\begin{figure}[t!]
	\begin{center}
	\begin{tabular}{c c }
		\includegraphics[scale=0.41]{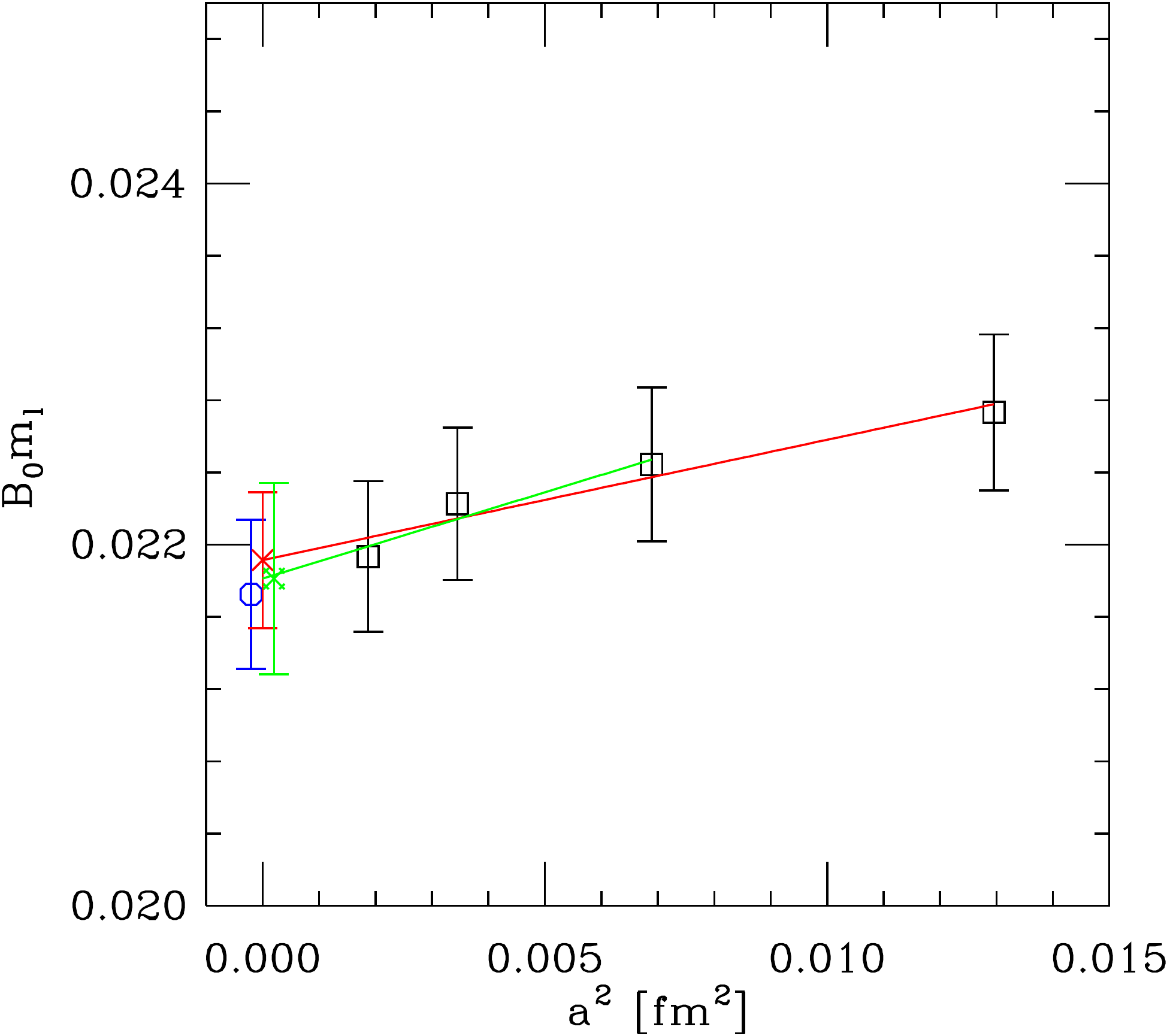}   		&
		\includegraphics[scale=0.41]{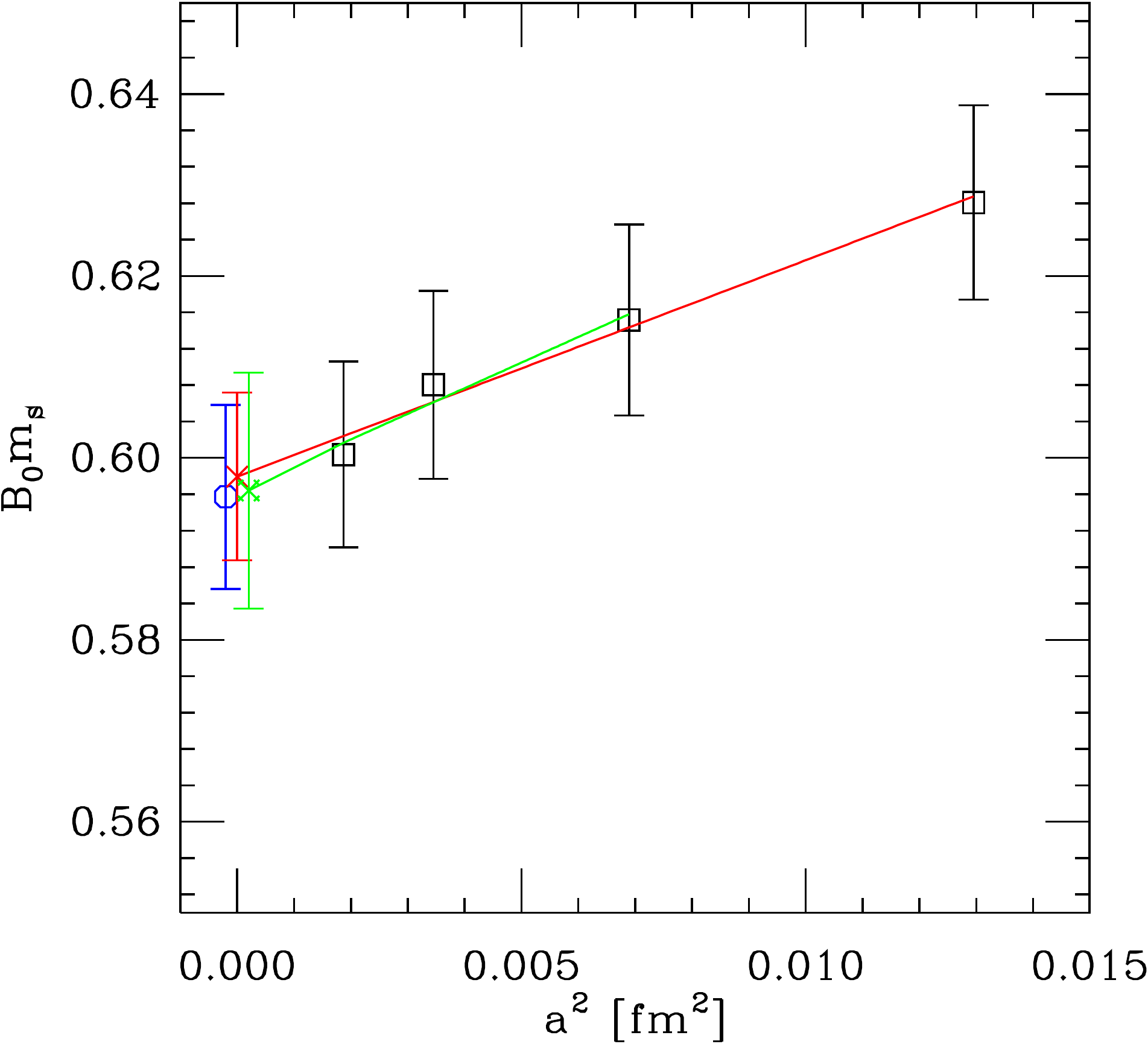}   		 \\
	\end{tabular}
	\end{center}
\caption{Products $B_0m_l$ (left) and $B_0m_s$ (right) versus $a^2$.  The black squares show our
data at nonzero lattice spacing, while the blue octagons show our continuum values.  Two simple
linear extrapolations are shown for comparison.  The red line in each plot is a fit
to all four lattice spacings, and the red cross is its extrapolation.  Similarly, the green line and fancy cross
in each plot comes from a fit that omits the coarsest lattice spacing.
 \label{fig:B0m}
}
\end{figure}

Both  $B_0$ and the quark masses need to be
renormalized before we can properly compare values at different lattice spacings and extrapolate to the
continuum.  We use the 1-loop renormalization from \cite{Aubin:2004ck}, \eq{QCD-mass-renorm}, to do the
extrapolation.   
As discussed in the context of EM mass renormalization in \secref{renormalization}, this means that
there are substantial errors from renormalization 
affecting the continuum values of $B_0$, $m_l$, and $m_s$ in \tabref{other-params}.   This is 
true even though we take out the renormalization factors, defined for the continuum to be the same
as those of the $\beta=7.08$, $m'_l/m'_s=0.0031/0.031$ ensemble.   Such errors would be important if we
wanted to extract quark masses or $B_0$ in a continuum scheme such as $\msbar$. However  the
renormalization errors
are irrelevant here and not included in \tabref{other-params} 
because only the renormalization-invariant 
products  $B_0 m_l$ and  $B_0 m_s$ enter into the results from our \chpt\ fits.
This is illustrated in
\figref{B0m}, which shows these products computed from the values in \tabref{other-params}, and
compares the continuum values (blue octagons) to the values that would
have been obtained by linear extrapolation in $a^2$ from all four of our lattice spacings (red lines and
crosses) or the three spacings with $a\ltwid0.09$ fm (green lines and fancy crosses).  Although the
continuum values of $m_l$, $m_s$, or $B_0$ were not obtained from such extrapolations,%
\footnote{The physical quark masses come from two-loop chiral fits described in \cite{RMP},
while $B_0$ comes from linear extrapolation of the values in \tabref{other-params} after
renormalization}
 the figure shows that the products have small discretization errors and
 smooth behavior with $a^2$. Renormalization factors, along with their large 1-loop errors, cancel out.

The values of quark masses and $B_0$ shown in the table may be used for any ensemble in the same
 group of approximate lattice spacings as the ones listed.  For example, 
  for the $a\approx0.06$ fm, $\beta=7.47$, $m'_l/m'_s=0.0036/0.018$ ensemble, which is not listed
  in \tabref{other-params}, one should just use the values listed for the 
  $a\approx0.06$ fm, $\beta=7.46$, $m'_l/m'_s=0.0018/0.018$ ensemble.   The small
changes in $\beta$, and hence in lattice scale, result in even smaller changes in
discretization effects and renormalization constants.   Thus even though the quantities shown 
are unrenormalized, their differences among a group of ensembles with 
the same group of approximate lattice spacings  are negligible.  Note from the table that even when the
lattice scale changes from $\approx\!0.12$ fm to $\approx\!0.045$ fm, the changes in the masses 
in $r_1$ units are less than 3\% and those in $B_0$ are less than 7\%.

The final needed input for our fits are the values of the taste-splittings $a^2\Delta_b$ in \eq{chi-def}.
\tabref{other-params} gives these splittings in $r_1$ units.  For unlisted  ensembles, the explicit factor of $a^2$
in the splittings results in  changes of a few percent from the listed ones. We include these
changes in our fitting routines,
even though they are smaller than the current statistical errors on the splittings.  One can make
the adjustment simply by multiplying the listed value by the ratio $(r_1/a)^2_{\rm listed}/(r_1/a)^2_{\rm unlisted}$,
with the $r_1/a$ values taken from \tabref{ensembles}.  For the two $a\approx0.06$ fm ensembles
mentioned in the previous paragraph, the adjustment in splittings is about 2\%.

\subsection{FV corrections to our data}
 \label{sec:FVcorr-data}

\begin{figure}
\vspace{-5mm}
\begin{center}\includegraphics[width=0.65\textwidth]{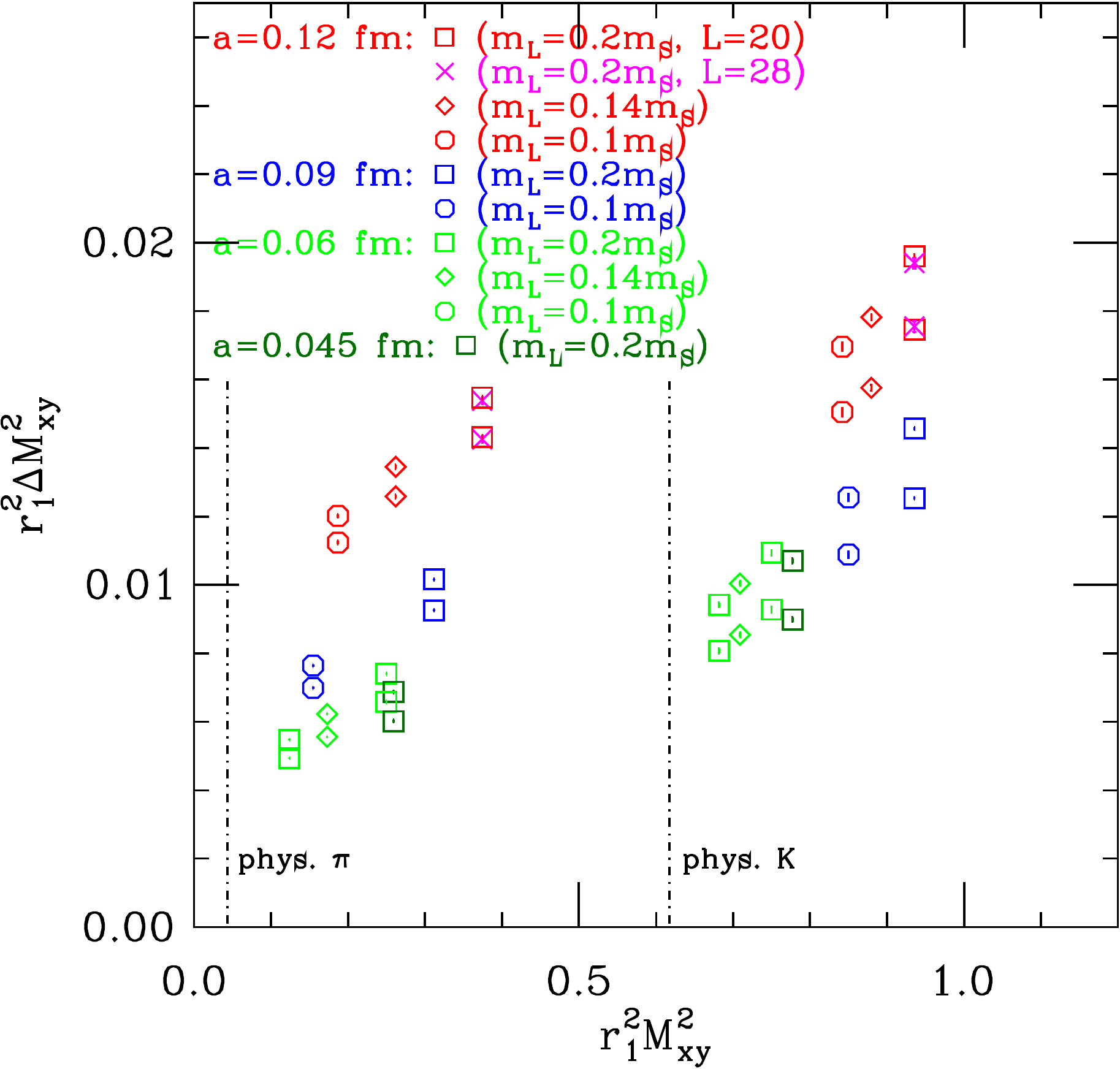}\end{center}
\vspace{-6mm}
\caption{Finite-volume corrections to $\Delta M_{xy}^2$ for a small subset of our data versus $M_{xy}^2$ itself, 
where both quantities are expressed in $r_1$ units.  For each pair 
of points with the same color and symbol, the lower point shows the raw datum, while the upper shows the result 
after correction for FV effects, \ie in infinite volume.
Colors and symbols identify the lattice spacing and light sea-quark mass, 
and (in one case) the spatial lattice
size, as shown in the legend.  Points in the left-hand cluster are pion-like and unitary ($m_x=m_y=m'_l$), while those in the right-hand cluster 
are kaon-like and almost always unitary  ($m_x=m'_l$, $m_y =m'_s$).  The exceptions are kaon-like points for
$a\!\approx0.12$ fm, which have $m_y=0.8m'_s$, which is closer to the physical strange mass than
$m'_s$ itself.   The locations of the physical pion and kaon masses are indicated  by the vertical dot-dashed black
lines.  All points shown are for mesons with charge $\pm e$.  \label{fig:FVdata}
}
\end{figure}

\Figref{FVdata} shows a small subset  of our data for $\Delta M^2_{xy}$, plotted as a function
of the meson mass, $r^2_1M^2_{xy}$, before
and after correction for FV effects. The subset consists of charge $\pm e$ unitary or nearly-unitary points, as described in more
detail in the figure caption.  Because the correction due to photon diagrams is proportional
to $M^2_{xy}$, see \eq{delta-photon},  the absolute FV effect is larger for kaon-like points (right-hand half
of the plot) than for pion-like points (left-hand half).  The correction ranges from 0.0013 to 0.0021 for kaons
and 0.0005 to 0.0009 for pions.  
Even the fractional correction is generally larger for kaons than for pions since
the LO contribution to $\Delta M^2_{xy}$ itself is the mass-independent quantity $\Delta_{EM}$ (the Dashen
term, \eq{dem}), which has no FV correction.  The  correction varies from  10\% to 16\% for kaons,
and from 6\% to 12\% for pions.    

Strictly speaking, the full FV correction to $\Delta M^2_{xy}$ depends on the chiral fit, because the
FV effect of the meson tadpole $\delta^{\rm meson}_{\rm FV}$, \eq{delta-meson}, depends on the
fit parameter $\Delta_{\rm EM}$.  
However, this dependence would not be visible in 
\figref{FVdata}, because the exponentially suppressed meson tadpole FV corrections are very small compared
with those from the photon diagrams, which are independent of the fit parameters.

Because the FV corrections depend, at least in principle, on the parameters of the fit, we fit uncorrected (raw)
data for $\Delta M^2_{xy}$ to a chiral fit form that includes the FV NLO adjustment $\delta_{FV}$ in 
\eq{deltaFV-def}.  However, we will always present the results of 
chiral fits after {\it  a posteriori}\ correction to infinite volume 
of both the data and the fit lines.  This allows us to present  results obtained from different volumes in an
accessible fashion, and also facilitates comparison to experiment.  


\subsection{Fits}
 \label{sec:fits}

We fit various subsets of the data to the 
chiral forms described in  \secref{schiptem}, with the FV corrections appropriate to each ensemble
added on.   The chiral forms include discretization effects, so from now
on we will refer to the fits as {\it chiral-discretization fits}.  The complete data set, which includes $a\approx0.12$, 0.09, 0.06,
and 0.045 fm ensembles, and quark charges 0,  $\pm e/3$, $\pm 2e/3$, $\pm e$, $\pm 4e/3$, $\pm 2e$,
 is based on a total of 11,654 configurations and 
has 2978 data points for  $\Delta M^2_{xy}$. Without the  $a\approx0.12$ fm ensembles, which are often 
omitted from our fits, the data set has 2166 points based on 6040 configurations.  Because points from the same 
ensemble but 
with different valence masses and/or quark charges are highly correlated, and because the number of points is 
not very much less than the number of configurations, the full covariance matrix 
is nearly singular and has many poorly determined low eigenvalues.  
Fits with acceptable $p$ values to the whole data set are therefore not possible. 
However, once the data is thinned to a more reasonable number of points in comparison to the number of 
configurations  ($\sim\! 250$ to 450 points),  acceptable fits are possible.   
For fits with up to about 350 points,
we are able to include the complete covariance matrix, with no modifications.  For fits with more points than that,
statistical and roundoff errors typically lead to a small number of negative eigenvalues (up to about 10) in the 
covariance matrix. We remove such eigenvalues with SVD when finding the inverse covariance matrix used
in the fitting procedure.  For every dropped eigenvalue, we reduce the number of degrees of freedom by 1 in
computing the $p$ value of the fit.  Our central fit, with 264 points, has no negative (and therefore no dropped) 
eigenvalues; the alternative fits used in estimating the errors of the chiral-continuum extrapolation do include
some with dropped negative eigenvalues.

When determining the $p$ value of a given fit, we take into account the fact that the sample
covariance matrix is used, rather than the exact covariance matrix that would be computed from an infinite
number of configurations in our ensembles.  We make the leading corrections in $1/N$, where
$N$ is the total number of (independent) configurations in our sample \cite{PVALUES}.

 We fit the thinned data to the 
LO+NLO \schpt\ form 
(6 parameters; \eqsthree{Delta-LO}{schpt-logs}{anal_NLO}),  plus generic discretization terms at LO and NLO 
(6 parameters; \eqs{generic_LO}{generic_NLO}), and
NNLO analytic terms (16 parameters; \eq{anal_NNLO}).   The higher order analytic terms, which include discretization terms, 
are necessary because our statistical errors in $\Delta M^2_{xy}$ are
as low as $0.2\%$, and always less than  3.3\%.%
\footnote{The smallest errors tend to occur either when both valence quarks have masses near the heavier end of our range, or when just one  quark is light, and it is uncharged.  Typically, relative errors for mesons with net charge are less than those for neutral mesons, because the total EM effect is smaller for the neutrals.  Approximately 80\% of the data points have
errors of 1\% or less.}
In addition, as described at the end of  \secref{schiptem}, we must include at least the N${}^3$LO term $\lambda_6$,
\eq{anal_NNNLO}, to obtain chiral-discretization  fits with acceptable $p$ values.
When we include data with charges greater than physical, other N${}^3$LO  analytical  terms are also 
necessary to
obtain acceptable fits.  

Our central fit includes data from the $a\approx0.09$, 0.06,
and 0.045 fm ensembles, and quark charges 0,  $\pm e/3$, and $\pm 2e/3$. 
As explained following 
\eq{anal_NLO},
we fix to zero the NLO analytic parameter $\kappa_2$, which describes sea-quark mass dependence at NLO, and 
leave the LO parameter \deltae\ unconstrained.  The generic discretization parameter corresponding
to $\kappa_2$, called $\psi_2$, is  also fixed to zero.  The fit thus has a total of 27 parameters.   

Except for
the NLO parameter $\kappa_5$,  all
NLO and NNLO parameters, as well as the  N${}^3$LO term $\lambda_6$, are constrained in the
central fit by Bayesian priors
with a Gaussian width of 3 around 0.  As discussed following \eq{anal_NLO}, the usual \chpt\ expectation would 
be that these parameters are $\cO(1)$; we believe constraining them with a prior width of 3 is reasonable
given that it is known that the size of the chiral corrections to $\epsilon$ are relatively large. The width for $\kappa_5$ is taken to be a factor of 10 larger still, in
recognition of the fact that it gets large unphysical contributions from EM quark-mass renormalization.  The width
for the generic discretization parameters $\psi_i$ is 0.044, which implies a 1-$\sigma$ deviation of 5.1\%
at $a\approx 0.09$ fm, 2.5\% at $a\approx0.06$ fm, and 1.4\% at $a\approx 0.045$ fm.

The purpose
of the Bayesian priors is to (loosely) enforce \chpt\ behavior, as well as to stabilize the fit to lattice-spacing 
dependence, for which there are many parameters and several directions in parameter space not well 
constrained by the data.   For the generic lattice spacing dependence, we can write the errors
as $(a\Lambda)^2$, where $\Lambda$ is a discretization scale, $\Lambda\approx 
540$ MeV,%
\footnote{Note that the actual lattice scales used, which we obtain from $r_1/a$ values in \tabref{ensembles} 
and $r_1=0.3117(22)$ fm \cite{Bazavov:2011aa},  are
somewhat smaller than their nominal values.}
 which we judge is large enough to be conservative.  In any case, the effects of increasing the 
prior widths by factors of 3 or 10 (or in many cases, removing the Bayesian constraints entirely) is included
in the systematic errors, as discussed in \secref{results}.

The central fit includes points with meson masses up to about 635 MeV. When masses significantly higher than 
that  are included, it is difficult to fit the data to \chpt\ forms, even with the NNLO analytic terms in the fit 
function.   Some alternative chiral-discretization  fits that are used to estimate systematic
errors include data up to about 660 MeV, but their $p$ values are rather poor ($10^{-4}$ to  $10^{-3}$).  
Other alternatives reduce the maximum meson mass included; the lowest maximum is about 540 MeV.
We do not go below this because, in order to be able to interpolate to the
physical kaon with controlled errors,  it is necessary to include the meson made from  one valence quark with mass near $m_s$ and the other the lightest valence quark .  We always include the lightest mesons available, which are ``pions'' with mass of about 250
MeV at $a\approx 0.09$ fm and about 225 MeV at $a\approx\! 0.06$ and 0.045 fm. 

We emphasize that the masses
mentioned in the previous paragraph 
all refer to taste-$\xi_5$ (Goldstone) mesons, which are the only mesons for which
we have a significant amount of data.  Mesons with other tastes can appear at one loop in the chiral expansion.
The minimum RMS mass of such mesons is about 330 MeV  at $a\approx 0.09$ fm, about 260 MeV at 
$a\approx\! 0.06$, and about 240 MeV  at $a\approx 0.045$ fm.   The taste splittings have less effect
on the maximum masses; for the central fit the maximum RMS mass goes from  about 670 MeV  at $a\approx 0.09$ fm to about 650 MeV  at $a\approx 0.045$ fm.

\begin{figure}
\begin{center}\includegraphics[width=0.65\textwidth]{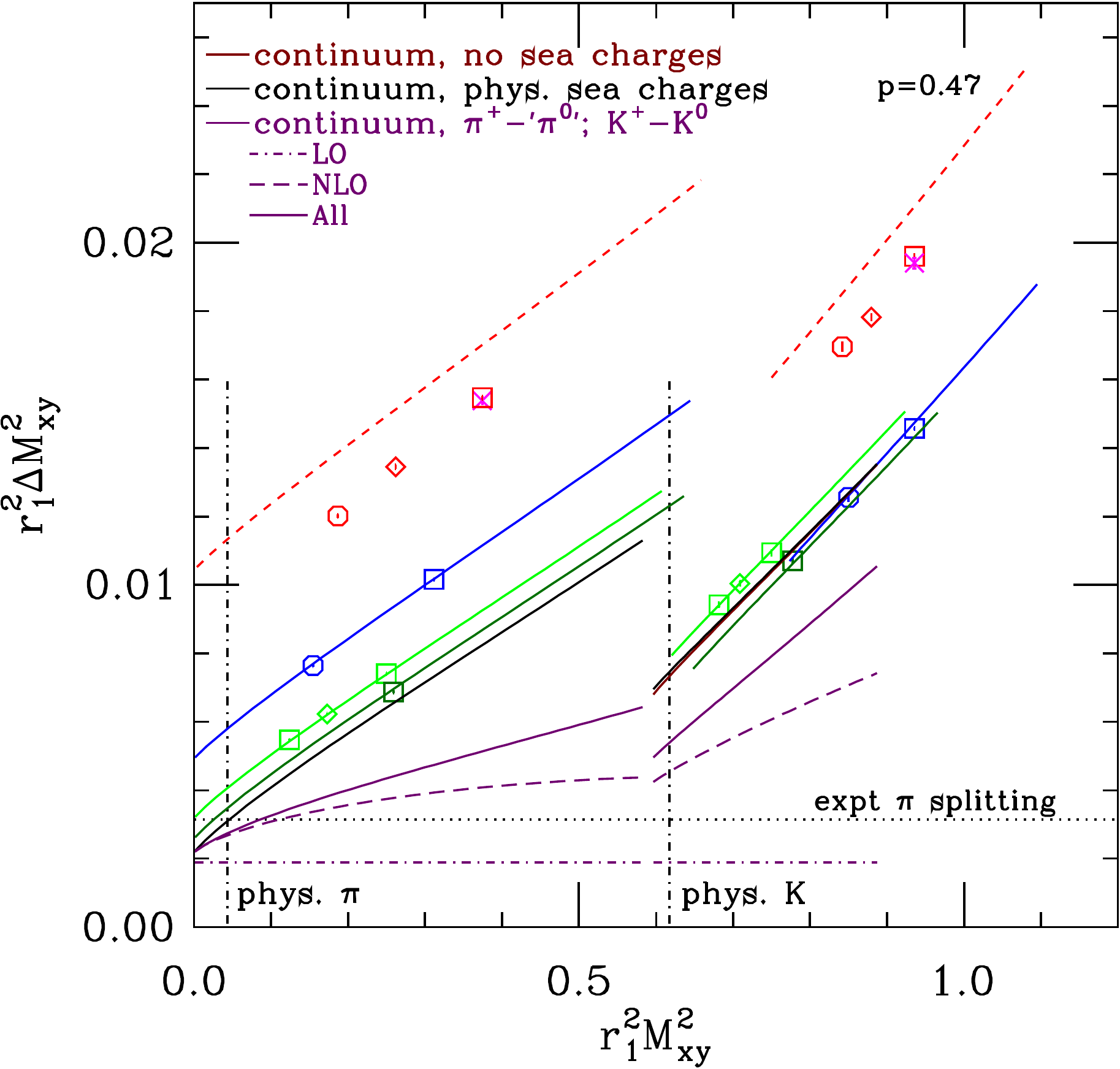}\end{center}
\caption{Central fit to  the EM splitting $\Delta M_{xy}^2$ \vs the
sum of the valence-quark masses, after correction for FV effects. The same small subset of the data as in 
\figref{FVdata} is shown. The blue, light green, and dark green lines show the fit to the $a\approx0.09$,
0.06, and 0.045 fm data, respectively.  The largest lattice spacing (red and magenta points, $a\approx 0.12$ fm) is not included in the fit; the dashed red lines show how the fit does in ``predicting'' these points. 
The horizontal dotted line shows the experimental value of the  $\pi^+$--$\pi^{0}$ splitting;
the vertical dashed-dotted lines show the quark mass values for physical $\pi$ and $K$ mesons. 
The black and brown curves are extrapolations of $\Delta M_{xy}^2$ to the continuum, with and without 
the NLO effects of sea quark charges, respectively.  (The brown curve is barely visible under the right hand black curve;
the curves are identical for the pions at left, and only the black curve is visible.)  The solid purple curves are
obtained from the black ones by subtracting $\Delta M_{xy}^2$ for the corresponding neutral mesons, $K^0$
and ``$\pi^0$.''  The dashed-dotted line and the dashed purple curves show the LO, and LO+NLO contributions
to the total solid purple lines, respectively.
 \label{fig:EMfit}
}
\end{figure}

\begin{figure}
\begin{center}\includegraphics[width=0.65\textwidth]{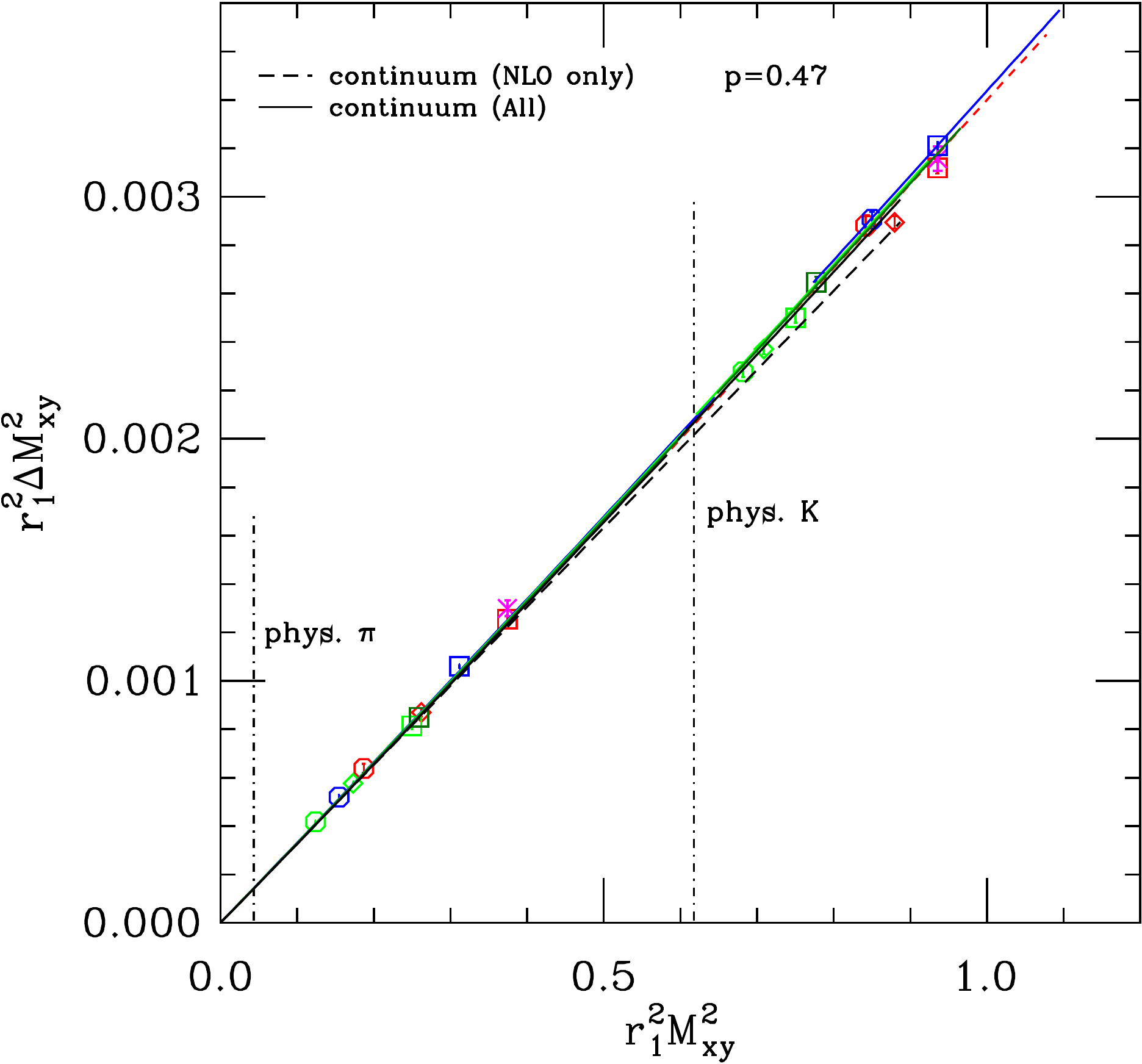}\end{center}
\caption{Same fit as \figref{EMfit}, but showing the neutral mesons made out of quarks with charges $\pm e/3$.
 The meaning of the blue, light green, dark green, and dashed red curves is  the same as in  \figref{EMfit}; note the difference in vertical scale between the two plots. 
The solid black curves  are extrapolations of $\Delta M_{xy}^2$ to the continuum.  
The dashed black lines show the NLO contribution
to the solid black curves; there are no LO contributions.   The discretization errors and sea-mass
dependence for neutral mesons are very small, as are the nonlinear contributions to the valence-mass dependence.
 \label{fig:EMfit_neutral}
}
\end{figure}

\Figref{EMfit} shows the same subset of our data as in \figref{FVdata} (charge $\pm e$, unitary or approximately unitary) after correction for FV effects, along with 
the central chiral-discretization  fit and its extrapolations.  The unitary (or approximately unitary) points from the
same fit for neutral mesons made out of $d$- or $s$-type quarks (charges $\pm e/3$) are shown in  \figref{EMfit_neutral}. The fit  has 264 data 
points, 237 degrees of freedom,   $\chi^2=248.0$, and 
 $p=0.47$.  (Without correction for the use of the sample covariance matrix, the $p$ value  of this fit would 
 have been 0.30.)  
Here and below, when we give $\chi^2$ or $p$ values without further qualification, they are
 the standard ones, where $\chi^2$ comes only from the difference of the data and the fit, and the degrees of
 freedom are equal to the total number of data points minus the number of parameters, without regard to whether
 those parameters are constrained by Bayesian priors.  We will specify when we actually mean the augmented values, where the priors are treated as additional data, contributing to $\chi^2$ as well as to the degrees of
 freedom.  Because the priors  are loose, in the sense that the parameters are
 to a great extent constrained by the data and not the priors, we expect that the augmented $p$ values 
 will be larger than the standard $p$ values.   In the case of the central fit,  $(\chi^2/{\rm d.o.f.})_{\rm aug}= 
 255.3/263$, giving $p_{\rm aug}=0.79$.  
 
 In \figref{EMfit}, the blue, light green, and dark green curves show the quality of the fit to the $a\approx0.09$,
0.06, and 0.045 fm data, respectively. The points at $a\approx 0.12$ fm (red and magenta) are not included
in the fit, but the dashed red curves show that the fit does reasonably well in predicting the data at this
lattice spacing.  It is  more difficult to extrapolate to larger lattice spacing than to smaller lattice spacing, since
larger lattice spacing may be sensitive to higher-order terms that are either not included in the fit or not well
determined on finer lattices.   

For the neutral mesons (\figref{EMfit_neutral}), the discretization errors,
as well as the sea-mass dependence, are quite small, since points from different lattice spacings and
sea-mass values line up very well.   Further, as required by chiral symmetry,  $\Delta M^2_{xy}$ vanishes
in the chiral limit.  It is also noteworthy that the curvature in the fit lines is small, as may
be deduced from the small difference between the curves and the dashed black line, which is straight.  
There are no chiral logarithms for neutral particles at NLO, and the NNLO logs are not included in
our fit function.  There is  a contribution from the NNLO analytic term that is quadratic in valence
masses and can contribute to neutral mesons (the $\rho_{11}$ term in \eq{anal_NNLO}), but it
is rather small:  $\rho_{11} = 0.52(3)$.  All our alternative chiral-discretization fits preserve these
simple features, which are enforced by the lattice data.  Therefore we only show the charged-meson plots
for the alternative fits below.

The black curves in \figref{EMfit} show the fit after setting valence and sea masses equal,
adjusting $m_s$ to its physical value, extrapolating to the continuum, and adjusting the sea charges to
their physical values using NLO \chpt. The last adjustment vanishes identically for pions and is very small for kaons.
The brown kaon curve (barely visible under the black kaon curve) shows the value before adjustment, \ie with vanishing
sea-quark charges.
From the black curves for the $\pi^+$ and $K^+$,
we subtract the corresponding black curves for the neutral mesons $``\pi^0"$ and $K^0$ 
shown in \figref{EMfit_neutral},%
\footnote{More precisely \figref{EMfit_neutral} only shows the $d\bar d$ contribution
to the ``$\pi^0$.''}  
giving
the solid purple curves, whose values at the physical point for each meson (indicated by 
the vertical dashed-dotted lines) are the physical results.    

The solid purple curve in \figref{EMfit}  includes all chiral terms
through NNLO (and with the N${}^3$LO term $\lambda_6$).  We also show the LO contribution alone
(the mass-independent
horizontal dashed-dotted purple line) and the LO+NLO contributions (the dashed purple curves).  
In this fit the LO contribution has the value $r_1^2\dem =  0.00189(12)$; this is about 60\% of the
value 0.00315 that would be necessary to give the full experimental pion splitting at LO.
As expected from the fact that $\epsilon$, which measures higher order contributions to $(M^2)^\gamma$,
is of order 1, the NLO contributions are relatively large, especially for the kaons or heavier-than-physical
pions.   The NNLO contributions are clearly much smaller than the NLO ones for physical kaons, and 
negligible, or nearly so, for physical pions.  Thus, after an anomalously large
NLO contribution, SU(3) \chpt\ appears to converge reasonably well.

One may wonder whether this picture of the convergence of \chpt\ is strongly influenced by the Bayesian
priors that constrain NLO and NNLO LECs in the fit.  In fact, the priors on physical LECs (those whose
contributions do not vanish in the continuum limit) have almost no effect on the convergence or the results.
If we remove all prior constraint on these physical%
\footnote{The LO physical LEC \deltae\ is never constrained unless nonzero $\kappa_2$ is allowed. We also include $\kappa_5$ and $\lambda_6$ among the ``physical'' LECs even though they get
unphysical contributions from EM renormalization, because they do not vanish in the continuum limit.}
LECs ($\kappa_2$, $\kappa_3$, $\kappa_4$, $\kappa_5$, $\rho_6$, $\rho_7$, $\rho_8$, $\rho_9$, 
$\rho_{10}$, $\rho_{11}$, $\rho_{12}$, $\rho_{13}$, $\rho_{14}$,  $\rho'_1$, $\rho'_2$, $\lambda_5$),
the fit and results are almost unaffected:  $\epsilon$ changes by only $0.03\%$.   This is however
dependent on our setting parameter $\kappa_2=0$ as discussed above; the separation
between LO and NLO contributions can be drastically altered if $\kappa_2$ is allowed to vary.  

\begin{figure}[t!]
	\begin{center}
	\begin{tabular}{c c }
		\includegraphics[scale=0.42]{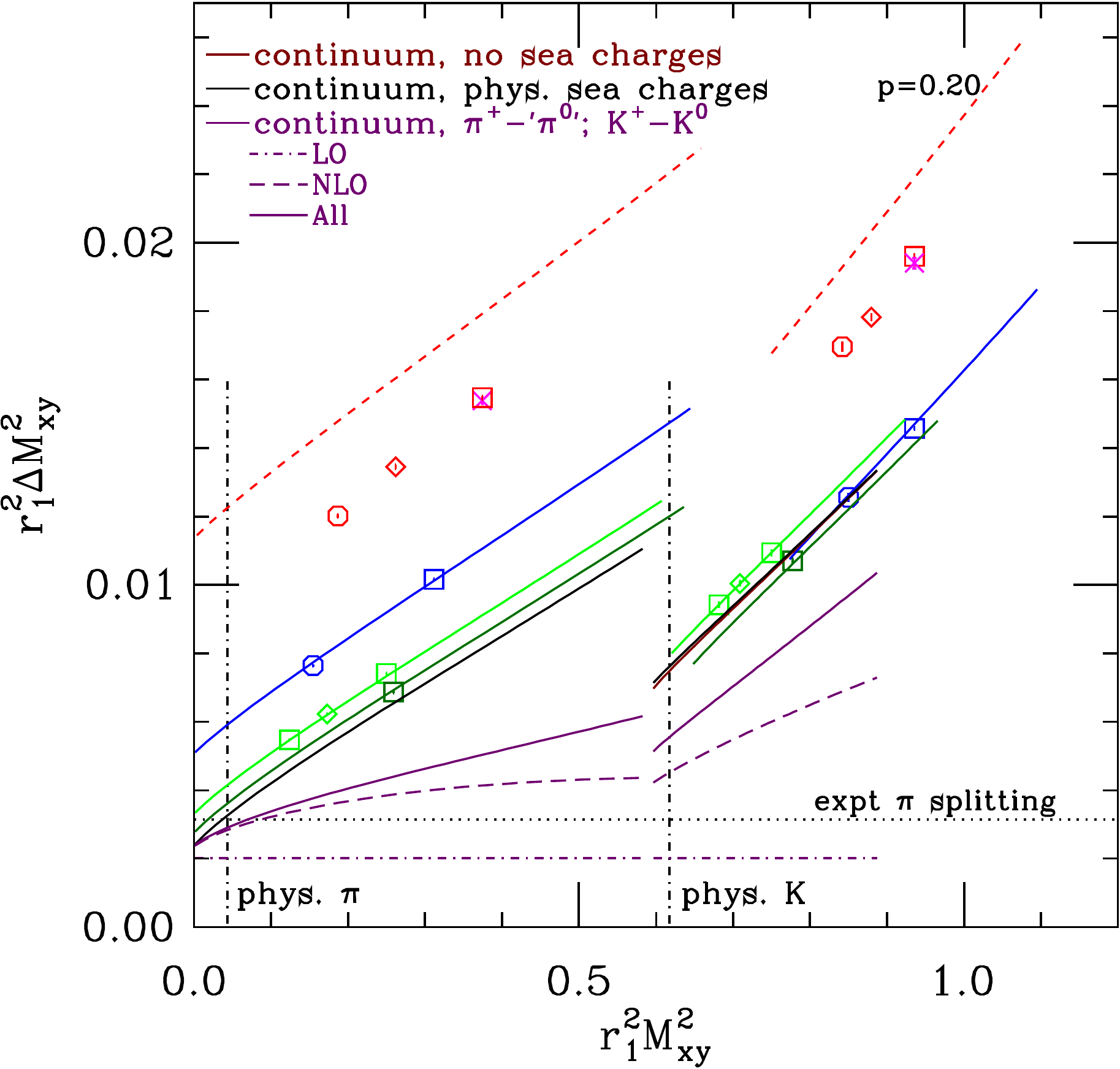}   
		&
		\includegraphics[scale=0.42]{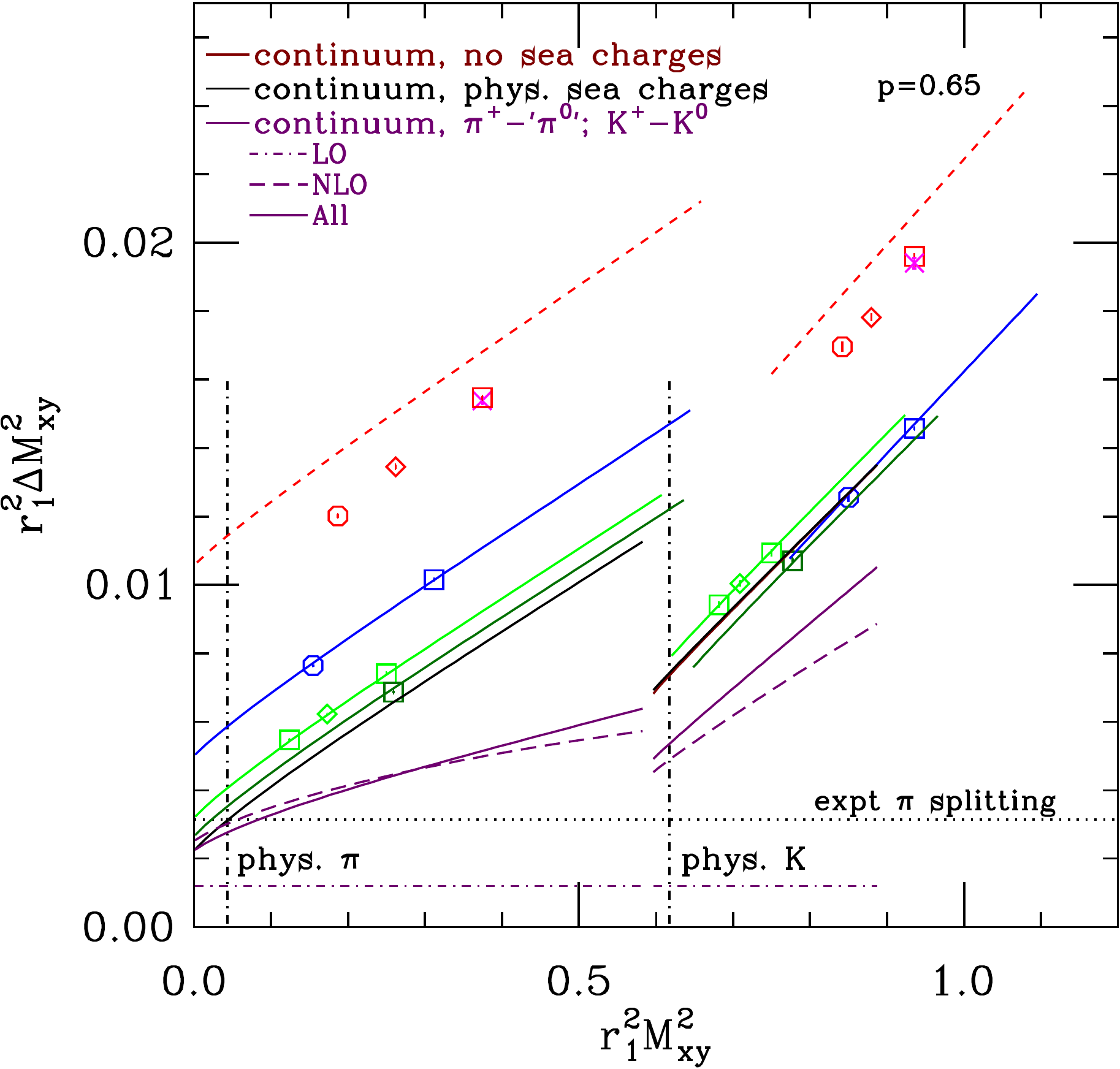}   
		 \\
		 (a)  &  (b)  
	\end{tabular}
	\end{center}
\caption{Two alternative chiral-discretization fits.  The data included in the fits, as well as the
meaning of symbols and curves is the same as for
the central fit, \figref{EMfit}.  
Fit (a) sets parameters  $\rho_6$ and $\rho_7$ to zero, as well as $\kappa_2$.  The differences with \figref{EMfit}
are small: all curves are slightly higher in the chiral limit, and the predictions for the $a\approx0.12$ fm points
(dashed red curves) are noticeably higher.
Fit (b)
does not fix the parameter $\kappa_2$ (or $\rho_6$ and $\rho_7$) to zero. The main difference from \figref{EMfit}
that is apparent in fit (b) is  the relative size of the contributions to the continuum result
of the LO contribution (horizontal, purple dashed-dotted line), and the LO+NLO contribution
(dashed purple curves).   The full continuum-extrapolated results (solid purple curves) are however very
close to those in  \figref{EMfit}.
 \label{fig:EMfit_alt}
}
\end{figure}

\Figref{EMfit_alt} shows two  alternative chiral-discretization  fits to the same set of data
points as the central fit. Both of these fits have
reasonable $p$ values. In (a) the NNLO parameters $\rho_6$ and 
$\rho_7$ are set to zero, in addition to the NLO parameter $\kappa_2$.   These NNLO parameters play
a role that is similar to $\kappa_2$: the corresponding analytic terms  depend on the sea-quark masses and
are nonzero in the chiral limit.  No major changes from the central fit are visible, but the LO continuum contribution 
(dashed-dotted purple curve) is slightly higher than for the central fit (here, $r_1^2\dem =  0.00202(4)$), and all the 
curves (fit lines as well as extrapolations) are correspondingly  higher in the chiral limit.   The value of $\epsilon$,
however, is only 0.002 below that of the central fit. 
One  somewhat more obvious change
is that the  predictions for the $a\approx0.12$ fm points
(dashed red curves) are somewhat worse than in the central fit.  

In \figref{EMfit_alt}(b),  $\kappa_2$ is not fixed, but is constrained
by our standard  prior for physical LECs, $0\pm3$.  The LO parameter  is now also constrained by priors
with central value $r_1^2\dem=0.003$ and width $0.001$.  The posterior value is  almost
two sigma lower, $r_1^2\dem=0.0012(3)$, now less than 40\% of the value that would be necessary to give the
experimental pion splitting at LO.  Nevertheless,  the final results (sum of all
chiral orders) from this fit are quite close to
those of the central fit, as can be seen by comparing the solid purple curves in \figrefs{EMfit}{EMfit_alt}(b).
Indeed, the value of $\epsilon$ coming from this alternative fit is just 0.02 below that in the central fit.
The fit lines to the data at fixed lattice spacings and sea masses are also very similar in \figrefs{EMfit}{EMfit_alt}(b).

\begin{figure}
\vspace{-5mm}
\begin{center}\includegraphics[width=0.65\textwidth]{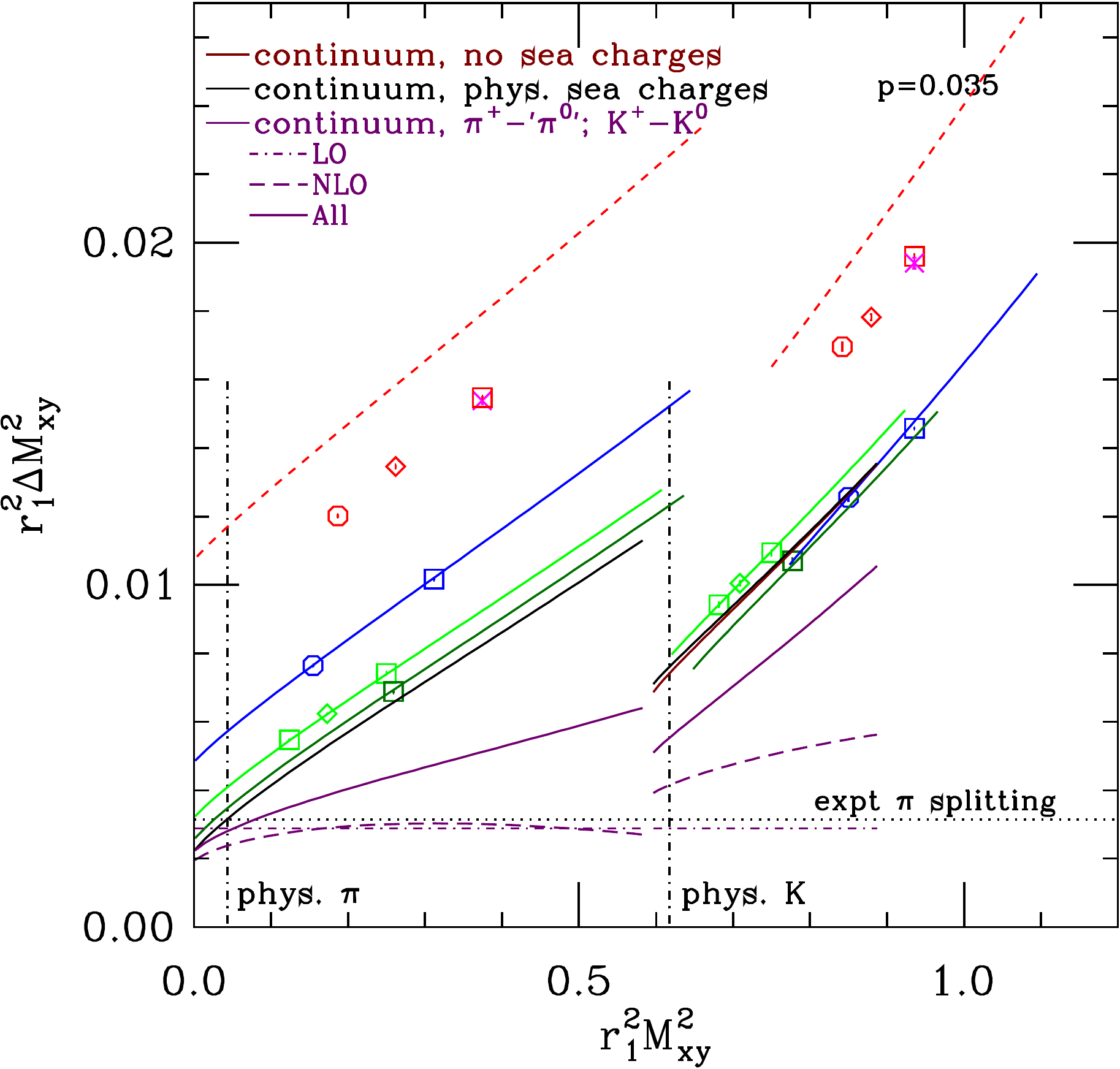}\end{center}
\vspace{-6mm}
\caption{An alternative chiral-discretization fit to the same data as for the central fit.  The
meaning of symbols and curves is the same as for
the \figref{EMfit}.   This fit puts a strict prior on $\deltae$ to force it to be very near to the value
that would give the physical pion splitting at LO.   Note that the higher chiral orders reduce the pion chiral
limit significantly below the LO contribution.
 \label{fig:EMfit_alt3}
}
\end{figure}

 In \figref{EMfit_alt3}, we show a third alternative fit to the same data points as the central fit.  
Here, we have put a very tight prior on $\deltae$, $r_1^2 \dem = 0.0031\pm 0.0001$) to force the
LO \chpt\ contribution to be close to the experimental pion splitting.  The posterior value is about two sigma
below this, $r_1^2 \dem = 0.0029(1)$.  The chiral LECs all have priors $0\pm3$, including $\kappa_2$, which
is allowed to vary, but now has a negative posterior value in order to reduce the pion chiral limit of the fit 
to something 
that is better tolerated by the data.  The $p$ value of the fit ($p=0.035$) is significantly less than the other
fits that we have considered so far, but is still acceptable.  The resulting value of $\epsilon$ is 0.024 higher
than that of our central fit, and is in fact the largest positive deviation from the central value of all the alternative
fits we have considered.  

The relative contributions in the continuum of various orders in the chiral expansion as predicted by the fits are 
also sensitive to the parameters that control 
lattice spacing dependence ($\kappa_1$, $\rho_1$, $\rho_2$, $\rho_3$, $\rho_4$, 
$\rho_{5}$, $\psi_0$, and $\psi_i$).  As mentioned above, the fit becomes unstable if these
parameters are completely unconstrained. If the prior widths are widened but not eliminated, the effects
on the results are controlled (and included in the systematic error estimates), but the division between LO and
NLO contributions can again be significantly changed.  Thus the division between orders in \chpt\ shown in 
\figref{EMfit} is at best very rough.  The final results are nevertheless  much more stable than the 
individual \chpt\ orders, as we have already seen in comparing the fits in \figrefthree{EMfit}{EMfit_alt}{EMfit_alt3}.  This remains  true even for the more extreme divisions between orders considered
below.

\begin{figure}[t!]
	\begin{center}
	\begin{tabular}{c c }
		\includegraphics[scale=0.42]{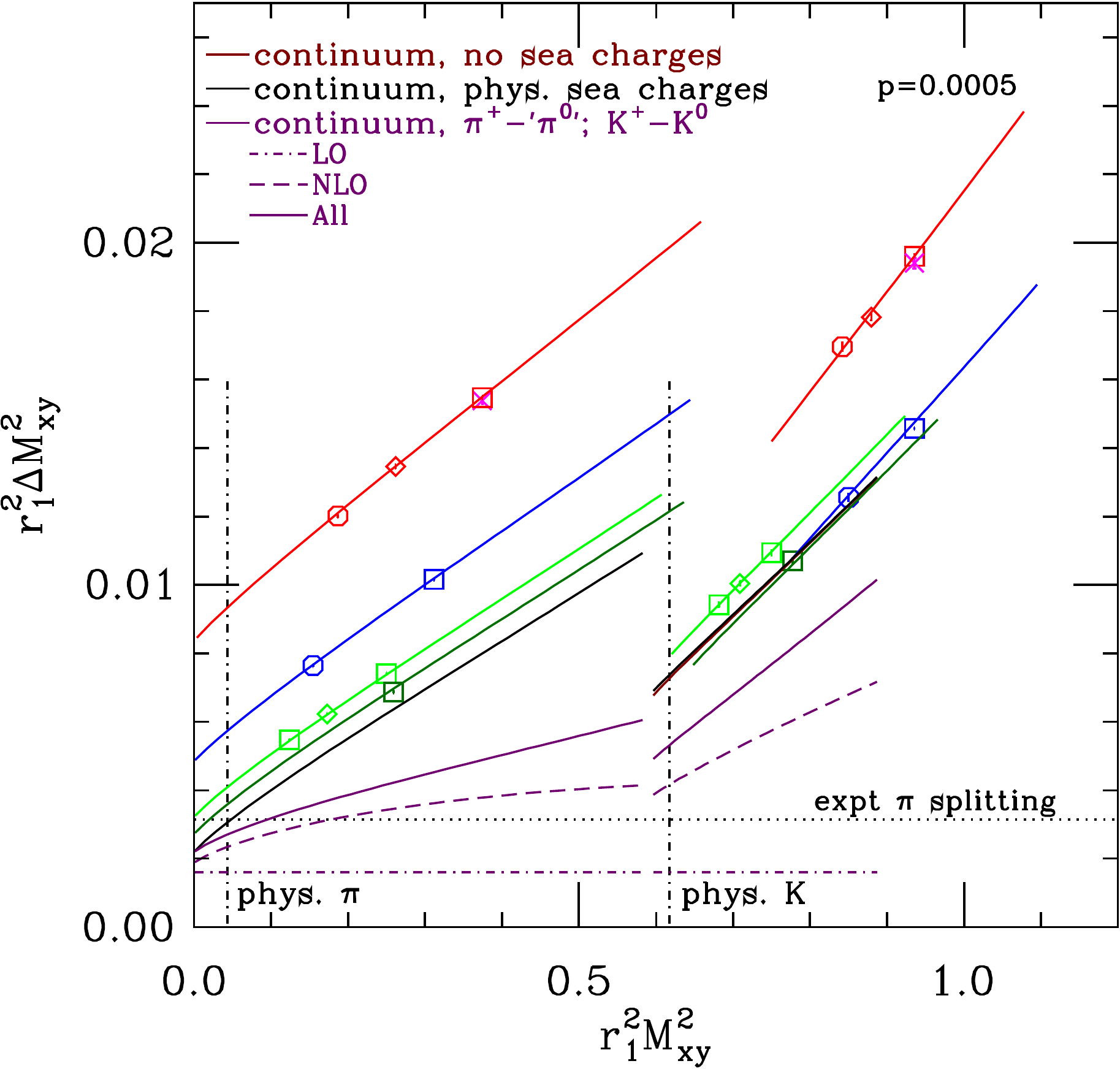}   
		&
		\includegraphics[scale=0.42]{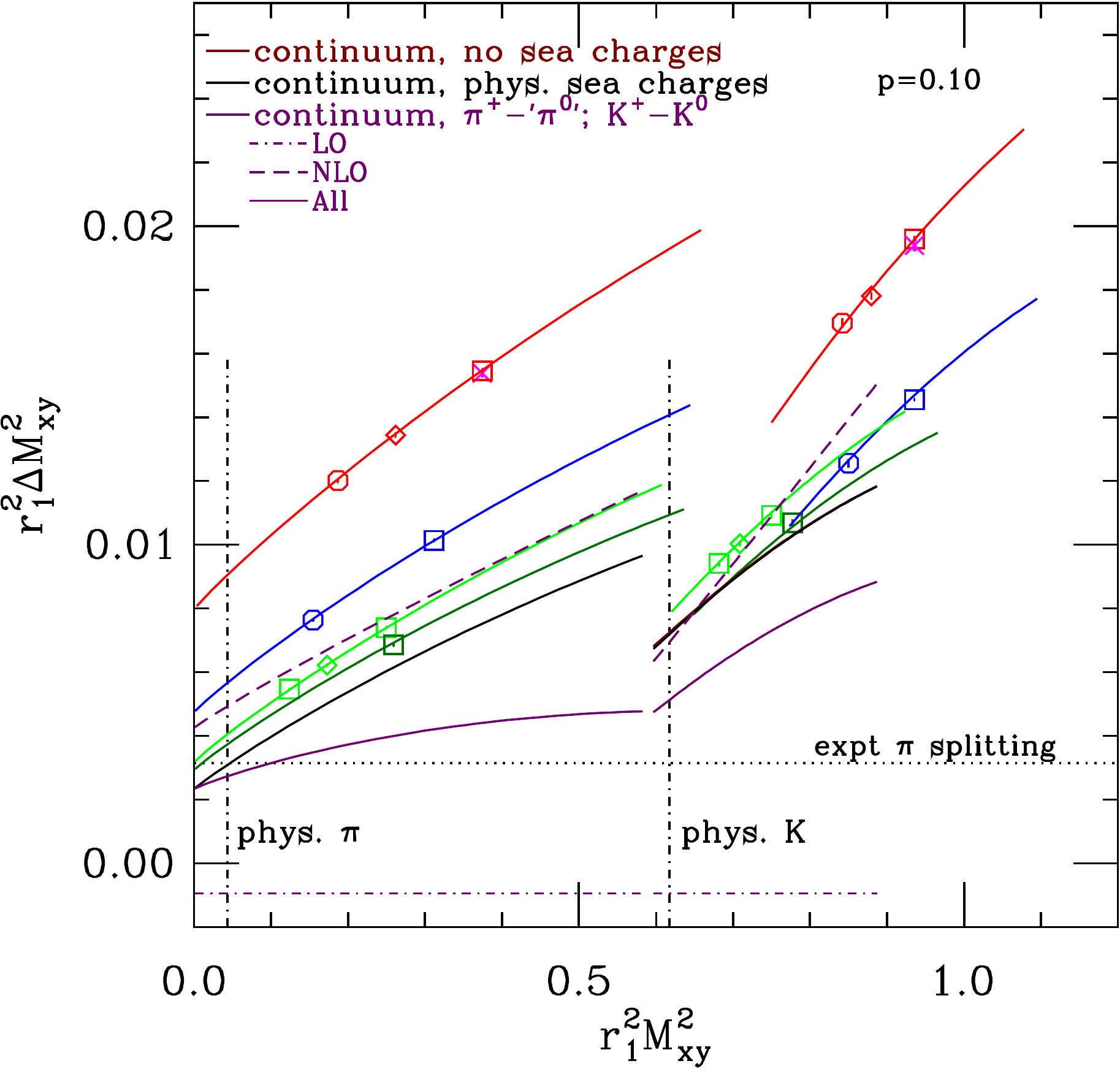}   
		 \\
		 (a)  &  (b)  
	\end{tabular}
	\end{center}
	\caption{ Two examples of chiral-discretization  fit and extrapolation like the central fit in \figref{EMfit}, but including the points from the $a\approx0.12$ fm ensembles.  Fit (a) is most similar to the central fit, in that $\kappa_2$ is fixed to zero
	and $\deltae$ is unconstrained.  Fit (b) allows $\kappa_2$ to vary (with a prior $0\pm40$), and imposes  a prior of $0.003\pm0.001$ on $r_1^2\dem$.  Nevertheless the fit value of $\deltae$ is negative. }
	\protect\label{fig:EMfits-coarse}
\end{figure}

It is not surprising that inclusion of the $a\approx0.12$ fm ensembles leads to difficulties
with the fits. The taste-breaking effects at this lattice spacing 
produce large discretization errors, and the fact that the
physical strange quark is about 35\% below the simulated strange mass gives further
problems.    The smallest meson-mass maximum that allows us to interpolate to the kaon is approximately 
645 MeV for the Goldstone meson, and  about 750 MeV for the RMS taste meson.  For low masses, the 
taste effects are even worse:  while the lowest available Goldstone mass is   about 275 MeV,
this corresponds to an RMS taste mass of  about 465 MeV.
Thus, when we add in the $a\approx0.12$ fm ensembles, the chiral-discretization
fits have various undesirable features.  \Figref{EMfits-coarse} shows two examples of such fits.  \Figref{EMfits-coarse}(a) is rather similar to the central fit:  $\deltae$ is unconstrained but $\kappa_2$ is fixed to zero.  Despite the fact that we have increased the
prior widths of the LECs controlling  lattice spacing dependence ($\kappa_1$, $\rho_1$, $\rho_2$, $\rho_3$, 
$\rho_4$, $\rho_{5}$) to 40, and the width of the generic variation parameters to $0.11$, the fit is poor 
($p=0.0005$).   Nevertheless $\epsilon$ is only 0.01 below that of the central fit. 

Fits with reasonable $p$ values that include the $a\approx0.12$ fm ensembles are 
possible. In \figref{EMfits-coarse}(b) we allow $\kappa_2$
to vary, and put a relatively loose prior on \deltae\ ($r_1^2\dem = 0.003\pm0.001$), as well as dramatically increasing the 
prior widths of the parameters that control lattice spacing dependence. We now obtain $p=0.098$.  However,
this fit has a negative value of \deltae, which implies an extreme breakdown of \chpt, as well as very large discretization
LECs ($\kappa_1\approx-34$, $\rho_1\approx-70$, $\rho_2\approx33$, $\lambda_2\approx16$).  It may very well be justified to drop this fit 
on these grounds.  To be conservative, however, we keep it in estimating the systematic errors.  Indeed, it is the fit
that gives a value of $\epsilon$ that is furthest away from our central value  (0.082 lower) out of all the chiral-discretization  alternatives we consider.

Adding in points with quark charges that are greater than the physical ones leads to problems with the 
chiral-discretization fits that are similar to those we have when adding in the
$a\approx\!0.12$ fm ensembles.  This may be because the higher charges bring in greater discretization
errors.  Indeed, there is evidence \cite{EM12} that taste violations from photon exchange begin to be important when the charges increase above their physical values.
For data with quark charges 0, $\pm 1/3$, $\pm 2/3$, $\pm 1$, and $\pm 4/3$, a fit like the central fit (but 
including all the $e^4 p^2$ LECs ($\lambda_i$ in \eq{anal_NNNLO}) and with
somewhat larger priors ($0 \pm 5$ on LECs) has $p=0.005$ and an $\epsilon$ that is 0.03 below that of the central fit.
A fit with $\kappa_2$ not fixed to zero, and very loose priors on LECs and generic discretization parameters,
has $p=0.21$. However $\deltae$ is negative, and 
 discretization terms are very large.  Both features are quite similar to those seen in \figref{EMfits-coarse}(b).
 In this case, $\epsilon$ is  $0.065$ below that of the central fit.

\section{Systematic Errors and Results of EM Calculation\label{sec:results}}

Our calculation has the following significant sources of systematic errors:
 (1) chiral-continuum uncertainties from the extrapolations to the physical light quark mass and to $a=0$, (2) finite-volume (FV) effects, (3) systematic issues
involved in the EM renormalization,  (4) effects of using the  ``$\pi^0$,'' which
does not include quark-line disconnected contributions, instead of the true  $\pi^0$,  (5) errors in the physical value of $r_1$, the quantity we use to set
the scale, (6) uncertainties in the physical values of the quark masses after extrapolation to the continuum, 
and (7) effects of EM quenching.  In the following subsections, we discuss each source of error in turn.  
For the two EM quantities we calculate,
$\epsilon$ and  \ek,
\tabref{errors}
lists   central values and statistical errors from the fit shown in \figref{EMfit},  and each systematic error.  

The separation between EM and isospin-violating effects is dependent on the scheme, which
enters through the EM renormalization of quark masses. Our calculation
is performed in the BMW scheme \cite{Borsanyi:2013lga}, described in \secref{renormalization}.  
For some purposes, it may be
useful to gauge the effects  of
changing to another reasonable scheme.  In \secref{scheme-dependence}, we estimate such
scheme dependence  for $\epsilon$ and \ek.

\begin{table}
\begin{center}
\begin{small}
\begin{tabular}{|c|r@{}l|r@{}l|}
\hline
source&\multicolumn{2}{|c|}{$\epsilon$}& \multicolumn{2}{|c|}{\ek\strike{\KEM}} \\
\hline
\hline
central value                  & \hspace{5mm}0&.776\hspace{4mm}                             & \hspace{5mm}0&.035\hspace{4mm}  \strike{44&.3}  \\
\hline
statistics              &0&.012                               &0&.003\strike{3&.2}   \\
\hline
\hspace{2mm}chiral-continuum \hspace{2mm}  &  \multicolumn{2}{|c|}{$\hspace{-1.5mm}{}^{+0.024}_{-0.082}$}       &0&.002\strike{2&.7}         \\
\hline
finite volume                   & 0 & .056                               &&--                                                                       \\
\hline
renormalization                  & 0 & .002                               &0&.012\strike{15&.2 }                                                                       \\
\hline
``$\pi^0$''                         & 0 & .034                               &&--                                                                        \\
\hline
absolute scale                   & 0 & .001                               &0&.000\strike{0&.3}                                                                       \\
\hline
quark masses                   & 0 & .009                               &0&.011\strike{13&.7}                                                                        \\
\hline
EM quenching                   & 0 & .040                               &0&.012\strike{14&.8}                                                                        \\
\hline
\end{tabular}
\end{small}
\caption{Central values and errors for $\epsilon$ and \ek. \label{tab:errors} }
\end{center}
\end{table} 

\subsection{Chiral-continuum error}
\label{sec:chiral-error}

We determine this error by considering a wide range of alternative chiral-discretization fits, with
various priors and/or parameters set to zero,  and apply them to various subsets of the data:
different maximum meson masses included, different thinning, omitting or including the 
coarsest ($a\approx0.12$ fm) ensembles, and omitting or including quark charges greater than the physical
charges.  Several of these fits have been presented in \secref{fits}.  To be conservative we include fits that
have $p$ values as small as $10^{-5}$, as well as ones that have very large discretization terms and/or
exhibit extreme breakdown of \chpt\ (\eg negative \deltae).  Altogether, 89 fits are included.  We take the
largest positive and negative differences from the value in the central fit as the error.  For $\epsilon$,
this gives a positive error of $+0.024$, coming from the fit in \figref{EMfit_alt3}, and a negative
error of $-0.082$, coming from the fit in \figref{EMfits-coarse}(b).  For \ek, the maximum positive and negative
differences are comparable, so we average them and quote a symmetric error of 
0.002\strike{2.7 (MeV)$^2$}.

\subsection{Finite-volume error}
\label{sec:FV-error}
To estimate the systematic error associated with the FV correction we use  (a {\it residual}\/ FV error), we examine the deviations of the fit lines from the data in \figref{FV}. By far the largest deviation 
occurs for the  ``pion'' (blue) curve at $L=16$.  Let $x$ be
the difference between the predicted infinite-volume value of the pion mass from the $L=16$ point alone and
the value from all the other points.   We take $x$ as the presumed absolute value of the error of our FV estimates,
and divide it by the size of the estimated FV correction at $L=20$, to find a fractional residual 
FV error of approximately 20\%.
We use the $L=20$
point because most of the data used in the central fit is from physical volumes of approximately that
 size or slightly
larger; using the $L=28$ point instead would make a negligible difference.   An error of 20\% is also reasonable 
because usual corrections from higher orders in SU(3) \chpt\ are of this size. 
For $\epsilon$ the net effect of the FV corrections we have made is 0.28, which we find simply by comparing our
central value with the value obtained by refitting the data with FV corrections turned off.  Our estimate for the
residual FV error is then 20\% of 0.28, or 0.056.  

For neutral mesons there are no chiral logarithms at NLO, and hence no FV effects at this order.  There will
be FV effects at higher orders, but they are very likely to be much smaller than our other systematic
errors in \ek, which are quite large.  We therefore do not include a residual FV error for \ek\ in \tabref{errors}.

\subsection{EM renormalization error}
\label{sec:renorm-error}
We use the BMW scheme, as defined by \eq{isospin-limit-def} and as implemented by
\eq{isospin-limit2}, to perform nonperturbative EM renormalization
of the $u$- and $d$-quark masses.  For $\epsilon$, this is sufficient, since the renormalization
of the $s$-quark mass cancels in the difference \KPZ.  However $s$-quark mass
renormalization is crucial for obtaining \ek.  We extend the renormalization to the $s$ quark using \eq{ms}.

We can implement \eq{isospin-limit2} to high accuracy from our chiral fits, so the only significant systematic
errors in the scheme come from the errors in our values of the derivatives of the squared meson masses with 
respect to quark mass:  $B\equiv \partial M^2_\pi/\partial m_l$, $B_l=  \partial M^2_K/\partial m_l$, 
and $B_s=  \partial M^2_K/\partial m_s$.   We only need these quantities for physical quark masses
and in the continuum limit, since
we perform the renormalization after the chiral-discretization fit and its extrapolations. 
For $B$, we have the SU(2) \chpt\ result, \eq{B}, which is
quite precise:  the error from $\bar\ell_3=2.81(64)$ \cite{Aoki:2016frl}   results in a 0.4\% error in B.  Corrections from NNLO should be even smaller, since the NLO
correction is already only 2\%.    For $B_l$ and $B_s$, SU(3) \chpt\ would be needed, but the higher order
corrections, as well as the uncertainty in the relevant LEC, are large.  Instead, we extract these quantities from our lattice data, and make a simple extrapolation (linear in $a^2$) to the continuum.
  We find $B_s/B=0.974(15)$, $B_l/B=0.946(19)$, where we give the results
in terms of the central value of $B$ (errors in $B$ should not be added on to these values).  
The error in $B_l$ is small enough that the resulting error in $\epsilon$ is small compared to other systematic
errors; $\epsilon$ is independent of $B_s$.  The total renormalization error on $\epsilon$ is 0.002.

For \ek, we find a renormalization error  of $0.012$\strike{$15.2\ ({\rm MeV})^2$}.  The error is dominated by the 
uncertainty coming
from $B_s$, and would therefore benefit from increased precision in this quantity.  A significant improvement 
in $B_s$ could be obtained from a dedicated 
pure QCD calculation with several closely spaced strange-quark masses 
around the physical value at each lattice spacing.    However,  the fact that \ek\ has an uncontrolled
quenched-EM error means that one cannot decrease the overall error very much without also going to
dynamical QED simulations (or equivalent approaches to include the effects of sea-quark charges
at order $\alpha_{\rm EM}$).

\subsection{Error from dropping disconnected $\pi^0$ diagrams}
\label{sec:pi0-error}
As described in \secref{pi0}, we simulate a ``$\pi^0$'' in which quark-line disconnected diagrams
are dropped, rather than the physical $\pi^0$.  The difference is $\cO(\alpha_{\rm EM}M_\pi^2)$.

We may estimate the  size of this effect by noting that the disconnected contributions are solely responsible
for the chiral logarithm term found in \rcite{Urech:1994hd}.  The connected contributions, which are equal
to  \uuEM\ or \ddEM, have no chiral logarithms at NLO.  Indeed, there are no NLO chiral logarithms for
any neutral meson that has only connected contributions, such as the neutral kaon.
In \secref{pi0}, we estimated  the chiral logarithm term in \pizEM\ as approximately 
$30\ ({\rm MeV})^2$. Using the result from our central fit for \dem\ instead of the value from 
\rcite{Bijnens:2006mk}, gives a smaller result of $25\ ({\rm MeV})^2$.    

In estimating the  error on $\epsilon$, we also need the pion splitting, which appears in the denominator. 
 The experimental value is $1261\ ({\rm MeV})^2$, but we can put the denominator on the same
 footing as  the error in the numerator by using instead the value obtained from LO, namely
 \dem.  With the value of \deltae\ from \rcite{Bijnens:2006mk}, we get
 about $900 \ ({\rm MeV})^2$ for the LO pion splitting. Taking this smaller
 value for the denominator and the larger estimate of the  error in numerator,  we get a conservative
 estimate of 0.034 for the error 
 in $\epsilon$.   This value is independent of what is assumed for $\deltae$, since it cancels between
 the numerator and denominator.   
 
 Another approach to estimating the ``$\pi^0$'' error would be to compare $\epsilon$ with $\epsilon'$, defined by
 \eq{epsp-def}.  In $\epsilon'$ the experimental value of the pion splitting appears in the numerator
 instead of the computed value of the pion EM splitting, so $\epsilon'$ is independent of how we treat the
 $\pi^0$.  From the discussion in \secref{intro}, we expect
 that, in the absence of statistical or other systematic errors, $\epsilon =\epsilon'+\epsilon_m$,
 where $\epsilon_m=0.04(2)$ \cite{Aoki:2016frl}.   However,
 chiral-discretization errors play a significant role here, since there is a partial cancelation of errors in $\epsilon$
 when we subtract \pisplitEM\ from \KPZ.  Indeed, the chiral-discretization error for $\epsilon'$ is a factor
 of about 4 larger than for $\epsilon$.  If we ignore this problem and just focus on the central fit,
 $\epsilon - (\epsilon'+0.04) = 0.089$.  This is slightly smaller  than the expected error (0.091) from
 the addition in quadrature of the chiral-discretization and ``$\pi^0$'' errors on $\epsilon$, and the 
 $\epsilon_m$ error.   Because there are also
 likely to be some residual FV errors in the difference $\epsilon - (\epsilon'+0.04)$,  the
 $\epsilon'$ result suggests that the errors we have already included are reasonable and do not need
 to be increased further.  
  
 The calculation of the EM effect for the $K^0$ is independent of the treatment of the $\pi^0$, so there
 is no corresponding error in \ek.

\subsection{Scale error}
\label{secs:scale-error}
The absolute scale of our ensembles is set by $r_1=0.3117(22)$ fm \cite{Bazavov:2011aa}.   To find the induced
error in our results, we rerun the analysis with $r_1$ changed by 1 $\sigma$.  In doing so,
 it is necessary
to include the changes, caused by the scale, in the physical quark masses in the continuum limit.  The scale
error in these masses is given in  \tabref{other-params}.  Note that the estimates of the quark masses
move in the same direction as $r_1$ because  the quark masses are adjusted to reproduce the experimental 
values of the meson masses multiplied by $r_1$.

The resulting scale errors are very small:  0.001 in $\epsilon$ and 0.0002 in \ek\strike{$0.3\ ({\rm MeV})^2$ in \KEM}.  
In each case, the effect of changing the scale itself is largely cancelled by the scale changes in
the quark masses.  \strike{For  $\epsilon$,} Only the denominators, which come\strike{s} from the squared experimental splitting
multiplied by $r_1^2$, are\strike{is} affected by the change in the scale itself, and only the numerators are\strike{is} affected by the
changes in quark masses.

\subsection{Quark mass  error}
\label{sec:quark-mass-error}

To find the errors coming from our values of the physical quark masses, we rerun the analysis with the continuum physical mass values%
\footnote{The values of physical quark masses at nonzero lattice
spacings affect our results only through the values of $r_1/a$, which are extrapolated to these quark masses in our mass-independent scale-setting scheme. The effects on the final results are negligible.}
 given in \tabref{other-params} changed
by 1 systematic $\sigma$  (not including scale errors).
  Because the nonscale errors arise largely from variations over the same set of pure QCD chiral fits, 
$m_l$ and $m_s$ are highly, positively correlated, and we change both in the same direction.  
We find quark mass errors of  0.009 in $\epsilon$ and  0.011 in \ek\strike{$13.7\ ({\rm MeV})^2$ in \KEM}.
Assuming instead that the errors on $m_l$ and $m_s$ were
uncorrelated would change the resulting errors only slightly because changes in one of the masses
always dominate:  $m_s$ dominates for $\epsilon$, while $m_l$ dominates for \ek.  

Errors arising from the other inputs in \tabref{other-params} are negligible and are not included in
\tabref{errors}.  Because only the products $B_0m_l$ and $B_0m_s$ enter our results, it is clear
that the errors in $B_0$ will have a negligible effect compared to the effect of the quark mass errors.
We bound the effects of the errors in the splittings by rerunning the analysis with all splittings at
a lattice spacing changed by
$1\,\sigma$ in the same direction, but with the direction varied randomly at different lattice spacings. 
 Because splitting errors are uncorrelated for different ensembles, and only somewhat correlated
for different splittings on the same ensemble, changes of this type provide an upper limit on the changes
we find if we change individual spacings randomly within their errors.  
 The maximum differences  we find are   0.007 in $\epsilon$ and 0.0004 in \ek,\strike{$0.5\ ({\rm MeV})^2$ in 
\KEM,} which are in each case smaller than statistical errors, and much smaller than the dominant systematic
errors.

\subsection{Quenched EM error}
\label{sec:quench-error}

For $\epsilon$, the effect of having quenched the EM interactions is controlled at NLO in SU(3) \chpt, per the 
argument of \rcite{Bijnens:2006mk}.   
This is because effects that depend on the sea-quark charges and unknown LECS are independent of
valence-quark charges and therefore
cancel in \KPZ\ and in \pisplitEM\ --- see \eq{list_NLO}.  Errors arise at NNLO, in which
cross terms between valence and sea charges can first appear in analytic terms, which have unknown LECs.
Examples of such terms are ones proportional to $q_{xy}(\mu_x-\mu_y)(q_u\mu_u+q_d\mu_d+q_s\mu_s)$ 
or $(q_x+q_y)(\mu_x+\mu_y)(q_u\mu_u+q_d\mu_d+q_s\mu_s)$.
From our central fit, the calculated effect of turning
on the sea quark charges is 0.040, or 8.2\% of the 0.486 NLO contribution for neutral sea quarks.  Assuming
the quenching effects on NNLO would be of this same size, an estimate of the electroquenching error is 8.2\%
of the   0.250 NNLO contribution, or 0.020.  It may be, however, that the electroquenching effect at NLO is
anomalously small.  In particular, there is no effect on ``pions'' (mesons with degenerate quarks) at this order.
We therefore follow a more conservative approach and take the full value
of the NLO sea-quark charge effect, namely 0.040, as the error estimate for $\epsilon$.

As explained at the end of \secref{chiptem},  the electroquenching error in \ek\ is uncontrolled, in the sense that it is not computable at lowest nontrivial
\chpt\ order.   We can get a rough handle on this error by $1/N_c$ counting. At $\cO(\alpha_{\rm EM})$,
the electroquenching effects come from diagrams with 
either (1) a photon that connects a sea-quark loop to a valence
line, (2) a photon that spans a single sea-quark loop, or (3) a photon that connects two sea-quark loops.
In all three cases, each loop must also have attached gluons,%
\footnote{This follows from Furry's theorem, which forbids loops with only  one photon vertex,
as well as the usual cancellation of vacuum bubbles that are completely unconnected to the rest
of the diagram.}
 so $1/N_c$ counting
applies.   Diagrams (1) and (2) are then suppressed by  $1/N_c$, while diagram (3) is suppressed by
$1/N_c^2$.  Diagram (1) is further suppressed by SU(3) flavor \cite{Borsanyi:2013lga}, since the sum of
sea-quark charges vanishes, and quark mass factors must be included to get a nonzero result.  Since diagram
(2) cancels for $\epsilon$, the double suppression of electroquenching effects may explain why the contribution
of sea-quark charges
is only 0.04 at NLO.   However, diagram (2) does not cancel for \ek, so we have only the $1/N_c$ suppression. 
We therefore take 1/3 of the central value, namely 0.012,\strike{$14.8\ ({\rm MeV})^2$,} as the electroquenching
error in \ek.

\subsection{Scheme dependence}
\label{sec:scheme-dependence}

It may be helpful for some purposes to estimate the changes that would be induced in our results
if we changed to a different scheme for EM renormalization.  For example, in a pure-QCD calculation
that relies on our results to remove EM effects from physical quantities that are used to set the quark masses
or scale, it would be useful to know how much the results could change in a different scheme for separating
EM from isospin-violating effects.    

In addition to the BMW scheme, we have tried renormalizing the
quark masses using the $\msbar$ scheme at scale $\mu=2$ GeV.  Unfortunately, we have only a
1-loop determination of the renormalization, and this may suffer from rather large perturbative errors,
as we remarked in \secref{renormalization}.  Nevertheless, comparison of the $\msbar$ scheme at 1 loop,
\eq{EM-mass-renorm}, with that
of the BMW scheme, \eq{ms}, gives at least  a rough estimate of how much the results may change over various choices
of ``reasonable'' schemes.    


With the $\msbar$ scheme and the central fit, we obtain $\epsilon=0.814(12)$, where the
error is statistical only.  This suggests a scheme dependence of $\sim\!0.04$ in $\epsilon$. 
The small dependence is in accordance with the general discussion at the beginning of
\secref{renormalization}. The corresponding
result for the neutral kaon is
\ek=0.365(2)\strike{$\KEM = 461(3)\ ({\rm MeV})^2$,} giving a scheme dependence of \strike{$\sim\! 420\ ({\rm MeV})^2$.}$\sim\!0.330$.  Note that, if the two-loop corrections from QCD make a comparable contribution to
the 1-loop EM renormalization as they do in the pure QCD, asqtad case \cite{Mason:2005bj}, the value of
\ek\ in the $\msbar$ scheme at 2 GeV would be reduced by a factor of order 3.

The large dependence on scheme for \ek\ is not surprising, since \ek\ is very sensitive to
the renormalization of the strange quark mass.    The fractional shift in the continuum of the strange mass under 
EM renormalization in the BMW scheme is 0.32\%, while in 1-loop $\msbar$ it is only 0.12\%.
Neither of these shifts is of an unreasonable size for an $\cO(\alpha_{\rm EM})$ effect. 
The fractional difference of 0.2\%
in the  strange mass would correspond to a change in \KEM\ of roughly  $0.002 M^2_K/(1+1/27)$, giving
a change of $\ek$ of approximately $0.37$, 
\strike{$\approx 470\ ({\rm MeV})^2$,} where we take $M_K \approx 495$ MeV. The factor of $1/(1+27)$ comes
from the fact that the light quark mass, which is not changing, is about 1/27 of the strange quark mass.  

Adding the systematic errors in \tabref{errors} in quadrature, we find
\begin{eqnarray}\eqn{eps-all-errrors}
\epsilon &=& 0.78(1)_{\rm stat}({}^{+\phantom{1}8}_{-11})_{\rm syst}\;. \eqn{epsilon-result} \\
\ek &=& 0.035(3)_{\rm stat}(20)_{\rm syst}\; .\eqn{K0-result}
\strike{ \KEM &=& 44(3)_{\rm stat}(25)_{\rm syst}\;({\rm MeV})^2 \; .}
\end{eqnarray}
The result for \ek\ implies   $\KEM = 44(3)_{\rm stat}(25)_{\rm syst}\;({\rm MeV})^2 .$
A preliminary value for $(M^2_{K^0})^\gamma$ was reported in \rcite{Basak:2013iw}.  That
result did not yet take into account EM quark-mass renormalization and is thus not reliable.

\section{Calculation  of $m_u/m_d$ \label{sec:mumd}}

Using the values of $\epsilon$ and \ek\ given in
\eqs{epsilon-result}{K0-result}, we can use the dependence of the kaon mass on the light quark
mass to find the quark mass ratio $m_u/m_d$.
Because  $\epsilon$ and \ek\ are physical parameters (albeit in a fixed scheme for
separating EM from strong isospin-violating effects), we need not use the same set of simulations
for this step.  Here we use the MILC HISQ (2+1+1)-flavor QCD ensembles, since these have
smaller lattice artifacts than the asqtad ensembles and contain ensembles with light quark
masses near their physical values.
Table \ref{tab:hisqensembles} shows the 2+1+1 flavor ensembles with approximately physical light
sea quark masses, which are used in this section.

\begin{table} \renewcommand{\arraystretch}{0.85}
\caption{
\label{tab:hisqensembles}
Ensembles used in the calculation of $m_u/m_d$.
The first column in this table is the approximate lattice spacing in fm.
The second column is the gauge coupling $\beta=10/g^2$, and the
next three columns are the sea-quark masses in lattice units.  The primes on the masses indicate that they
are the values used in the runs, and in general differ slightly from the physical values because of
tuning errors.
}
\begin{tabular}{|l|l|lll|l|l|llll|}
\hline
key & $\beta$ & $am'_l$ & $am'_s$ & $am'_c$ & $(L/a)^3\times (T/a)$      & $N_{lats}$ & $a$ (fm)& $L$ (fm) & $M_\pi L$ & $M_\pi$ (MeV) \\ \hline
0.15 & 5.80  & 0.00235 & 0.0647  & 0.831 & $32^3\times 48$  & 1000  & 0.15079(17) & 4.83 & 3.2 & 130 \\
0.12 & 6.00  & 0.00184 & 0.0507  & 0.628 & $48^3\times 64$  & 999   & 0.12111(10) & 5.82 & 3.9 & 133 \\
0.09 & 6.30  & 0.0012  & 0.0363  & 0.432 & $64^3\times 96$  & 1031  & 0.08772(12) & 5.62 & 3.7 & 130 \\
0.06 & 6.72  & 0.0008  & 0.022   & 0.260 & $96^3\times 192$ & 895 & 0.05673(5) & 5.44 & 3.7 & 135 \\
0.04 & 7.00  & 0.000569  & 0.01555   & 0.1827 & $144^3\times 288$ & 470  & 0.04254(5) & 6.12 & 4.17 & 134 \\
\hline
\end{tabular} 
\end{table}

The procedure for finding $m_u/m_d$ is described in detail in \rcite{Bazavov:2014wgs}.
Very briefly, the essential steps are:
\begin{enumerate}
\item{ Use the pion mass and decay constant to fix the lattice spacing and average light quark
mass, $m_l=(m_u+m_d)/2$, on each ensemble.  Here we use the physical $\pi^0$ mass, since this has
small electromagnetic contributions. This mass is also adjusted for QCD finite size effects.}
\item{ Find the tuned strange quark mass on each ensemble by matching $2M_K^2-M_\pi^2$.  In this step
the lattice masses use the average light quark mass computed in the first step, and the
input $M_K$ is the average of the $K^0$ and $K^+$ masses after subtracting the electromagnetic
contributions parameterized by $\epsilon$ and \ek.}
\item{Use the derivative of the lattice $M_K^2$ with respect to the  light  valence quark mass and the
difference between the $K^0$ and $K^+$ masses after removing electromagnetic contributions to find
$m_d-m_u$ and hence $m_u/m_d$ on each ensemble. }
\item{Fit the values of $m_u/m_d$ on each ensemble to a smooth function of the lattice spacing,
and evaluate the fit at $a=0$.  For our central fit we fit the points with $a \le 0.12$ fm
to a quadratic in $\alpha_s a^2$, using the strong coupling constant $\alpha_s$ determined from
the plaquette. Alternative fits to estimate systematic errors from the continuum extrapolation include
a quadratic fit including the 0.15 fm data, a linear fit excluding the 0.15 fm ensemble,
 and a linear fit excluding both the 0.12 and 0.15 fm ensembles.
}
\end{enumerate}

\begin{figure}
\vspace{-1.35in}
\begin{center}\includegraphics[width=.70\textwidth]{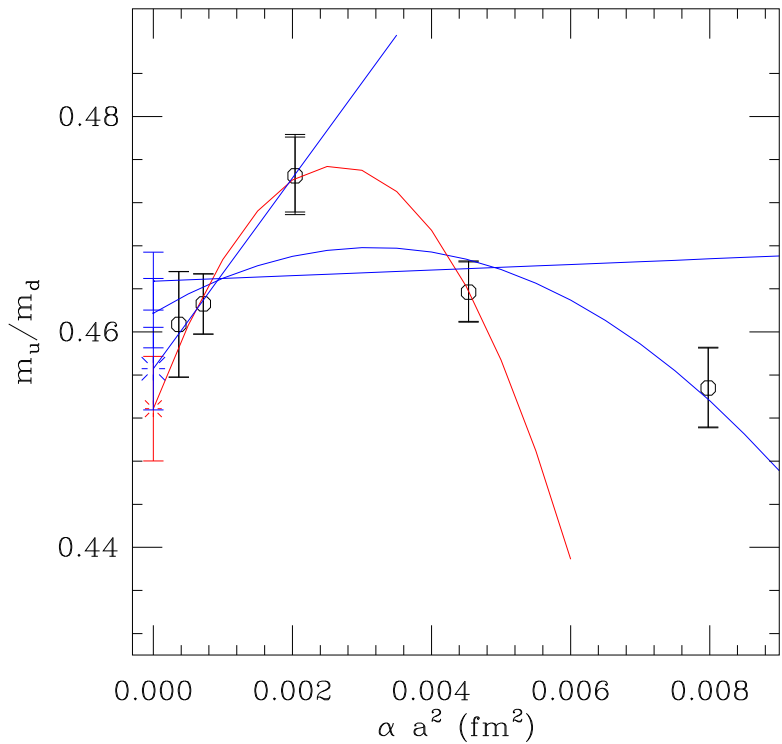}\end{center}
\vspace{-1.35in}
\caption{
\label{fig:udratio_june2018}
$m_u/m_d$ on the physical quark mass HISQ ensembles, and the continuum extrapolation.
The red line is the fit used for our central value, and the blue lines three of the alternative
fits used for estimating systematic error from the continuum extrapolation.
These alternate fits are a quadratic fit including all the data points, a linear fit omitting the 0.15 fm. data, and a
linear fit omitting both the 0.15 and 0.12 fm data.
}
\end{figure}

The most important differences between this analysis and that of \rcite{Bazavov:2014wgs} are
the extension of the 0.06 fm ensemble to 895 lattices and the addition of an ensemble with $a \approx 0.04$ fm.
Figure~\ref{fig:udratio_june2018} shows the values of $m_u/m_d$ for each ensemble, and the continuum
extrapolation.
With the addition of this data at small lattice spacings, we now choose to omit the 0.15 fm data
from our central fit, and use the fit including this ensemble as one of our alternative fits
for estimating the systematic error due to the choice of continuum extrapolation.
We take the range of all of these continuum extrapolations as our estimate of the systematic
error coming from the choices made in our continuum extrapolation.

Using the MILC HISQ (2+1+1)-flavor QCD ensembles and the values of $\epsilon$ and \ek\ given in 
\eqs{epsilon-result}{K0-result}, and following the approach described in \rcite{Bazavov:2014wgs}, we obtain
\begin{equation}
\label{eq:udratio}
m_u/m_d = 0.4529(48)_\mathrm{stat}(\null_{-0}^{+118})_\mathrm{cont.}( \null_{-66}^{+91})_\mathrm{\epsilon}
(0)_\mathrm{\ek}(4)_\mathrm{FV_{QCD}}(13)_\mathrm{\Delta M_K(exp.)}\ \ \ .
\end{equation}
The errors on the quantity are, in order, the statistical error and the errors from choices in the continuum extrapolation, from $\epsilon$, from \ek, from finite volume in the pure QCD calculation,
and from the error in the experimental value of $M_{K^0}-M_{K^+}$ \cite{Olive:2016xmw}. 
The finite-volume effects are taken to be the difference between a NLO staggered chiral perturbation theory calculation
and a nonstaggered calculation at NNLO for $M_\pi$ and $F_\pi$ and NLO for $M_K$ and $F_K$.
We note that the result in \eq{udratio} should be considered an update to the result quoted in
\rcite{Bazavov:2017lyh}, $m_u/m_d =   0.4556 (55)_\text{stat}(\null_{-67}^{+114})_\text{syst}(13)_{\Delta M_K}$.  The current result includes newly generated 2+1+1 HISQ configurations at $a\approx0.06$ fm and $0.04$ fm, as well as all our configurations at $a\approx0.09$ fm. \Rcite{Bazavov:2017lyh}, which
focused on physics for quarks heavier than $m_c$, included only the subset of configurations at 
$a\approx0.09$ fm for which we have generated propagators for those heavy quarks.
The smaller statistical error of the current result reflects the larger data set used.  Our
procedures for estimating systematic errors, however, actually give slightly larger values in the current
analsyis than in \rcite{Bazavov:2017lyh}.

To this level of precision, and within the scheme we are using, our EM errors in  $m_u/m_d$ come only from $\epsilon$
and not from \ek, despite the large relative error in the latter quantity.
The errors in \ek\ do, however, have an effect on the errors in $m_s$ and
 in ratios such as $m_s/m_l$ \cite{Bazavov:2017lyh,Bazavov:2018omf}.

\section{Conclusions and Outlook}
\label{sec:conclusions}

\begin{figure}
	\centering
    \includegraphics[height=0.5\textwidth]{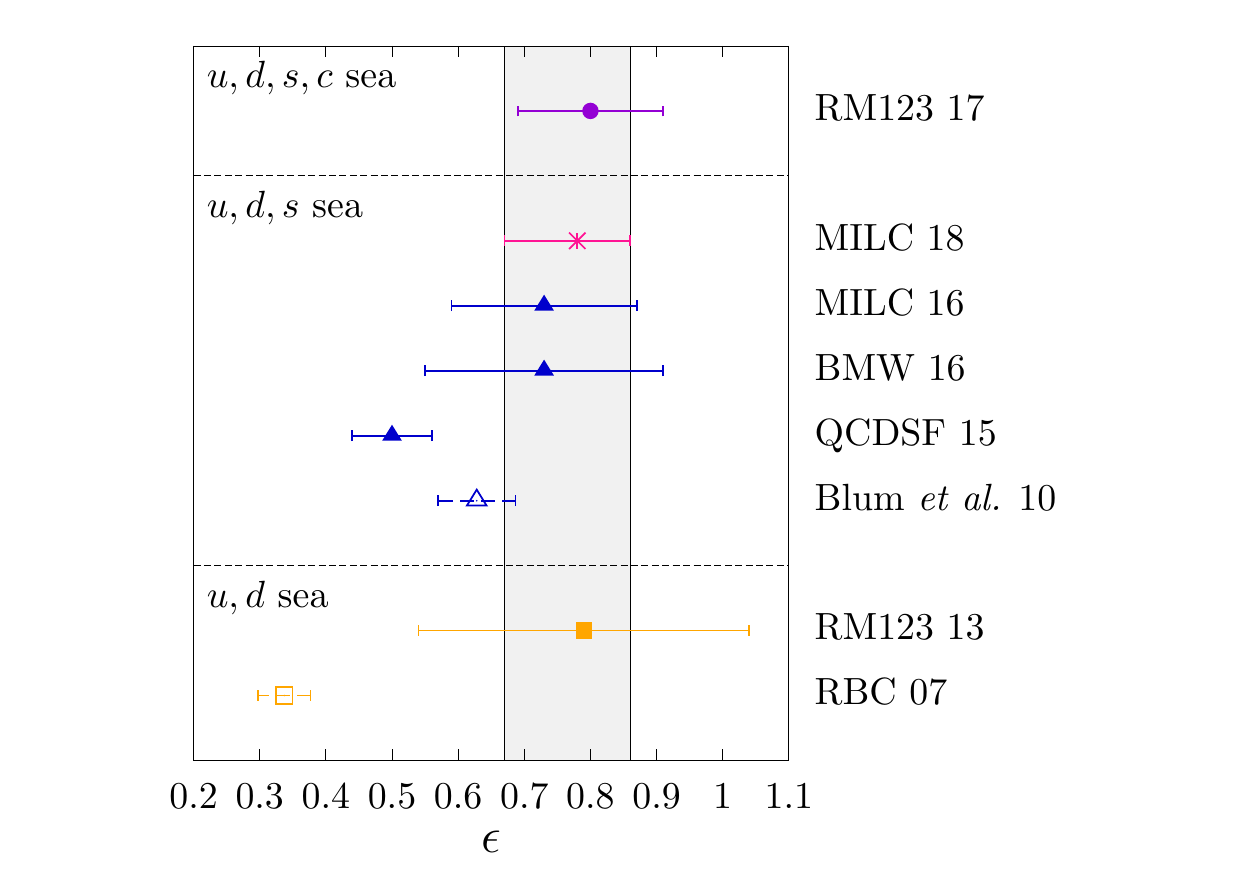}
	\caption{	Comparison of $\epsilon$ in Eq.~(\ref{eq:epsilon-result}) (magenta burst) with previous unquenched lattice-QCD
        calculations.  The open symbols with dashed error bars represent early work, with only statistical errors
        quoted.  The references are RM123 17 \cite{Giusti:2017dmp}, MILC 16 \cite{Basak:2016jnn} (a preliminary result), BMW 16 \cite{Fodor:2016bgu}, QCDSF 15 \cite{Horsley:2015eaa}, Blum {\it et al.}\ 10 \cite{Blum:2010ym},
        RM123 13 \cite{deDivitiis:2013xla}, and RBC~07~\cite{Blum:2007cy}.
      	\label{fig:epsilon_summary}}
\end{figure}

Using the three-flavor MILC asqtad configurations, we have computed the EM quantities $\epsilon$ and $\ek$, 
which parameterize the EM contribution to the $K^+$--$K^0$ mass splitting, and to the $K^0$ mass itself, respectively.  Our results are given in \eqs{epsilon-result}{K0-result}.  A comparison of our result for $\epsilon$
with those of other groups (and our preliminary result, labeled as MILC 16) is shown in \figref{epsilon_summary}.  We note that different groups in general use different schemes for separating
electromagnetic and strong isospin-violating effects.  Nevertheless, the scheme-dependence of
$\epsilon$ is likely to be small --- see the discussion in \secrefs{renormalization}{scheme-dependence}. With the exceptions of the early result
in RBC 07, which quotes statistical errors only, and the result from QCDSF 15, from which we
differ by about 2 sigma,  the agreement with the work  of other groups is good.

\begin{figure}
	\centering
    \includegraphics[height=0.5\textwidth]{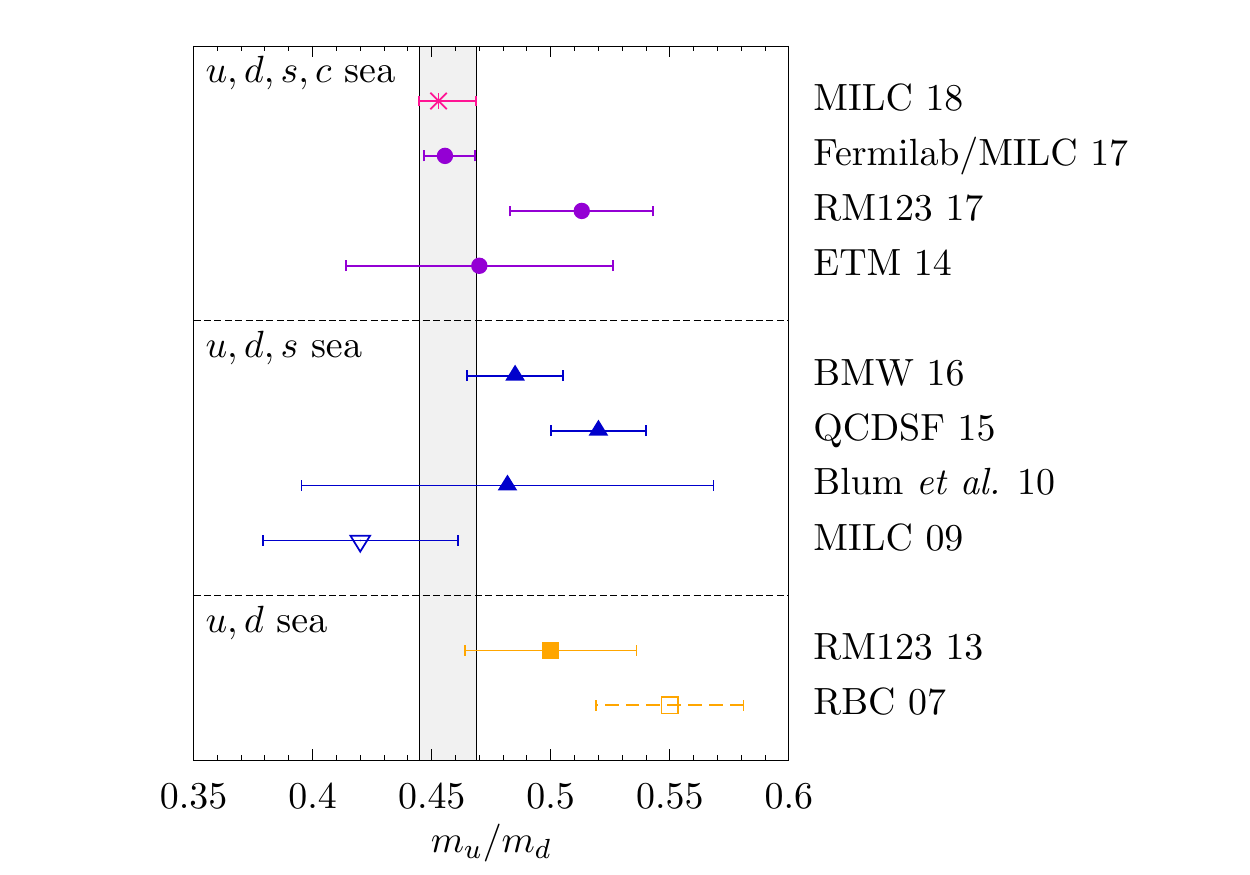}
	\caption{Comparison of $m_u/m_d$ in Eq.~(\ref{eq:udratio}) (magenta burst) with previous unquenched lattice-QCD
        calculations that include a lattice evaluation of the EM effects.  For comparison,
        we also show, with an open triangle, the MILC 09 result, which uses a phenomenological
        estimate of the EM effects.  An early result, RBC 07, just quotes statistical errors and is shown
        with dashed error bars and an open symbol.  The current (MILC 18) result 
        should be considered an update
        of the Fermilab/MILC 17 result (see discussion  following \eq{udratio}).        
         The references are Fermilab/MILC 17 \cite{Bazavov:2017lyh}, RM123 17 \cite{Giusti:2017dmp}, ETM 14 \cite{Carrasco:2014cwa}, BMW 16 \cite{Fodor:2016bgu}, QCDSF 15 \cite{Horsley:2015eaa}, Blum {\it et al.}\ 10 \cite{Blum:2010ym}, MILC 09 \cite{RMP}, 
        RM123 13 \cite{deDivitiis:2013xla}, and RBC 07 \cite{Blum:2007cy}.
	\label{fig:mu_md_summary}}
\end{figure}

With the EM contributions in hand, we have proceeded to compute the quark mass ratio $m_u/m_d$ in QCD, using the four-flavor MILC HISQ configurations. \Figref{mu_md_summary} compares
our work with that of other lattice groups.  In general, we only show results that employ a lattice evaluation of the EM effects; however we have included  for comparison the MILC 09 \cite{RMP}
result (shown with an open symbol), which relies on a phenomenological estimate 
of $\epsilon$.  With our new results of the EM effects, our estimate for the EM uncertainty in $m_u/m_d$ has been reduced by 
more than a factor of 5 from our error in MILC 09.  Other systematic errors are comparable between
MILC 09 and MILC 18, so the total error is reduced by a factor of about 3.5.

Note that our current value for $m_u/m_d$ is plotted in \figref{mu_md_summary} with the $u$, $d$, $s$, $c$ sea results.  The pure QCD
HISQ ensembles that are used in finding $m_u/m_d$  indeed have 2+1+1 dynamical flavors.
On the other hand, our EM calculation giving $\epsilon$ and \ek\ employs the asqtad 2+1 ensembles.
The error from omitting the dynamical charm quark, however, is expected to be at most a few percent.
An error of that size would be small compared to the other errors in the EM calculation, so should not
effect the final value for $m_u/m_d$ significantly. 

Our result for $m_u/m_d$ is consistent with those from most other groups, but lies on the low side
of the range of results. From
\figref{udratio_june2018} one can see that the low continuum value from our data set is due to the
results from the two finest lattice spacings, $a\approx 0.06$ fm and $a  \approx 0.04$ fm.  The latter
is finer than the finest of the ensembles used by the other groups, which has $a\approx 0.054$ fm. 
Because discretization errors depend on the lattice action,   however,  it is unclear at this
point whether the difference in available lattice spacings is relevant to the apparent differences
seen in \figref{mu_md_summary}.

While the electroquenching errors for $\epsilon$ are under control, these errors are uncontrolled for most quantities, for example, \ek.  To move beyond the electroquenched
approximation, we have developed a dynamical EM code \cite{Zhou:2014gga} and are
beginning to generate unquenched QCD+QED ensembles.  These	ensembles will be crucial
to our efforts to obtain precise results for the hadronic contributions to $(g-2)_\mu$, as well as for
calculations such as the proton-neutron mass difference and improvements in the result for
\ek.

\begin{acknowledgments} 
We thank Laurent Lellouch, Antonin Portelli, and Francesco Sanfilippo for useful discussions. 
The spectrum running was done on computers at the National Center for
Supercomputing Applications, Indiana University, the Texas Advanced
Computing Center (TACC), and the National Institute for Computational
Science (NICS). Configurations were generated with resources provided by the 
USQCD
Collaboration, the Argonne Leadership Computing Facility, and the National Energy Research Scientific 
Computing Center, which are funded by the Office of Science of the U.S.
Department of Energy; and with resources provided by the National Center for Atmospheric
Research,  NICS, the Pittsburgh Supercomputer
Center, the San Diego Supercomputer Center, and TACC,
which are funded through the National Science Foundation's XSEDE Program,
and Indiana University. This work
was supported in part by the U.S. Department of Energy
under Grants DE-FG02-91ER-40628, 
DE-FG02-91ER-40661, DE-FG02-04ER-41298, DE-FC02-06ER41446, DE-SC0010120, and DE-FC02-06ER41443; and by
and the National Science Foundation 
under Grants PHY07-57333, PHY07-03296, PHY07-57035,
PHY07-04171, PHY09-03571, PHY09-70137, PHY10-67881, PHY14-14614,
PHY17-19626,  PHY05-55234, PHY05-55235, and PHY12-12389.
This research was supported in part by Lilly Endowment, Inc., through
its support for the Indiana University Pervasive Technology Institute,
and in part by the Indiana METACyt Initiative. The Indiana METACyt
Initiative at IU was also supported in part by Lilly Endowment, Inc.
Fermilab is operated by Fermi
Research Alliance, LLC, under Contract No. DE-AC02-07CH11359 with the U.S. Department of Energy. 
For this work we employ QUDA \cite{quda}.   
\end{acknowledgments}

\appendix*
\section{Obtaining the mass from the self-energy at finite $T$}
\label{AppendixA}

For infinite $T$, the standard procedure to get the correction
to the squared mass is to evaluate the Euclidean self energy $\sigma(p_0)$ at $p_0=im$. In this appendix, we show that that method does not in general
give the right answer at finite $T$.  In particular, $\sigma(im)$ is dependent on the
routing of the loop momentum through the diagram.  Nevertheless, we show that the particular
momentum routing chosen in \secref{FVChPT} does  allow us to extract the mass correction from
$\sigma(im)$ because the natural continuation of  $\sigma(p_0)$ away from the Matsubara frequencies $2\pi n/T$ happens to be particularly simple.
 
To introduce our notation and approach, we first review the usual procedure when the time extent $T$
is infinite.  The momentum-space Euclidean propagator has the form
\begin{equation}\eqn{p-prop}
\tilde G_\infty(p_0) = \frac{1}{p_0^2+m^2+\sigma_{\infty}(p_0)} ,
\end{equation}
where $m$ is the Lagrangian mass, $\sigma_{\infty}$ is the self energy, the subscript $\infty$ indicates that $T$ is infinite, and we have
taken the case of vanishing spatial momentum, $p=(p_0,\vec 0)$, for simplicity. 

To find the physical mass, we Fourier transform to position space
\bea
G_\infty(t) &=& \int\frac{dp_0}{2\pi}\; e^{ip_0t} \tilde G_\infty(p_0)\eqn{FT-infT}\\
&=& Ce^{-Mt} +\cdots \qquad [t>0],
\eqn{result-infT}
\eea
where $C$ is a constant, and $p_0^2=-M^2$ is the location of the single-particle pole 
\be\eqn{mass-infT}
M^2 = m^2 +\sigma_\infty(iM) \approx m^2 + \sigma_\infty(im),
\ee
and $\cdots$ in \eq{result-infT} represents the contributions of excited and multiparticle states.
  \Equation{result-infT} follows from \eq{FT-infT} by completing the contour in the upper half plane using Jordan's lemma,
which requires only that 
\be\eqn{Jordan_infT}
\lim\limits_{|p_0|\to\infty} \frac{1}{p_0^2+m^2+ \sigma_{\infty}(p_0)} = 0
\ee
in the upper half plane.  
From \eq{mass-infT} we read off the standard answer: the first order
 correction to the
 squared mass is simply the self-energy evaluated at $p_0=im$.

 When $T$ is finite, the calculation of the mass correction changes in two crucial ways.  First of all, the integral
 over $p_0$ in \eq{FT-infT} becomes a sum over $p_0= 2\pi \ell/T$, where $\ell$ runs over the integers.   The
  self energy $\sigma(p_0)$ and hence $\tilde G(p_0)$ are moreover only
 well-defined for these discrete values of $p_0$.  We may continue   these functions to 
  $\sigma_{{\rm cont}}(p_0)$ and $\tilde G_{\rm cont}(p_0)$, defined on the full complex $p_0$ plane, but the 
 continued functions are not unique.
  Second, the internal loop energy (for example, $k_0$ in \eq{sunset}) in the determination of 
 $\sigma(p_0)$ is itself discrete, so that $\sigma(p_0)$ is not the same function of $p_0$ as $\sigma_{\infty}(p_0)$, even on the discrete points $p_0= 2\pi \ell/T$.  These two changes interact in interesting ways, with
 the result that the procedure to obtain the squared-mass correction by evaluating $\sigma_{{\rm cont}}(p_0)$ at $p_0=im$
 is not valid in general.  

We discuss the discrete sum over $p_0$ first.  To extract the mass, we need to compute
\be
G(t) = \frac{1}{T} \sum\limits_{p_0=2\pi\ell/T} e^{ip_0t}\; \tilde G(p_0)\eqn{FT-finiteT}.
\ee
$G(t)$ is a periodic function of $t$ with period $T$.
The standard technique is to use the Poisson summation formula to rewrite $G(t)$ as a sum of nonperiodic
 propagators to each periodic image of the fundamental domain $0\le t\le T$. Usually these nonperiodic propagators are just the known $T\!=\!\infty$ propagators, but here that is not the case, since we are keeping
 $T$ finite for the internal energy sums in $\sigma$.  We instead simply use a continuation 
  $\tilde G_{\rm cont}(p_0)$ of $\tilde G(p_0)$, which defines $G_{\rm cont}(t)$ by Fourier transformation.  
    The Poisson formula then gives
  \bea
G(t) &=& \sum\limits_{n=-\infty}^{+\infty}G_{\rm cont}(t+nT),\eqn{Poisson-p0}\\
G_{\rm cont}(\tau) &=&  \int\frac{dp_0}{2\pi}\; e^{ip_0t} \;\tilde G_{\rm cont}(p_0)
.\eqn{Gcont}
\eea

Although $\tilde G_{\rm cont}(p_0)$ is not unique, it is straightforward to check that another
continuation constructed by adding a function that vanishes at $p_0=2\pi\ell/T$,
such as $\sin(p_0T/2)$,
will not change $G(t)$, although it does of course change $G_{\rm cont}(\tau)$.  This still leaves open the
question of how $\tilde G_{\rm cont}(p_0)$ should be chosen.  For now, we simply state that we should
choose $\tilde G_{\rm cont}(p_0)$ so that $G_{\rm cont}(\tau)$ is strongly damped for large $\tau$. 
By a standard theorem of Fourier transformations, 
we can accomplish this if  $\tilde G_{\rm cont}(p_0)$ and all its derivatives are continuous and
absolutely integrable over the real $p_0$ line \cite{Gasquet:1999}.

If $G_{\rm cont}(t)$  is exponentially damped for $mt\gg 1$, we can, in practical situations, neglect most or
all of the $n\not=0$ terms in \eq{Poisson-p0}.  A standard approach is  just to include $n=-1$ in addition to $n=0$,
so that we include a backward propagating meson in our fit Ansatz for $G(t)$:
\be
G(t) \sim C \left(e^{-Mt}+e^{-M(T-t)}\right)\eqn{G-fit}.
\ee
The fit for $mt\gg 1$ will then effectively isolate the first contribution, from $G_{\rm cont}(t)$, and extract the corrected 
mass $M$ from its exponential decay.    If Jordan's lemma applies to the Fourier transform and
if the only single-particle pole in $\tilde G(p_0)$ is the usual one near $p_0=im$, then the correction to the squared mass is indeed just 
$\sigma_{{\rm cont}}(im)$.   We will see below, however,
that this will not be true in general.

So we are led to consideration of the finite-$T$ self energy in momentum space $\sigma(p_0)$, and 
how it may be continued away from the special values $p_0=2\pi\ell/T$ to $\sigma_{{\rm cont}}(p_0)$.
A natural choice for $\sigma_{{\rm cont}}(p_0)$  is simply the result
 of doing the loop energy/momenta sums for arbitrary external $p_0$, instead of only
for the special values.  For  example, we can perform  the sum in the first term
on the right-hand side of \eq{sunset} for any $p_0$.  Because the resulting self-energy function and its
derivatives obey the 
continuity and integrability conditions mentioned above,  $G_{\rm cont}(t)$ will automatically be exponentially damped%
\footnote{More  precisely, it will decrease faster than any power of $1/t$ for large $t$
\cite{Gasquet:1999}.}    as desired.

An undesirable, but unavoidable,
feature of this continuation $\sigma_{{\rm cont}}(p_0)$ is that it depends on the routing of the external momentum $p_0$
through the diagram.  The dependence on routing vanishes when $p_0=2\pi\ell/T$ because the loop energy
may be shifted by this amount.  But away from these special points, there is no reason for  $\sigma_{{\rm cont}}
(p_0)$ to be independent of the routing; we have checked this dependence numerically for 
\be\eqn{self-energy-example}
\sigma_{{\rm cont}}(p_0) = \frac{1}{L^3T} \sum\limits'_{k_0,\vec k}\; \cI_{\hat {\rm s}} - \int \frac{d^4 k}{(2\pi)^4}\; \cI_{\hat {\rm s}},
\ee
with $\cI_{\hat {\rm s}}$ the photon-sunset integrand given by \eq{sunset-integrand}.  Here we have considered
the difference between the sum and the integral, rather than the sum itself, to avoid having to cut off the
sum over $\vec k$, which is irrelevant to the current discussion.%
\footnote{From now on we use the term ``mass correction'' to mean the finite-$L$ and finite-$T$ contribution to the 
mass correction.  The additional correction when $T$ and $L$ are infinite will not affect any of the following 
discussion, as long as that correction is small enough that it does not violate the perturbative expansion.}
Further, the dependence on momentum routing
persists when $\sigma_{{\rm cont}}$ is evaluated at $p_0=im$, which indicates that the rule relating the
mass correction to $\sigma_{{\rm cont}}(im)$ cannot be true in general.  We emphasize that this is a problem 
with the rule, rather than some fundamental problem with the definition of the mass correction itself:  The finite-$T$ 
propagator $G(t)$ is of course completely independent of the routing.

To examine this issue further, we consider the two obvious possible momentum routings in the sunset diagram.
Routing A, which we used in \secref{FVChPT}, has $p-k$ on the photon line and $k$ on the internal meson line. 
  Routing B has $k$ on the photon line and $p-k$ on the internal meson line.
With $p=(p_0,\vec 0)$ and $\vec k \not= 0$,
\bea\eqn{sunset-integrand-A}
\cI^A_{\hat {\rm s}} &=& \frac{\vec k^2+ m^2 - p_0^2}{\vec k^2(k_0^2+\vec k^2 +m^2)}, \\
\cI^B_{\hat {\rm s}} &=& \frac{\vec k^2+ m^2 +2p_0k_0- 3p_0^2}{\vec k^2((k_0-p_0)^2+\vec k^2 +m^2)}, 
\eqn{sunset-integrand-B}
\eea
where \eq{sunset-integrand-A} is copied from \eq{sunset-integrand}.  In both cases
we have added on the 00 component of the tadpole, which is independent of the external momentum.
The linear term in $k_0$ in the numerator of  $\cI^B_{\hat {\rm s}}$ cannot be dropped since the denominator
is not symmetric under $k_0\to-k_0$.   When $p_0=2\pi\ell/T$, $\cI^A_{\hat {\rm s}}$ and $\cI^B_{\hat {\rm s}}$
clearly give the same result for $\sigma(p_0)$, as can be seen by shifting the summation variable
$k_0\to k_0+p_0$ in $\cI^B_{\hat {\rm s}}$ (and then dropping a linear term in $k_0$ in the numerator).

Extracting the mass correction is easy for routing A.
We can see from \eq{sunset-integrand-A} that $\sigma^A_{{\rm cont}}(p_0)$ has the simple form
$\alpha+\beta p_0^2$, where $\alpha$ and $\beta$  are independent of $p_0$.  Therefore, $\tilde G^A_{\rm cont}(p_0)$ has a simple pole close to $p_0=im$, and Jordan's lemma
allows us to close the contour  as usual in the upper half plane (for $t>0$) for the Fourier transform of  
$\tilde G^A_{\rm cont}(p_0)$.   This determines the squared-mass correction 
to be $\sigma^A_{{\rm cont}}(im)$,
as was assumed in \secref{FVChPT}.  
 
Extracting the mass correction in the case of routing B is more subtle.
To see the relation between the self energy from $\cI^B_{\hat {\rm s}}$ and $\cI^A_{\hat {\rm s}}$ when $p_0$ is
not at a special point, we  use the Poisson summation formula to write
\be\eqn{Poisson-B}
\sigma^B_{{\rm cont}}(p_0)=\frac{1}{L^3}\sum\limits'_{\vec k}\sum\limits_n \int \frac{dk_0}{2\pi}\; e^{inTk_0}\; \cI^B_{\hat {\rm s}} -  \int \frac{d^4 k}{(2\pi)^4}\; \cI^B_{\hat {\rm s}},
\ee
where $n$ runs over the integers. We can now make the shift $k_0\to k_0+p_0$ in both integrals, converting
$\cI^B_{\hat {\rm s}}$ into $\cI^A_{\hat {\rm s}}$. Differences remain, however, 
from the resulting phase $e^{inTp_0}$
and from the term in the numerator linear in $k_0$, which gives a nonvanishing contribution when $n\not=0$.
The difference between the self-energies is then 
\be\eqn{BA-diff}
\Delta\sigma(p_0)=
 -\frac{2}{L^3}\sum\limits_{n\ge1}\sum\limits'_{\vec k}\frac{e^{-\omega_{k}nT}}{\vec k^2}\Bigg[\sin^2(nTp_0/2)
\frac{\vec k^2+m^2-p_0^2}{\omega_{k}}+p_0\sin(nTp_0)\Bigg],
\ee
where $\Delta\sigma \equiv\sigma^B_{{\rm cont}}-\sigma^A_{{\rm cont}}$, and $\omega_{k}\equiv\sqrt{\vec k^2 +m^2}$. 

\begin{figure}[t!]
  \centering
  {\includegraphics[width=0.32\linewidth]
  {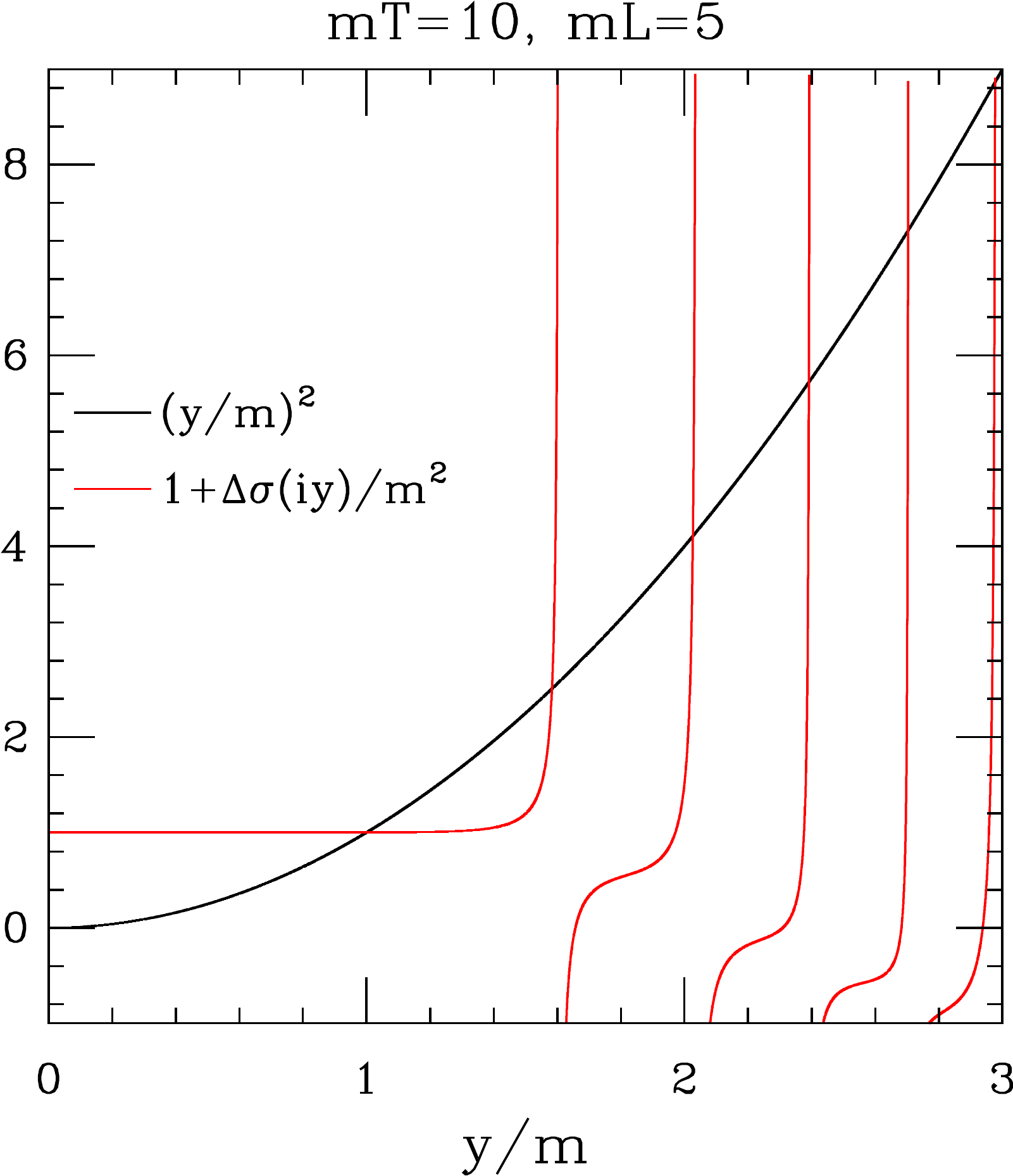}}
  \
  {\includegraphics[width=0.32\linewidth]
  {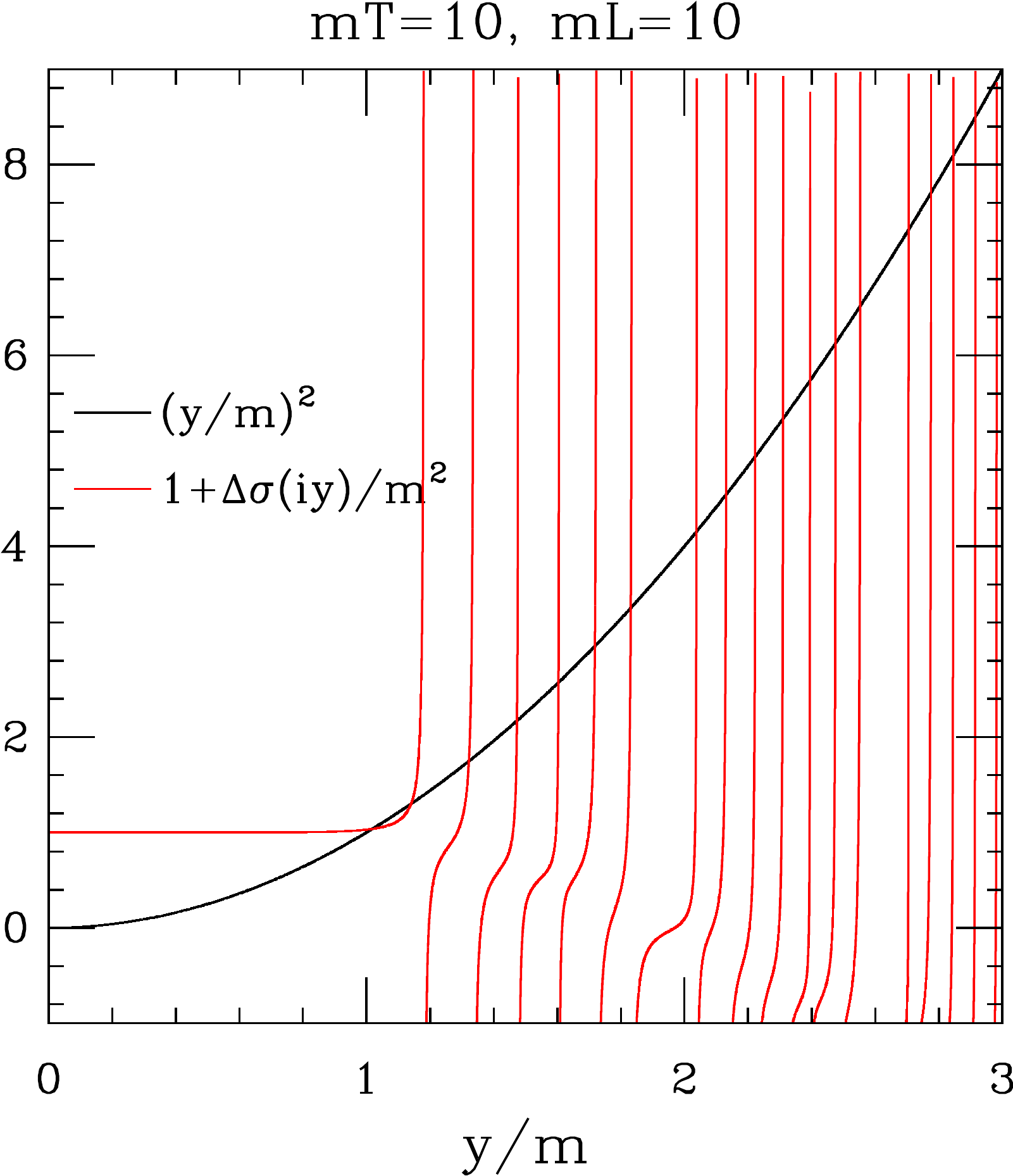}}
  \
  {\includegraphics[width=0.32\linewidth]
  {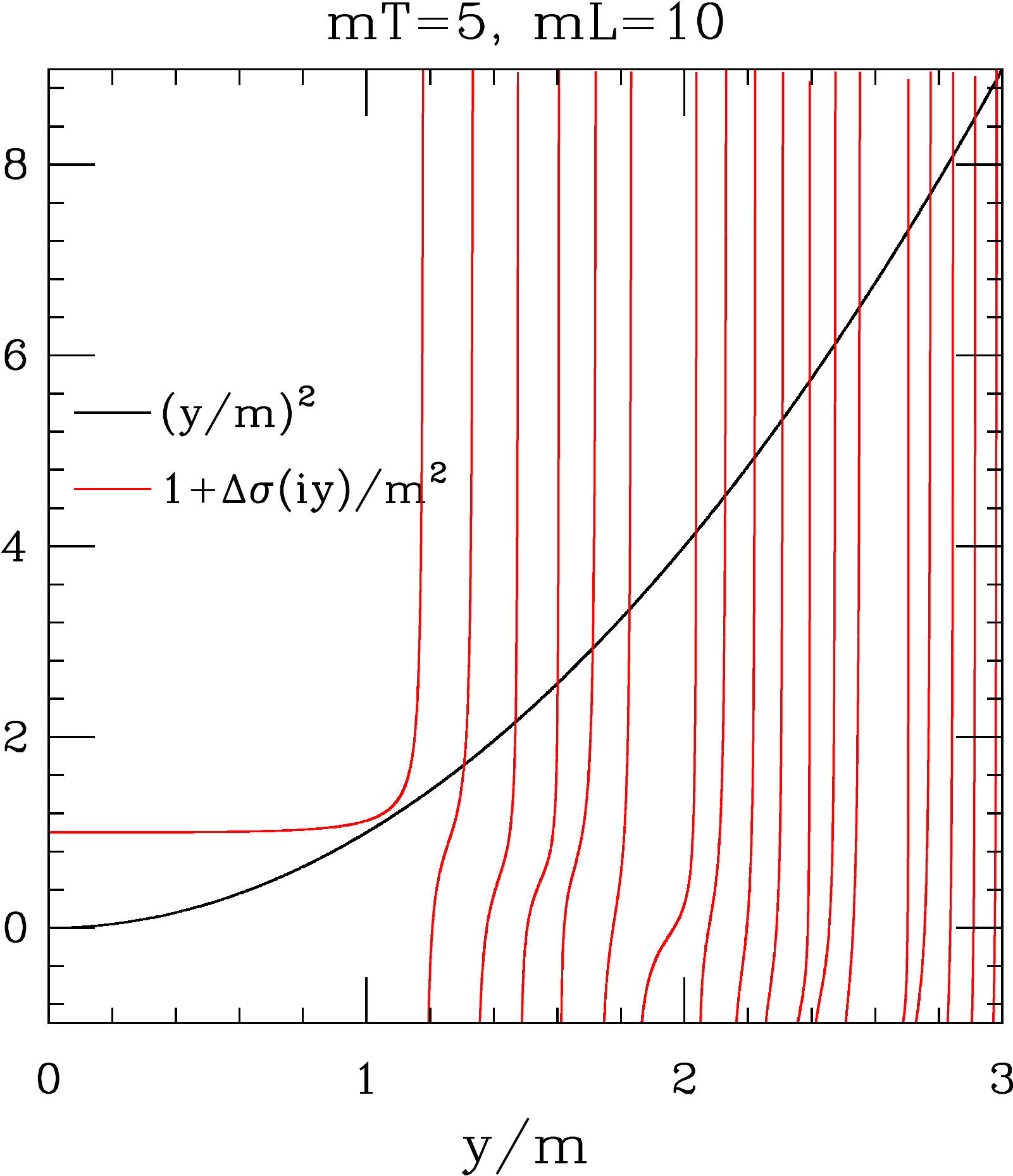}}
  \\
\vspace{-2mm}
\caption{\label{fig:B-poles}
Location of the poles of the momentum space propagator $\tilde G_{\rm cont}(p_0)$ for routing B, for three different values of
$mT$ and $mL$.  The quantities $(y/m)^2$ (black lines) and $1+\Delta\sigma(iy)/m^2$ (red lines) are shown
as a function of $y/m$, where $y$ is the imaginary part of the Euclidean energy $p_0$.  The $y/m$ values
of the poles are given by the locations of the crossings of the two curves.}
\vspace{-0.08in}
\end{figure}

 Because of the additional factors of  $\sin^2(nTp_0/2)$ and 
$\sin(nTp_0)$,   which blow up for large imaginary $p_0$, the analytic 
properties of  $\sigma^B_{{\rm cont}}$ are not standard, and we must reexamine the usual assumptions
that go into finding the mass.  To simplify the  algebra we take $mT\gg 1$ and consider
$p_0=iy$ with $y \gtwid m$.  We can therefore neglect the exponentially falling terms $\exp(-ynT)$ 
from the sine  functions and keep only the growing ones.  The sum over $n$  then  immediately gives
\be\eqn{diff-summed}
\Delta\sigma(iy)
= \frac{1}{2L^3}\sum\limits'_{\vec k}
\frac{(\omega_k + y)^2}{\vec k^2\; \omega_{k}\;(e^{T(\omega_k-y)}-1)}\ .
\ee
The values of $\omega_k$ for each discrete value of $\vec k$ therefore determine singularities  
in  $\Delta\sigma$.
As $y$ approaches a discrete value of $\omega_{k}$ from below, $\Delta\sigma$ goes to $+\infty$, and then
comes up from $-\infty$ as $y$ increases above $\omega_{k}$.  Because the self-energy 
varies over the full range
$(-\infty,\infty)$, $\tilde G^B_{\rm cont}$ will have a  
pole near each of the singularities in $\Delta\sigma$. 
\Figref{B-poles} shows how this occurs for three choices of $mT$ and $mL$.  The equation for the poles is $y^2 = m^2 + \sigma^B_{{\rm cont}}(iy)$, which we find from the crossings of the curves
$(y/m)^2$ and $1+\Delta\sigma(iy)/m^2$, where we have neglected the difference between $\Delta\sigma$
and $\sigma^B_{{\rm cont}}$.  This difference is  $\sigma^A_{{\rm cont}}$, which just gives a relatively
small correction to the terms $(y/m)^2$ and $1$, and does not change the qualitative picture.  In the plots, we have included all the $\vec k$ values in the sum
in \eq{diff-summed} that contribute significantly in the region of $y/m$ shown.

The left-hand plot ($mT=10$,
$mL=5$) shows that, in addition to the ``normal'' pole close to $y=m$, there are anomalous poles close
the singular values in $\Delta\sigma$ where $y=\omega_k$, for some $\vec k$.  As $L$ increases, the possible
values of $k$ get closer, and the poles get denser.  We observe this feature in the middle plot ($mT=10$, 
$mL=10$).  As $L\to\infty$, the poles pile up at $y=m$.  
On the other hand, as $mT$ gets smaller, the residues of the singularities in $\Delta\sigma$ increase like $1/(mT)$.  This can lead to the
particularly strange situation where the normal pole in $\tilde G_{\rm cont}$ close to $y=m$ disappears, as shown in the right-hand
plot ($mT=5$, $mL=10$). 

Because there are many poles in the propagator, and often many of them are close  to $p_0=im$,
the  quantity  
$\sigma^B_{{\rm cont}}(im)$  has no direct relation to the mass correction.
Nevertheless, the finite-T propagator $G(t)$ is always well-defined, and in principle one could always
extract the mass from $G(t)$ numerically.  The relation between the self-energy
at $p_0=im$ and the mass correction is however problematic, and it seems unlikely in most cases that the
dependence of the self-energy on $p_0$ will be simple enough to
relate the mass-correction to the self energy at $p_0=im$, as we did for  $\sigma^A_{{\rm cont}}(im)$.

It is worth making contact here with the argument given in \rcite{Borsanyi:2014jba} about the effect of finite $T$.
They write the difference between finite and infinite $T$ for arbitrary momentum routing as in \eq{Poisson-B}
\be\eqn{Poisson-BMW}
\delta\sigma_{{\rm cont}}(p_0)=\frac{1}{L^3}\sum\limits'_{\vec k}\sum\limits_n' \int \frac{dk_0}{2\pi}\; e^{inTk_0}\; \cI(k_0,\vec k,p_0) ,
\ee
where the prime on the sum on $n$ indicates that $n=0$ should be omitted; it is cancelled by the infinite-$T$
subtraction.  They then argue that  $\cI(k_0,\vec k,im)$ has no poles on the real $k_0$ axis and
is infinitely differentiable, with all of its derivatives integrable, which implies that  
$\delta\sigma_{{\rm cont}}(im)$ vanishes faster than any power of $1/T$ as $T\to\infty$. This argument explains
why the \qedl\ FV correction $\delta^{\gamma,\qedl}_{\rm FV}$ shows negligible dependence on $T$
for the values of $mT$ relevant to \figref{dgamma}.  Indeed, using routing A, the unique single-particle pole 
in $\tilde G_{\rm cont}$ near $p_0=im$ implies that the leading $T$-dependence in \qedl\ is suppressed by
a factor of $\exp(-mT)$.\footnote{Note that the extra term
$\delta^{{\rm t},+}_{\rm FV}(mL,mT)$ for \qedtl, given in \eq{tadpole+}, is  not negligible for any
of our data.}  

However, the Borsanyi {\it et al.}\ argument does not apply in general for routings that generate complicated $p_0$ dependence away
from the discrete points $2\pi\ell/T$.  In particular, the argument cannot be use to conclude that routing-dependent 
differences in  $\sigma_{{\rm cont}}(im)$ are similarly suppressed by $\exp(-mT)$ and therefore negligible for values  of $mT$ used in our computation.
To see this, we look at a simple example with
\be\eqn{integrand-example}
\cI(k_0,\vec k,p_0)  = \frac{1}{(k_0+p_0)^2+\vec k^2 +m^2}. \\
\ee
With $p_0=iy$ and $y \le \omega_k$ (for fixed $\vec k$), there is a simple 
pole  in the upper half plane at $k_0=i(\omega_k-y)$.
 When $y=m$, the $n\ge 1$
terms in the sum give contributions (after integration over $k_0$) proportional to $\exp(-nT(\omega_k-m))$, while the
$n\le-1$ contributions are more highly suppressed for large $T$ since the pole in the lower half-plane is further from the axis.
Because $\vec k\not= 0$, all the terms in the sum over $\vec k$ indeed decay exponentially with $T$.  The
rate of decay, however, can be very small for large $L$, because the lowest momenta have magnitude
$2\pi/L$.  Thus it is not obvious that the difference between
finite $T$ and infinite $T$ can be neglected, even if one just focusses on $\sigma_{{\rm cont}}(im)$.
More importantly, there are poles in $\sigma_{{\rm cont}}(iy)$
for $y=\omega_k$ (for some $\vec k$) as the $k_0$ pole in $\cI$ moves down to the real axis.  These poles
mean that $-p_0^2 = y^2 = m^2 + \sigma_{{\rm cont}}(iy)$ can have multiple solutions, so there are
multiple poles in the momentum space propagator $\tilde G(p_0)$, as we have seen
in \figref{B-poles}.  If the higher poles are close to $y=m$ (as in \figref{B-poles} (middle)), or the
$y\approx m$ pole is absent entirely (as in \figref{B-poles} (right)), $\sigma_{{\rm cont}}(im)$ will have little
to do with the finite-$T$ mass correction.      Unfortunately,
it is likely that the generic case will be like routing B rather than routing A --- it seems to be an accident that
with routing A no $p_0$ dependence appears in the denominator of our integrand, so that 
 $\sigma^A_{{\rm cont}}(p_0)$ is a simple (quadratic) polynomial in $p_0$.  

\begin{figure}[h!]
  \centering
  {\includegraphics[width=0.6\linewidth]
  {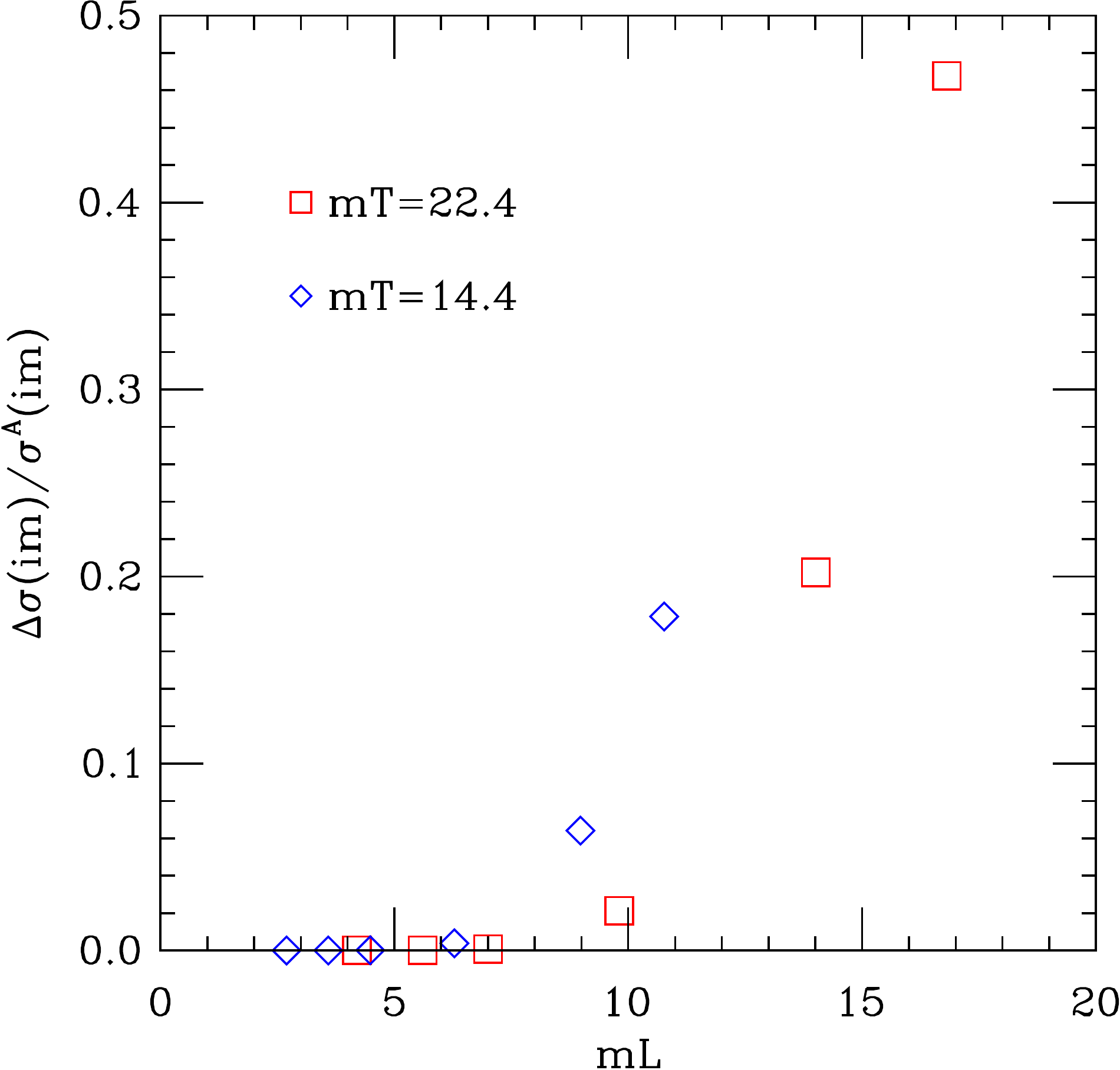}}
   \\
\vspace{-2mm}
\caption{\label{fig:shift-actual}
Relative size of the routing dependence of the self-energy contribution at $p_0=im$ of the sunset graph, 
for the same data at 0.12 fm that was presented in \protect{\figref{FV}}. The red squares are for ``kaon'' points; the 
blue diamonds, for ``pion'' points.   For the three leftmost diamonds and the three leftmost squares, where no
deviation from 0 is visible, the actual deviation varies between $\approx\!5\times10^{-5}$ and
$\approx\!1\times10^{-4}$.
}
\vspace{-0.08in}
\end{figure}

Finally, as an estimate of how important these effects are for the actual simulation data, we study
the routing dependence of the self energy at $p_0=im$, coming from the photon sunset graph and 00
component of the photon tadpole.   
(As elsewhere in this Appendix, the spatial part of the photon tadpole is not included
because it has no routing dependence.) 
In \figref{shift-actual},  we plot the ratio of $\Delta\sigma(im)/\sigma^A_{{\rm cont}}(im)$ {\it vs}\/.\ $mL$ 
for the data shown in 
\figref{FV} above.    As expected from the above discussion, the dependence 
increases with $mL$ for fixed $mT$, and 
decreases with $mT$ for fixed $mL$.  Note that, even though $mT$ is large, the routing dependence
is not negligible for much for our data, and approaches 50\% for the largest values of $mL$. 

\begin{thebibliography}{99}

\bibitem{Dashen:1969eg}
  R.~F.~Dashen,
  Phys.\ Rev.\  {\bf 183}, 1245 (1969).
  



 \bibitem{Aoki:2016frl} 
  S.~Aoki {\it et al.} [FLAG Collaboration],
  Eur.\ Phys.\ J.\ C {\bf 77}, no. 2, 112 (2017)
  doi:10.1140/epjc/s10052-016-4509-7
  [arXiv:1607.00299 [hep-lat]].
 
\bibitem{continuum-models}
See, for example, J.\ Bijnens and J.\ Prades, Nucl.\
Phys.\ B {\bf 490} (1997) 239; J.F.\ Donoghue and
A.F.\ Perez, Phys.\ Rev.\ D {\bf 55} (1997) 7075; and B.\
Moussallam, Nucl.\ Phys.\ B {\bf 504} (1997) 381.


  
\bibitem{RMP} 
A. Bazavov {\it et al.}, 
Rev. Mod. Phys. {\bf 82}, 1349 (2010) [arXiv:0903.3598].

  
\bibitem{qrat}
A.\ Bazavov et al., 
PoS CD09 (2009) 007, [arXiv:0910.2966]. 
A. Bazavov {\it et al}.
[MILC], PoS LATTICE {\bf 2009}, 079 (2009) [arXiv:0910.3618]. 


  \bibitem{Laiho:2011np} 
  J.~Laiho and R.~S.~Van de Water,
  PoS LATTICE {\bf 2011}, 293 (2011)
  [arXiv:1112.4861 [hep-lat]].
  
\bibitem{Aubin:2004fs} 
  C.~Aubin {\it et al.} [MILC Collaboration],
  Phys.\ Rev.\ D {\bf 70}, 114501 (2004)
  doi:10.1103/PhysRevD.70.114501
  [hep-lat/0407028].
 


\bibitem{Duncan:1996xy} 
  A.~Duncan, E.~Eichten and H.~Thacker,
  Phys.\ Rev.\ Lett.\  {\bf 76}, 3894 (1996)
  doi:10.1103/PhysRevLett.76.3894
  [hep-lat/9602005].

\bibitem{Blum:2007cy} 
  T.~Blum, T.~Doi, M.~Hayakawa, T.~Izubuchi and N.~Yamada,
  Phys.\ Rev.\ D {\bf 76}, 114508 (2007)
  doi:10.1103/PhysRevD.76.114508
  [arXiv:0708.0484 [hep-lat]].


  \bibitem{Blum:2010ym} 
  T.~Blum {\it et al.},
  Phys.\ Rev.\ D {\bf 82}, 094508 (2010)
  [arXiv:1006.1311 [hep-lat]].
  
  \bibitem{BMW11}
  A.~Portelli {\it et al.}, [BMW],
  PoS LATTICE {\bf 2011}, 136 (2011)
  [arXiv:1201.2787] and
  PoS LATTICE {\bf 2010}, 121 (2010)
  [arXiv:1011.4189].


    \bibitem{EM12}
  S.~Basak {\it et al.}  [MILC],
  PoS LATTICE {\bf 2012}, 137 (2012)
  [arXiv:1210.8157].

\bibitem{Basak:2013iw} 
  S.~Basak {\it et al.}  [MILC],
  PoS CD {\bf 12}, 030 (2013)
  [arXiv:1301.7137 [hep-lat]].


   
   

  \bibitem{Basak:2014vca} 
  S.~Basak {\it et al.} [MILC Collaboration],
  PoS LATTICE {\bf 2014}, 116 (2014)
  [arXiv:1409.7139 [hep-lat]];
  J.\ Phys.\ Conf.\ Ser.\  {\bf 640}, no. 1, 012052 (2015)
  doi:10.1088/1742-6596/640/1/012052
  [arXiv:1510.04997 [hep-lat]].
  
  \bibitem{Basak:2016jnn} 
  S.~Basak {\it et al.} [MILC Collaboration],
  PoS LATTICE {\bf 2015}, 259 (2016)
  doi:10.22323/1.251.0259
  [arXiv:1606.01228 [hep-lat]].
  
  \bibitem{Horsley:2015eaa} 
  R.~Horsley {\it et al.},
  J.\ Phys.\ G {\bf 43}, no. 10, 10LT02 (2016)
  doi:10.1088/0954-3899/43/10/10LT02
  [arXiv:1508.06401 [hep-lat]];
  JHEP {\bf 1604}, 093 (2016)
  doi:10.1007/JHEP04(2016)093
  [arXiv:1509.00799 [hep-lat]].
  
 
  
  \bibitem{Fodor:2016bgu} 
  Z.~Fodor {\it et al.},
  Phys.\ Rev.\ Lett.\  {\bf 117}, no. 8, 082001 (2016)
  doi:10.1103/PhysRevLett.117.082001
  [arXiv:1604.07112 [hep-lat]].
    
  \bibitem{deDivitiis:2013xla} 
  G.~M.~de Divitiis {\it et al.} [RM123 Collaboration],
  Phys.\ Rev.\ D {\bf 87}, no. 11, 114505 (2013)
  [arXiv:1303.4896 [hep-lat]].
  
 \bibitem{Giusti:2017dmp} 
  D.~Giusti, V.~Lubicz, C.~Tarantino, G.~Martinelli, F.~Sanfilippo, S.~Simula and N.~Tantalo,
  Phys.\ Rev.\ D {\bf 95}, no. 11, 114504 (2017)
  doi:10.1103/PhysRevD.95.114504
  [arXiv:1704.06561 [hep-lat]].

\bibitem{Portelli:2015wna} 
  A.~Portelli,
  PoS LATTICE {\bf 2014}, 013 (2015)
  [arXiv:1505.07057 [hep-lat]].




\bibitem{Bijnens:2006mk}
  J.~Bijnens and N.~Danielsson,
  Phys.\ Rev.\  D {\bf 75}, 014505 (2007)
  [arXiv:hep-lat/0610127].


  
\bibitem{Urech:1994hd} 
  R.~Urech,
  Nucl.\ Phys.\ B {\bf 433}, 234 (1995)
  doi:10.1016/0550-3213(95)90707-N
  [hep-ph/9405341].  
  



  \bibitem{Gasser-Leutwyler}
  J.\ Gasser and H.\ Leutwyler, 
  Nucl.\ Phys.\ B250 (1985) 465.

\bibitem{Bazavov:2012xda} 
  A.~Bazavov {\it et al.} [MILC Collaboration],
  Phys.\ Rev.\ D {\bf 87}, no. 5, 054505 (2013)
  doi:10.1103/PhysRevD.87.054505
  [arXiv:1212.4768 [hep-lat]].


\bibitem{Bernard:2010qd} C.~Bernard and E.~D.~Freeland,
  PoS LATTICE {\bf 2010}, 084 (2010)
  [arXiv:1011.3994].


\bibitem{Bazavov:2014wgs} 
  A.~Bazavov {\it et al.} [Fermilab Lattice and MILC Collaborations],
  Phys.\ Rev.\ D {\bf 90}, no. 7, 074509 (2014)
  doi:10.1103/PhysRevD.90.074509
  [arXiv:1407.3772 [hep-lat]].
  
\bibitem{basak} S. Basak {\it et al.} [MILC], PoS LATTICE {\bf 2008}, 127 (2008) [arXiv:0812.4486];
A.~Torok {\it et al.} [MILC]
  PoS LATTICE {\bf 2010}, 127 (2010).

 \bibitem{Durr:2010vn} 
   S.~Durr {\it et al.},
  Phys.\ Lett.\ B {\bf 701}, 265 (2011)
  [arXiv:1011.2403 [hep-lat]];
  JHEP {\bf 1108}, 148 (2011)
  [arXiv:1011.2711 [hep-lat]].


\bibitem{Gottlieb:1987mq} 
  S.~A.~Gottlieb, W.~Liu, D.~Toussaint, R.~L.~Renken and R.~L.~Sugar,
  Phys.\ Rev.\ D {\bf 35}, 2531 (1987).
  doi:10.1103/PhysRevD.35.2531

\bibitem{Duane:1985ym} 
  S.~Duane,
  Nucl.\ Phys.\ B {\bf 257}, 652 (1985).
  doi:10.1016/0550-3213(85)90369-4


\bibitem{Duane:1986iw} 
  S.~Duane and J.~B.~Kogut,
  Nucl.\ Phys.\ B {\bf 275}, 398 (1986).
  doi:10.1016/0550-3213(86)90606-1

\bibitem{Clark:2006fx} 
  M.~A.~Clark and A.~D.~Kennedy,
  Phys.\ Rev.\ Lett.\  {\bf 98}, 051601 (2007)
  doi:10.1103/PhysRevLett.98.051601
  [hep-lat/0608015].

\bibitem{Duane:1987de} 
  S.~Duane, A.~D.~Kennedy, B.~J.~Pendleton and D.~Roweth,
  Phys.\ Lett.\ B {\bf 195}, 216 (1987).
  doi:10.1016/0370-2693(87)91197-X

\bibitem{Sexton:1992nu} 
  J.~C.~Sexton and D.~H.~Weingarten,
  Nucl.\ Phys.\ B {\bf 380}, 665 (1992).
  doi:10.1016/0550-3213(92)90263-B

\bibitem{Kennedy:1998cu} 
 I.~Horvath,  A.~D.~Kennedy,  and S.~Sint,
  Nucl.\ Phys.\ Proc.\ Suppl.\  {\bf 73}, 834 (1999)
  doi:10.1016/S0920-5632(99)85217-7
  [hep-lat/9809092].

\bibitem{Hasenbusch:2001ne} 
  M.~Hasenbusch,
  Phys.\ Lett.\ B {\bf 519}, 177 (2001)
  doi:10.1016/S0370-2693(01)01102-9
  [hep-lat/0107019].

\bibitem{Omelyan:2002E1}
I.P.~Omelyan, I.M.~Mryglod, and R.~Folk,
2002. Phys.\ Rev.\ E {\bf 66}, 026701 (2002).

\bibitem{Takaishi:2005tz} 
  T.~Takaishi and P.~de Forcrand,
  Phys.\ Rev.\ E {\bf 73}, 036706 (2006)
  doi:10.1103/PhysRevE.73.036706
  [hep-lat/0505020].
    
\bibitem{quda} M.A.\ Clark {\it et al.}, Comput.\ Phys.\ Commun.\ 181, 1517 (2010) [arXiv:0911.3191]; R.\ Babich {\it et al.}, 
Intl.\ Conf.\ for High Perf.\ Computing, Networking, Storage and Analysis (SC), 2011 
[arXiv:1109.2935].

\bibitem{shi:ipdps11}
G.~Shi, S.~Gottlieb, A.~Torok and V.~Kindratenko, {\it {Design of MILC lattice
  QCD application for GPU clusters }},  in {\em {Proceedings of the 2011 IEEE
  Parallel and Distributed Processing Symposium, IPDPS '11. IEEE, 2011 }},
  2011.

\bibitem{shi:ipdps12}
G.~Shi, R.~Babich, M.~Clark, B.~Jo\'o, S.~Gottlieb and V.~Kindratenko, {\it
  {The Fat-Link Computation On Large GPU Clusters for Lattice QCD}},  in {\em
  {Proceedings of the 2012 IEEE Parallel and Distributed Processing Symposium,
  IPDPS '12. IEEE, 2012 }}, 2012.

\bibitem{Babich:2011np}
R.~Babich, M.~A. Clark, B.~Joo, G.~Shi, R.~C. Brower and S.~Gottlieb, {\it
  {Scaling Lattice QCD beyond 100 GPUs}},  in {\em {SC11 International
  Conference for High Performance Computing, Networking, Storage and Analysis
  Seattle, Washington, November 12-18, 2011}}, 2011
[arXiv:1109.2935].

\bibitem{Babich:2011zz}
R.~Babich, R.~Brower, M.~Clark, S.~Gottlieb, B.~Joo and G.~Shi, {\it {Progress
  on the QUDA code suite}},  {\em PoS} {\bf LATTICE2011} (2011) 033.

\bibitem{Bernard:2000gd} 
  C.~W.~Bernard {\it et al.},
  Phys.\ Rev.\ D {\bf 62}, 034503 (2000)
  doi:10.1103/PhysRevD.62.034503
  [hep-lat/0002028].
  
 \bibitem{Sommer:1993ce} 
  R.~Sommer,
  Nucl.\ Phys.\ B {\bf 411}, 839 (1994)
  doi:10.1016/0550-3213(94)90473-1
  [hep-lat/9310022].

\bibitem{Bazavov:2011aa} 
  A.~Bazavov {\it et al.} [Fermilab Lattice and MILC Collaborations],
  Phys.\ Rev.\ D {\bf 85}, 114506 (2012)
  doi:10.1103/PhysRevD.85.114506
  [arXiv:1112.3051 [hep-lat]].


\bibitem{Marinari:1981qf}
  E.~Marinari, G.~Parisi and C.~Rebbi,
  Nucl.\ Phys.\ B {\bf 190}, 734 (1981).

\bibitem{Follana:2004sz}
E.~Follana, A.~Hart, and C.T.H. Davies.
Phys.\ Rev.Lett.\ {\bf 93}, 241601 (2004) [hep-lat/0406010].

\bibitem{Durr:2004as}
S.\ D{\"u}rr, C.\ Hoelbling, and U.\ Wenger.
Phys.\ Rev.\ D{\bf 70}, 094502 (2004)  [hep-lat/0406027].

\bibitem{Shamir:2004zc}
Y.\ Shamir.
Phys.\ Rev.\ D{\bf 71}, 034509 (2005)  [hep-lat/0412014].



\bibitem{Durr:2004ta}
S.\ D{\"u}rr and C.\ Hoelbling.
Phys.\ Rev.\ D{\bf 71}, 054501 (2005).

\bibitem{Wong:2004nk}
K.\ Y.\ Wong and R.M.\ Woloshyn.
Phys.\ Rev.\ D{\bf 71}, 094508 (2005).



\bibitem{Prelovsek:2005rf}
S.\ Prelovsek.
Phys.\ Rev.\ D{\bf 73}, 014506 (2006).

\bibitem{Bernard:2006zw}
C.~Bernard,
Phys.\ \ Rev.\ D{\bf 73}, 114503 (2006) [hep-lat/0603011].

\bibitem{Bernard:2006ee}
C.\ Bernard, M.\ Golterman, and Y.\ Shamir.
Phys.\ Rev.\ D{\bf 73}, 114511 (2006) [hep-lat/0604017].

\bibitem{Durr:2006ze}
S.\ D{\"u}rr and C.\ Hoelbling.
Phys.\ Rev.\ D{\bf 74}, 014513 (2006).

\bibitem{Shamir:2006nj}
Y.\ Shamir.
Phys.\ Rev.\ D{\bf 75}, 054503 (2007).


\bibitem{Bernard:2007qf}
C.\ Bernard, C.\ DeTar, Z.\ Fu, and S.\ Prelovsek.
Phys.\ Rev.\ D{\bf 76}, 094504 (2007).


\bibitem{Donald:2011if}
G.\ Donald, C.T.H.\ Davies, E.\ Follana, and A.S.\
  Kronfeld.
Phys.\ Rev.\ D{\bf 84}, 054504 (2011).

\bibitem{Lee:1999zxa} 
  W.~J.~Lee and S.~R.~Sharpe,
  Phys.\ Rev.\ D {\bf 60}, 114503 (1999)
  doi:10.1103/PhysRevD.60.114503
  [hep-lat/9905023].

\bibitem{Aubin:2003mg}
  C.~Aubin and C.~Bernard,
  Phys.\ Rev.\ D {\bf 68}, 034014 (2003)
  [hep-lat/0304014].
   
  \bibitem{Bernard:1993sv} 
  C.~W.~Bernard and M.~F.~L.~Golterman,
  Phys.\ Rev.\ D {\bf 49}, 486 (1994)
  doi:10.1103/PhysRevD.49.486
  [hep-lat/9306005].

\bibitem{Sharpe:2000bc} 
  S.~R.~Sharpe and N.~Shoresh,
  Phys.\ Rev.\ D {\bf 62}, 094503 (2000)
  doi:10.1103/PhysRevD.62.094503
  [hep-lat/0006017].
  
  
  
\bibitem{Sharpe:2001fh} 
  S.~R.~Sharpe and N.~Shoresh,
  Phys.\ Rev.\ D {\bf 64}, 114510 (2001)
  doi:10.1103/PhysRevD.64.114510
  [hep-lat/0108003].

  
\bibitem{Damgaard:2000gh}
  P.~H.~Damgaard and K.~Splittorff,
  Phys.\ Rev.\  D {\bf 62}, 054509 (2000)
  [arXiv:hep-lat/0003017].
  
   \bibitem{AUBIN-BERNARD-REPLICA}
 C.\ Aubin and C.\ Bernard,
Nucl.\ Phys.\ {\bf B} (Proc.\ Suppl.) {\bf 129-130}, 182 (2004) [arXiv:hep-lat/0308036].

\bibitem{Sharpe_quarkflow}
  S.~R.~Sharpe,
  Phys.\ Rev.\  D {\bf 46}, 3146 (1992)
  [arXiv:hep-lat/9205020];
  S.~R.~Sharpe,
  Nucl.\ Phys.\ Proc.\ Suppl.\  {\bf 17}, 146 (1990).
   
  
\bibitem{Aubin:2004ck} 
  C.~Aubin {\it et al.} [HPQCD and MILC and UKQCD Collaborations],
  Phys.\ Rev.\ D {\bf 70}, 031504 (2004)
  doi:10.1103/PhysRevD.70.031504
  [hep-lat/0405022].
  
  \bibitem{Lepage:1992xa} 
  G.~P.~Lepage and P.~B.~Mackenzie,
  Phys.\ Rev.\ D {\bf 48}, 2250 (1993)
  doi:10.1103/PhysRevD.48.2250
  [hep-lat/9209022].
  

    \bibitem{Mason:2005bj} 
  Q.~Mason {\it et al.} [HPQCD Collaboration],
  Phys.\ Rev.\ D {\bf 73}, 114501 (2006)
  doi:10.1103/PhysRevD.73.114501
  [hep-ph/0511160].

 \bibitem{Borsanyi:2013lga} 
  S.~Borsanyi {\it et al.} [Budapest-Marseille-Wuppertal Collaboration],
  Phys.\ Rev.\ Lett.\  {\bf 111}, no. 25, 252001 (2013)
  doi:10.1103/PhysRevLett.111.252001
  [arXiv:1306.2287 [hep-lat]].

\bibitem{Borsanyi:2014jba} 
  S.~Borsanyi {\it et al.},
  Science {\bf 347}, 1452 (2015)
  doi:10.1126/science.1257050
  [arXiv:1406.4088 [hep-lat]].

   \bibitem{Hayakawa:2008an} 
S.~Uno and  M.~Hayakawa,
  Prog.\ Theor.\ Phys.\  {\bf 120}, 413 (2008)
  [arXiv:0804.2044 [hep-ph]].
  

\bibitem{Davoudi:2014qua}
  Z.~Davoudi and M.~J.~Savage,
  Phys.\ Rev.\ D {\bf 90}, no. 5, 054503 (2014)
  doi:10.1103/PhysRevD.90.054503
  [arXiv:1402.6741 [hep-lat]].

  \bibitem{Fodor:2015pna} 
  Z.~Fodor, C.~Hoelbling, S.~D.~Katz, L.~Lellouch, A.~Portelli, K.~K.~Szabo and B.~C.~Toth,
  Phys.\ Lett.\ B {\bf 755}, 245 (2016)
  doi:10.1016/j.physletb.2016.01.047
  [arXiv:1502.06921 [hep-lat]].
  
  \bibitem{Lee:2015rua} 
  J.~W.~Lee and B.~C.~Tiburzi,
  Phys.\ Rev.\ D {\bf 93}, no. 3, 034012 (2016)
  doi:10.1103/PhysRevD.93.034012
  [arXiv:1508.04165 [hep-lat]].
  
  \bibitem{Bernard:2001yj} 
  C.~Bernard [MILC Collaboration],
  Phys.\ Rev.\ D {\bf 65}, 054031 (2002)
  [hep-lat/0111051].
  
  
   \bibitem{Basel}
  See, for instance, the {\it Wikipedia} article on the ``Basel problem,'' 
  {\tt https://en.wikipedia.org/wiki/Basel\_problem}.
  
     \bibitem{VEGAS}
  G.P.\ Lepage, 
  J.\  Comp.\ Phys.\ {\bf27}, 192 (1978).
  
  \bibitem{Aubin:2004he} 
  C.~Aubin {\it et al.} [MILC Collaboration],
  Nucl.\ Phys.\ Proc.\ Suppl.\  {\bf 140}, 231 (2005)
  doi:10.1016/j.nuclphysbps.2004.11.174
  [hep-lat/0409041].

  \bibitem{Wolfram}
  See, for instance, the  {\it Wolfram MathWorld} entry on Jacobi theta functions, 
  {\tt http://mathworld.wolfram.com/JacobiThetaFunctions.html}.

\bibitem{Endres:2015gda} 
  M.~G.~Endres, A.~Shindler, B.~C.~Tiburzi and A.~Walker-Loud,
  Phys.\ Rev.\ Lett.\  {\bf 117}, no. 7, 072002 (2016)
  doi:10.1103/PhysRevLett.117.072002
  [arXiv:1507.08916 [hep-lat]].

\bibitem{PVALUES}
See for example D.F.~Morrison, ``Multivariate Statistical Methods'', McGraw-Hill, 1967;
R.A.~Johnson and D.W.~Wichern, ``Applied Multivariate Statistical Analysis'', Prentice-Hall, 1982;
D.~Toussaint and W.~Freeman, arXiv:0808.2211.

\bibitem{Olive:2016xmw}
C.~Patrignani {\it et al.},
(Review of Particle Physics),
Chin. Phys. {\bf C40}, 100001 (2016).

\bibitem{Bazavov:2017lyh}
  A.~Bazavov {\it et al.},
  arXiv:1712.09262 [hep-lat].
  
  \bibitem{Bazavov:2018omf} 
  A.~Bazavov {\it et al.} [Fermilab Lattice and MILC and TUMQCD Collaborations],
  arXiv:1802.04248 [hep-lat].

 \bibitem{Carrasco:2014cwa} 
  N.~Carrasco {\it et al.} [European Twisted Mass Collaboration],
  Nucl.\ Phys.\ B {\bf 887}, 19 (2014)
  doi:10.1016/j.nuclphysb.2014.07.025
  [arXiv:1403.4504 [hep-lat]].

\bibitem{Zhou:2014gga} 
  R.~Zhou {\it et al.}\ [MILC Collaboration],
  PoS LATTICE {\bf 2014}, 024 (2014)
  doi:10.22323/1.214.0024
  [arXiv:1411.4115 [hep-lat]]; Y.~Liu {\it et al.}\ [MILC Collaboration], Proceedings of Lattice 2018, to appear.

\bibitem{Gasquet:1999}
See, for instance, C.\ Gasquet and P.\ Witomski, {\it Fourier Analysis and Applications} (translated by
R.\ Ryan), Springer, 1999, p. 172.
  


\end{thebibliography}
\end{document}